\documentclass[prd,preprintnumbers,aps,preprint,amsmath,amssymb,showkeys,superscriptaddress,
showpacs,floatfix,nofootinbib,unsortedaddress]{revtex4}

\usepackage{amssymb, bm, amsmath, graphicx, subfigure, physymb}
\usepackage[colorlinks=true,linktocpage=true,linkcolor=blue,citecolor=blue]{hyperref}
\usepackage[usenames,dvipsnames]{color}
\usepackage{subfigure}
\usepackage{slashed}
\usepackage{multirow,array}
\usepackage[utf8]{inputenc}
\usepackage{array}
\usepackage{enumerate}
\usepackage{amsfonts}
\usepackage{dsfont}
\usepackage{mathtools}
\usepackage{mathrsfs}
\usepackage{slashed}
\usepackage{latexsym}
\usepackage{hyperref}
\usepackage{epstopdf} 
\usepackage{bbm}

\pdfoutput=1
\newcommand{\beq}{\begin{equation}}
\newcommand{\eeq}{\end{equation}}
\newcommand{\bqa}{\begin{eqnarray}}
\newcommand{\eqa}{\end{eqnarray}}
\newcommand{\ben}{\begin{eqnarray*}}
\newcommand{\een}{\end{eqnarray*}}
\newcommand{\bes}{\begin{subequations}}
\newcommand{\ees}{\end{subequations}}

\newcommand{\beal}{\begin{align}}

\newcommand{\amu}{\alpha_\mu}

\newcommand{\dhd}{{\textstyle d}
\lower.03ex\hbox{\kern-0.38em$^{\scriptstyle-}$}\kern-0.05em{}}
\newcommand{\dbar}{{\textstyle \delta}
\lower.03ex\hbox{\kern-0.38em$^{\scriptstyle-}$}\kern-0.05em{}}
\newcommand{\half}{{1\over 2}}

\newcommand{\barc}{{\bar c}}

\newcommand{\bark}{{\bar k}}

\newcommand{\barq}{{\bar q}}

\newcommand{\barz}{{\bar z}}
\newcommand{\bary}{{\bar y}}

\newcommand{\barD}{{\bar D}}

\newcommand{\gPM}{g^{+-}}

\newcommand{\thalf}{\tfrac{1}{2}}
\newcommand{\Dtwo}[2]{D_2 ( \ul #1 , \ul #2 )}
\newcommand{\Dtwohat}[2]{\hat{D}_2 ( \ul #1 , \ul #2 )}

\newcommand{\Dfourhat}[4]{\hat{D}_4 ( \ul #1 , \ul #2 , \ul #3 , \ul #4 )}

\newcommand{\Dsixhat}[6]{\hat{D}_6 ( \ul #1 , \ul #2 , \ul #3 , \ul #4, \ul #5 , \ul #6 )}

\newcommand{\Deighthat}[8]{\hat{D}_8 ( \ul #1 , \ul #2 , \ul #3 , \ul #4, \ul #5 , \ul #6, \ul #7, \ul #8 )}
\newcommand{\dtwo}{\mathbbm{d}^2}

\newcommand{\ul}[1]{\underline{ #1}}
\newcommand{\ord}[1]{\mathcal{O} \left( #1 \right)}
\newcommand{\tr}{\mathrm{tr}}
\newcommand{\pp}{ {\prime \prime} }

\begin{document}

\title{Multiparticle Production at Mid-Rapidity in the Color-Glass Condensate}
\author{Mauricio Martinez}
\email[Email: ]{mmarti11@ncsu.edu}
\affiliation{Department of Physics, North Carolina State University, Raleigh, NC 27695, USA}
\author{Matthew D. Sievert}
\email[Email: ]{sievertmd@lanl.gov}
 \affiliation{Theoretical Division, Los Alamos National Laboratory, Los Alamos, NM 87545, USA}
\author{Douglas E. Wertepny}
\email[Email: ]{douglas.wertepny@usc.es}
\affiliation{Departamento de F\'isica de Part\'iculas and IGFAE, Universidade de Santiago de Compostela, 15782
Santiago de Compostela, Galicia-Spain}

\begin{abstract}
In this paper, we compute a number of cross sections for the production of multiple particles at mid-rapidity in the semi-dilute / dense regime of the color-glass condensate (CGC) effective field theory.  In particular, we present new results for the production of two quark-antiquark pairs (whether the same or different flavors) and for the production of one quark-antiquark pair and a gluon.  We also demonstrate the existence of a simple mapping which transforms the cross section to produce a quark-antiquark pair into the corresponding cross section to produce a gluon, which we use to obtain various results and to cross-check them against the literature.  We also discuss hadronization effects in the heavy flavor sector, writing explicit expressions for the production of various combinations of $D$ and $\barD$ mesons, $J/\psi$ mesons, and light hadrons.  The various multiparticle cross sections presented here contain a wealth of information and can be used to study heavy flavor production, charge-dependent correlations, and ``collective'' flow phenomena arising from initial-state dynamics.
\end{abstract}

\date{\today}
\maketitle

%
\section{Introduction}
\label{sec:intro} 
%

Correlations in the production of multiple soft or semi-hard particles in the mid-rapidity region of hadronic collisions are important probes of novel phenomena in quantum chromodynamics (QCD).  Whether in proton-proton (pp), proton-nucleus (pA), or heavy-ion (AA) collisions, multiparticle production reflects the many-body correlations generated by QCD.  In pp collisions, such correlations may be produced by quantum evolution through Dokshitzer-Gribov-Lipatov-Altarelli-Parisi (DGLAP) evolution~\cite{Dokshitzer:1977sg,Gribov:1972ri,Altarelli:1977zs} or by a range of higher-order corrections to the hard part (see, e.g. \cite{Ellis:1978ty,Curci:1980uw}) or small-$x$ evolution, including linear Balitsky-Fadin-Kuraev-Lipatov (BFKL) evolution~\cite{Kuraev:1977fs,Balitsky:1978ic}, nonlinear Balitsky-Kovchegov (BK) evolution~\cite{Kovchegov:1999yj,Balitsky:1995ub} and Jalilian-Marian-Iancu-McLerran-Weigert-Leonidov-Kovner (JIMWLK) evolution \cite{JalilianMarian:1996xn,JalilianMarian:1997gr,Iancu:2001ad}.  The resulting correlations are sensitive probes of the perturbative hard vertex and of the strongly-ordered emission structure of the evolution equations.  In pA collisions, these higher-order and evolution corrections are augmented by a new set of dynamical correlations arising from the enhancement of multiple scattering in the high charge densities of the heavy nucleus, characterized by the color-glass condensate (CGC) effective field theory (see e.g. \cite{Kovchegov:2012mbw} and references therein).  The resulting correlations are sensitive probes of the multiple scattering dynamics, including the significant effects of Bose enhancement in the strong gluon fields \cite{Altinoluk:2015uaa, Altinoluk:2018hcu}.  Finally, in AA collisions (as well as potentially in high-multiplicity pp and pA collisions), all these initial state correlations are modified and complemented by the final-state dynamics of a strongly-coupled quark-gluon plasma (QGP) phase.

A detailed characterization of multiparticle production in the strong color fields of the CGC is especially important in trying to differentiate intial-state effects from the final-state dynamics of the QGP, where strongly-coupled interactions lead to substantial many-body correlations among soft and semi-hard particles.  The canonical measures of this collective flow are the cumulants of azimuthal anisotropies $v_n \{ m \}$\cite{Luzum:2013yya}, with correlations among increasing numbers of particles reflected in higher values $m$ of the cumulants.  Other landmark properties believed to be possible in the QGP phase include the onset of novel transport mechanisms associated with the axial anomaly: the chiral magnetic effect, the chiral separation effect, the chiral vortical effect, and the chiral magnetic wave~\footnote{For a review on chiral magnetic and vortical effects in high-energy nuclear collisions we refer to the reader to Ref.~\cite{Kharzeev:2015znc}.}.  Signatures for all of these novel chiral dynamics are encoded in multiparticle correlations, often charge dependent, such as the same-sign and opposite-sign correlators $\gamma_{112}$ and $\gamma_{123}$  \cite{Sirunyan:2017quh}.  For all of these critical signatures of the quark-gluon plasma, it is essential to disentangle the ``background'' contributions coming from initial-state mechanisms to better quantify the properties of the QGP and improve the chances of discovering such novel anomalous dynamics.  Accordingly, a substantial effort has been made in recent years to compute multiparticle production in the CGC framework.

The purest realization of the CGC formalism is in the ``dilute / dense'' framework, in which density-enhanced effects of the ``dilute projectile'' are kept only to lowest order, while density-enhanced corrections in the ``dense target'' are resummed to all orders.  Few-particle production has been studied in the dilute / dense framework from the earliest days of the CGC formalism, starting with the inclusive single-gluon production cross section $d\sigma^G$ \cite{Kovchegov:1998bi,Blaizot:2004wu,Kovchegov:2001sc,Dumitru:2001ux,Kopeliovich:1998nw,JalilianMarian:2005jf} at mid-rapidity and followed shortly thereafter by the inclusive cross section $d\sigma^{q \barq}$ of a single $q \barq$ pair via gluon pair production \cite{Levin:1991ry,Blaizot:2004wv, Kovchegov:2006qn,Gelis:2003vh,Fujii:2006ab,Gelis:2005pb,Gelis:2004jp,Gelis:2015eua,Tanji:2017xiw}.  Corrections to these production channels were also considered in the form of small-$x$ evolution corrections \cite{Kovchegov:2006qn,Fujii:2006ab}. However, a detailed computation of higher multiparticle production cross sections in the dilute / dense framework becomes increasingly difficult due to the proliferation of ways another soft particle could be radiated from a pre-existing one.

A significant step toward overcoming this barrier was made through the development of the ``semi-dilute / dense'' framework \cite{Kovchegov:2012nd}.  This regime is designed to fill the gap between the dilute / dense regime, in which the projectile charge density is kept only to lowest order, and the dense / dense regime, where both projectile and target densities must be simultaneously resummed to all orders.  The semi-dilute / dense framework is appropriate for ``heavy-light ion collisions'' intermediate to, say, pPb and PbPb collisions.  For a collision between one light ion and one heavy ion, such as CuAu collisions, it is possible to construct a regime in which the large target density is resummed to all orders while corrections from the projectile density are calculated order by order in perturbation theory.  In the semi-dilute / dense framework, higher-order corrections which are enhanced by the projectile density are more important than genuine quantum corrections.  Formally, for a dense target nucleus with $A$ nucleons and a semi-dilute projectile nucleus with $a$ nucleons, the semi-dilute / dense regime can be quantified by the hierarchy of scales
\begin{subequations}
\begin{align}
\alpha_s^2 A^{1/3} \sim \ord{1} \\
\alpha_s \ll \alpha_s^2 a^{1/3} \ll 1 ,
\end{align}
\end{subequations}
or equivalently, in term of the saturation momenta $Q_{s, p}$ and $Q_{s, t}$ of the projectile and target respectively,
\begin{align} \label{e:heavylight}
\Lambda_{QCD}^2 \ll Q_{s, p}^2 \ll Q_{s, t}^2 .
\end{align}

With the help of the semi-dilute / dense framework, a number of significant steps have been taken in recent years toward the calculation of genuine multiparticle production in the CGC framework.  The key simplification that makes this possible in the semi-dilute / dense framework is that the {\it independent} emission of new soft particles from the high-density projectile becomes dominant over emission from the pre-existing system of soft particles.  As such, the first observable computed in the semi-dilute / dense framework was the production cross section $d\sigma^{G G}$ for two soft gluons \cite{Kovchegov:2012nd, Kovner:2012jm}.  A similar effort was made toward determining the production cross section $d\sigma^{q q}$ for two quarks coming from separate $q \barq$ pairs, with a partial calculation having been performed in Ref.~\cite{Altinoluk:2016vax} emphasizing the new role played by Fermi-Dirac quantum statistics among the two pairs.  This calculation later formed the basis of the partial calculation of the cross section $d\sigma^{q q G}$ for two quark / antiquark pairs plus a gluon, with the intent of studying the CGC contribution to the same-sign correlators $\gamma_{112} \, , \, \gamma_{123}$ \cite{Kovner:2017gab}.  It should be emphasized that in this important calculation \cite{Kovner:2017gab}, only correlations generated at the level of the wave functions were taken into account, without including the effects of multiple scattering that translate these wave functions into actual production cross sections.  And very recently, a new attempt has been made to extend these calculations to the third order in the projectile charge density through the computation of the triple-gluon production cross section $d\sigma^{G G G}$ \cite{Altinoluk:2018ogz}.  Other notable developments in soft multiparticle production include the identification of Bose enhancement as a driving mechanism of the Ridge~\cite{Altinoluk:2015uaa} in double-gluon production \cite{Altinoluk:2015uaa, Altinoluk:2018hcu}, the calculation of the soft double-photon cross section $d\sigma^{\gamma \gamma}$ \cite{Kovner:2017vro}, and the realization that soft double-pair production can be used to probe the gluon Wigner distribution with Weizs{\"a}cker-Williams gauge structure \cite{Boussarie:2018zwg}.  A variant of the semi-dilute / dense power counting can be found in the form of the lowest-order ``glasma graph'' calculations, which have been used to calculate multiparticle correlations such as the triple-gluon cross section \cite{Dusling:2009ar}.

Other important developments in the calculation of multiparticle production in the CGC formalism have emphasized production in the forward regime, where the ``hybrid factorization'' framework makes it possible to rigorously relate the particle production cross sections to collinear parton distribution functions in the (semi-)dilute projectile, dressed with the effects of multiple scattering in the dense target~\cite{Dumitru:2005gt,Altinoluk:2011qy,Altinoluk:2014eka,Altinoluk:2015vax}.  In this approach, observables such as forward double valence-quark production cross sections $d\sigma^{q_v q_v}$ \cite{Kovner:2017ssr}, forward triple valence-quark production cross sections $d\sigma^{q_v q_v q_v}$\cite{Kovner:2018vec}, and forward valence-quark + photon + gluon production \cite{Altinoluk:2018uax} have been calculated.  Similar studies of quadruple valence-quark production cross sections $d\sigma^{q_v q_v q_v q_v}$ have also been considered in a ``parton model'' description \cite{Dusling:2017dqg, Dusling:2017aot} without the benefit of an underlying hybrid factorization.  Other recent work on the subject also includes the demonstration \cite{Kovchegov:2018jun} that two-gluon correlations can break the ``accidental'' back-to-back symmetry which occurs at lowest order and related phenomenology  \cite{Mace:2018yvl, Mace:2018vwq}.  And finally, in a recent work \cite{Martinez:2018ygo}, we have considered single- and double-pair production $d\sigma^{q \barq}$ and $d\sigma^{(q \barq) (q \barq)}$ in coordinate space as a means of initializing spatial corrections of conserved charges in the quark-gluon plasma.

In this paper, our primary goal is to systematically extend the calculation of multiparticle production at mid-rapidity in the semi-dilute / dense framework to higher orders.  One of the key results we will derive here for the first time is the complete expression at exact $N_c$ for the double-pair production cross section $d\sigma^{(q \barq) (q \barq)}$ in momentum space, as written in Eqs.~\eqref{e:dblxsec}, \eqref{e:saus2}, and \eqref{e:Pac2}.  This expression significantly generalizes the result obtained in Ref.~\cite{Altinoluk:2016vax} by including contributions that were intentionally omitted there, by working with exact $N_c$, and by keeping the multiple scattering corrections to all orders.  In a key conceptual development, we show in Eq.~\eqref{e:Gmap2} that it is possible to map the amplitude (and therefore, the cross section) for producing $q \barq$ pairs into the corresponding quantities for producing gluons.  Thus, we are able to directly map the double-pair cross section into the corresponding cross section $d\sigma^{(q \barq) G}$ to produce a quark / antiquark pair and a gluon.  This expression, as written in Eq.~\eqref{e:mixedxsec}, is also a new result.  We also perform a number of validations of this gluonic mapping, verifying explicitly that it correctly reproduces the known results for single- and double-gluon production from the literature.  

While the preceding results all reflect the final-state production of multiple partons, they also open the door to a substantial program of computing hadronic-level observables derived from them.  By convoluting the partonic-level results with the appropriate fragmentation functions or projection operators and long-distance matrix elements, we can translate these partonic-level cross sections to full hadronic cross sections.  The resulting hadronic observables can be used to rigorously study the correlations among same-sign and opposite-sign charged hadrons, open and hidden heavy-flavor hadrons, heavy-flavor vs light hadrons, and more.  The phenomenology based on these hadronic obserables will provide critical new insight into initial-state mechanisms for collective flow, quarkonium correlations, and charge-dependent correlations which form the background to anomalous chiral dynamics in the QGP.

This paper is organized as follows.  In Sec.~\ref{sec:setup} we construct the scattering amplitudes for the production of soft particles in momentum space, starting with the quark/antiquark production amplitude in Sec.~\ref{sec:qqbaramp} and deriving the mapping to the gluon production amplitude in Sec.~\ref{sec:Gamp}.  Then in Sec.~\ref{sec:singlexsec} we compute the production cross section for a single $q \barq$ pair in Sec.~\ref{sec:singlepair} and map it in Sec.~\ref{sec:singleG} to the well-known gluon production cross section to validate the gluonic mapping.  Then in Sec.~\ref{sec:dblxsec} we proceed to calculate the new cross sections for the production of two sets of soft particles: double $q \barq$ pair production in Sec.~\ref{sec:dblpair}, mixed $q \barq G$ production in Sec.~\ref{sec:pairgluon}, and double gluon production in Sec.~\ref{sec:dblgluon}.  The successful cross-check against the double-gluon production cross section in Sec.~\ref{sec:dblgluon} represents another validation of the gluonic mapping derived in Sec.~\ref{sec:singleG}.  In Sec.~\ref{sec:Heavy} we utilize the techniques enumerated in Ref.~\cite{Ma:2018bax} to translate our partonic-level cross sections into hadronic cross sections for the production of open and hidden heavy flavor as an illustration of how to straightforwardly apply the results derived here to hadronic observables.  Finally, we conclude in Sec.~\ref{sec:concl} by reiterating the primary new results and exploring the many opportunities for phenomenological applications and further theoretical development which this work provides.  In Appendix~\ref{app:Averaging} we provide details of the Gaussian color averaging used for the (semi-)dilute projectile, and in Appendix~\ref{app:WLines} we formulate some useful algebraic properties of Wilson lines in momentum space.

Throughout this paper, we denote longitudinal momenta in light-front coordinates $v^\pm \equiv \sqrt{\frac{\gPM}{2}} (v^0 \pm v^3)$ and transverse vectors by $\ul{v} \equiv (v_\bot^1 , v_\bot^2)$ with magnitudes $v_T \equiv | \ul{v} |$.  Different authors use different conventions for the light-front metric $\gPM$; we will use $\gPM = 1$, but it is also common to encounter $\gPM = 2$.

%
\section{Production Amplitudes for (Anti)Quark Pairs and Gluons}
\label{sec:setup}
%

%
\subsection{Quark / Antiquark Pair Production Amplitude}
\label{sec:qqbaramp} 
%

%
\begin{figure}
\includegraphics[width=0.7\textwidth]{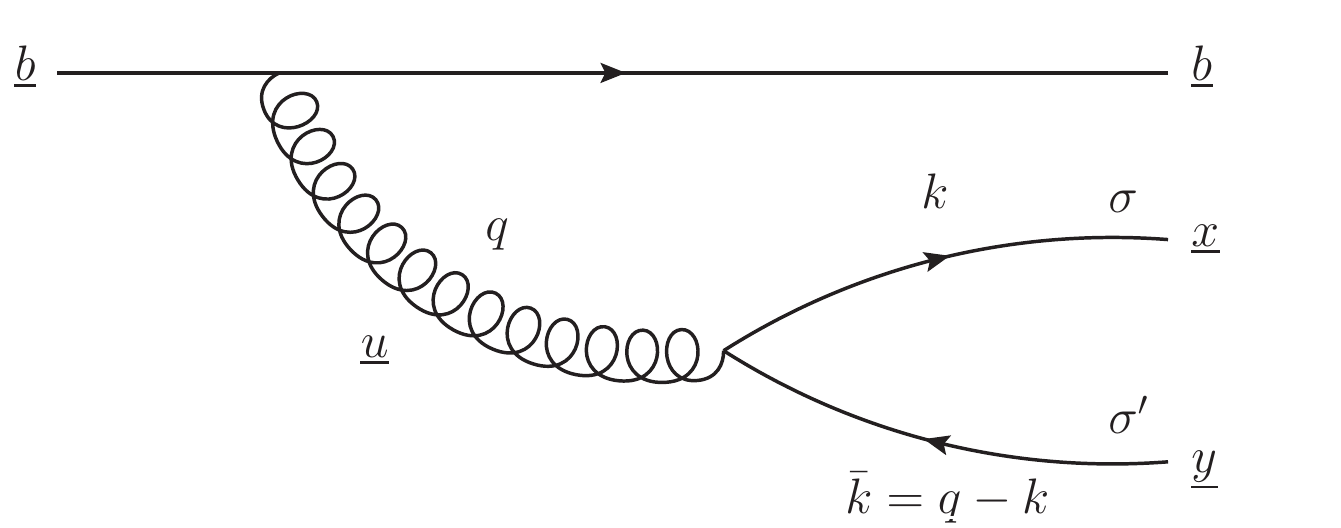}
\caption{The light-front wave functions to radiate a soft $q \barq$ pair at mid-rapidity from a valence source, shown here as a quark.} 
\label{f:Pair_WFs}
\end{figure}
%

The amplitude to radiate a soft quark/antiquark pair at mid-rapidity has been derived many times in the literature \cite{Blaizot:2004wv, Kovchegov:2006qn}.  In the notation of our previous work \cite{Martinez:2018ygo} as illustrated in Fig.~\ref{f:Pair_WFs}, we denote the light-front wave functions \cite{Brodsky:1997de, Lepage:1980fj} to radiate a soft $q \barq$ pair as $\psi_1 , \psi_2 , \psi_3$ corresponding to the various time orderings of the scattering in the target fields.  The term $\psi_1$ corresponds to scattering after the pair is created, $\psi_2$ to scattering after the gluon is emitted but before the pair is created, and $\psi_3$ to scattering before the pair is created.  The three wave functions are not all independent, but satisfy $\psi_1 + \psi_2 + \psi_3 = 0$, and the explicit expressions are given by
\begin{subequations} \label{e:WFmom}
\begin{align}
\psi_1 (\ul{q} , \ul{k} - \alpha \ul{q}) &= - 2 g \frac{\sqrt{\alpha (1-\alpha)}}{( \ul{k} - \alpha \ul{q})_T^2 + m^2 + \alpha(1-\alpha) q_T^2}
\notag \\ & \hspace{1cm} \times
\bigg\{ \delta_{\sigma \, , \, -\sigma'} \left[ \alpha + (1-2\alpha) \frac{\ul{q} \cdot \ul{k}}{q_T^2} 
- i \sigma' \, \frac{\ul{q} \times \ul{k}}{q_T^2} \right] 
-m \sigma' \, \delta_{\sigma \sigma'} \left[ \frac{\ul{q}_\bot^1}{q_T^2} 
- i \sigma' \, \frac{\ul{q}_\bot^2}{q_T^2} \right] \bigg\}
\\ \notag \\ 
\psi_2 (\ul{q} , \ul{k} - \alpha \ul{q}) &= -2 g \frac{\sqrt{\alpha (1-\alpha)}}{( \ul{k} - \alpha \ul{q})_T^2 + m^2}
\bigg\{ \delta_{\sigma \, , \, -\sigma'} \left[ - (1-2\alpha) \frac{\ul{q} \cdot (\ul{k} - \alpha \ul{q})}{q_T^2} 
+ i \sigma' \, \frac{\ul{q} \times (\ul{k} - \alpha \ul{q})}{q_T^2} \right] 
\notag \\ & \hspace{4cm} +
m \sigma' \, \delta_{\sigma \sigma'} \left[ \frac{\ul{q}_\bot^1}{q_T^2} 
- i \sigma' \, \frac{\ul{q}_\bot^2}{q_T^2} \right] \bigg\}
\\ \notag \\ 
\psi_3 (\ul{q} , \ul{k} - \alpha \ul{q}) &= - \psi_1 (\ul{q} , \ul{k} - \alpha \ul{q}) - 
\psi_2 (\ul{q} , \ul{k} - \alpha \ul{q}) ,
\end{align}
\end{subequations}
where $\alpha \equiv \frac{k^+}{k^+ + \bark^+}$ is the fraction of the
pair longitudinal momentum carried by the quark and $\ul{q} \equiv
\ul{k} + \ul{\bark}$ is the center-of-mass transverse momentum of the
$q \barq$ pair.  In \eqref{e:WFmom}, we have omitted the explicit
dependence of the wave functions on the momentum fraction $\alpha$ for
brevity.  Note also that, in comparison to Eqs.~(21) of
\cite{Martinez:2018ygo}, we have removed a factor of the coupling $g$
from the definition of the wave functions.  This corresponds to
absorbing this coupling constant into the scale $\mu^2$ defined in
\eqref{e:avgs} characterizing the sources of soft gluons.  

In terms of these wave functions, the single-pair amplitude summed over all time orderings is given in coordinate space by (see Eqs.~(30 - 31) of \cite{Martinez:2018ygo})
\begin{align} \label{e:coord1}
\mathcal{A} (\ul{x}, \ul{y}, \ul{b}) &= (V_{\ul b} t^a) \left[ \left( V_{\ul x} t^a V_{\ul y}^\dagger - V_{\ul b} t^a V_{\ul b}^\dagger \right) \psi_1 (\ul{u} - \ul{b} , \ul{x} - \ul{y}) \right.
\notag \\ &+
\left.
\left( V_{\ul u} t^a V_{\ul u}^\dagger - V_{\ul b} t^a V_{\ul b}^\dagger \right) \psi_2 (\ul{u} - \ul{b} , \ul{x} - \ul{y})
\right] ,
\end{align}
where, as labeled in Fig.~\ref{f:Pair_WFs}, $\ul{x}$, $\ul{y}$, and $\ul{b}$ are the final-state positions of the quark, antiquark, and valence quark, respectively, and $\ul{u} \equiv \alpha \ul{x} + (1-\alpha) \ul{y}$ is the center-of-mass position of the $q \bar q$ pair (equal to the gluon position).  The scattering of partons in the color fields of the target are described by Wilson lines in the fundamental or adjoint representations,
\begin{subequations} \label{e:Wdef}
\begin{align}
V_{\ul x} &\equiv \mathcal{P} \exp \left[ i g \int dx^+ \, A^- (x^+ , 0^- , \ul{x}) \right]
\\
U_{\ul x}^{a b} &\equiv \left( \mathcal{P} \exp \left[ i g \int dx^+ \, A_{adj}^- (x^+ , 0^- , \ul{x}) \right] \right)^{a b} ,
\end{align}
\end{subequations}
where we work in the $A^+ = 0$ light cone gauge.  With \eqref{e:coord1} written this way, the Wilson line $V_{\ul b}$ associated with the valence quark will always cancel against a corresponding one in the complex-conjugate amplitude.  

It is convenient to translate the specific model of the projectile as a distribution of valence quarks into a generic continuous charge density.  This can be accomplished by introducing the quantity $\rho^a (\ul{b})$, which loosely corresponds to the wave function of a color source in the projectile at position $\ul{b}$ which radiates a soft gluon with color $a$.  We can translate from the valence quark model of the projectile to the continuous color charge density by effectively replacing $(V_{\ul b} \, t^a) \rightarrow \rho^a (\ul{b})$.  (For another discussion of the translation between discrete and continuous charge distributions, see e.g. \cite{Kovchegov:2018jun}.)  With this change of notation, we can Fourier transform the buildling block \eqref{e:coord1} into momentum space to obtain
\begin{align} \label{e:coordFT}
\mathcal{A} (k , \bark) &= \int d^2 x \, d^2 y \, d^2 b \: e^{- i \ul{k} \cdot \ul{x}} \:
e^{- i \ul{\bark} \cdot \ul{y}} \: \rho^a (\ul{b})
\notag \\ & \hspace{-1cm} \times
\left[ \left( V_{\ul x} t^a V_{\ul y}^\dagger - V_{\ul b} t^a V_{\ul b}^\dagger \right) \psi_1 (\ul{u} - \ul{b} , \ul{x} - \ul{y}) +
\left( V_{\ul u} t^a V_{\ul u}^\dagger - V_{\ul b} t^a V_{\ul b}^\dagger \right) \psi_2 (\ul{u} - \ul{b} , \ul{x} - \ul{y})
\right] ,
\end{align}
where the momenta of the final-state quark and antiquark are $\ul{k}$ and $\ul{\bark}$, respectively.  Note that the Fourier factor for the valence quark cancels because its position $\ul{b}$ is the same in the initial and final states under the eikonal approximation.
\footnote{At first glance, the amplitude \eqref{e:coordFT} may appear to be problematic, because it contains an impact over impact parameters $\ul{b}$ of the source at the amplitude level, leading to two such impact parameter integrals in the cross section.  This is true; however, when averaged over color states of the projectile as in \eqref{e:avgs}, the correlator of two $\rho$'s possesses a delta function which sets these two positions equal.  Thus the continuous charge distribution leads to one integral over $d^2 b$ per source at the cross section level, as with the model of discrete valence quarks.}

Inserting the inverse transformation of the wave functions \eqref{e:WFmom}, Wilson lines, and source density
\begin{subequations}
\begin{align}
\psi (\ul{u} - \ul{b} , \ul{x} - \ul{y}) &= \int\frac{d^2 k'}{(2\pi)^2} \frac{d^2 q'}{(2\pi)^2} \:
e^{i (\ul{k'} - \alpha \ul{q'}) \cdot (\ul{x} - \ul{y})} \: e^{i \ul{q'} \cdot (\ul{u} - \ul{b})} \:
\psi (\ul{q'} , \ul{k'} - \alpha \ul{q'}) 
\\
V_{\ul x} &= \int\frac{d^2 \kappa}{(2\pi)^2} \, e^{i \ul{\kappa} \cdot \ul{x}} \, V(\ul{\kappa})
\\
V^\dagger_{\ul y} &= \int\frac{d^2 \kappa'}{(2\pi)^2} \, e^{-i \ul{\kappa'} \cdot \ul{y}} \, V^\dagger (\ul{\kappa'})
\\
\rho(\ul{b}) &= \int\frac{d^2 \ell}{(2\pi)^2} \, e^{i \ul{\ell} \cdot \ul{b}} \, \rho(\ul{\ell})
\end{align}
\end{subequations}
gives
\begin{align}
\mathcal{A} & (k , \bark) = 
\int \frac{d^2 k'}{(2\pi)^2} \, \frac{d^2 q'}{(2\pi)^2} \: \rho^a (\ul{q'}) \: 
\left[ V(\ul{k} - \ul{k'}) \, t^a \, V^\dagger (\ul{k} - \ul{k'} - \ul{q} + \ul{q'}) \right]
\: \psi_1 (\ul{q'}, \ul{k'} - \alpha \ul{q'}) 
\notag \\ & +
\int \frac{d^2 \kappa}{(2\pi)^2} \, \frac{d^2 q'}{(2\pi)^2} \: \rho^a (\ul{q'}) \: 
\left[ V(\ul{\kappa}) \, t^a V^\dagger (\ul{\kappa} - \ul{q} + \ul{q'}) \right]
\: \psi_2 (\ul{q'}, \ul{k} - \alpha \ul{q})
\notag \\ & 
- \int \frac{d^2 \kappa}{(2\pi)^2} \, \frac{d^2 \kappa'}{(2\pi)^2} \: 
\rho^a (\ul{q} - \ul{\kappa} + \ul{\kappa'}) \: 
\left[ V (\ul{\kappa}) \, t^a \, V^\dagger (\ul{\kappa'}) \right]
\left( \psi_1 (\ul{q}, \ul{k} - \alpha \ul{q}) + \psi_2 (\ul{q}, \ul{k} - \alpha \ul{q}) \right) ,
\end{align}
with $\ul{q} = \ul{k} + \ul{\bark}$ for brevity.  We can combine all three diagrams by redefining the dummy integration variables, obtaining the compact form
\begin{align} \label{e:ampl1}
\mathcal{A} (k , \bark) &= 
\int \frac{d^2 k'}{(2\pi)^2} \, \frac{d^2 \bark'}{(2\pi)^2} \: \rho^a (\ul{k'} + \ul{\bark'}) \: 
\left[ V(\ul{k} - \ul{k'}) \, t^a \, V^\dagger (\ul{\bark'} - \ul{\bark}) \right] \, 
\Psi (\ul{k} , \ul{\bark} ; \ul{k'} , \ul{\bark'}) ,
\end{align}
where the differences among the three diagrams are all encoded in the combined wave function
\begin{align} \label{e:WF1}
\Psi (\ul{k} , \ul{\bark} ; \ul{k'} , \ul{\bark'}) &\equiv 
\psi_1 \Big( \ul{k'} + \ul{\bark'} \: , \: (1-\alpha) \ul{k'} - \alpha \ul{\bark'} \Big)
+ \psi_2 \Big( \ul{k'} + \ul{\bark'} \: , \: (1-\alpha) \ul{k} - \alpha \ul{\bark} \Big)
\notag \\ &
- \psi_1 \Big( \ul{k} + \ul{\bark} \: , \: (1-\alpha) \ul{k} - \alpha \ul{\bark} \Big) 
- \psi_2 \Big( \ul{k} + \ul{\bark} \: , \: (1-\alpha) \ul{k} - \alpha \ul{\bark} \Big) .
\end{align}
With this expression, it is easy to do the manipulations over all diagrams at once and particularly to study their color structure, since the Wilson lines enter in exactly the same form for all diagrams.  As such, when we construct cross sections for the production of multiple pairs, we will only have to perform one calculation per diagrammatic topology, rather than having to repeat the calculation for many possible time orderings.  These various topologies will correspond to different ways to contract the diagrams, including both the color matrix $V t^a V^\dagger$ and the wave function $\Psi$, which is a matrix in the $2 \times 2$ spin space of the pair.

%
\subsection{Gluon Production Amplitude}
\label{sec:Gamp} 
%

%
\begin{figure}
\includegraphics[width=0.5\textwidth]{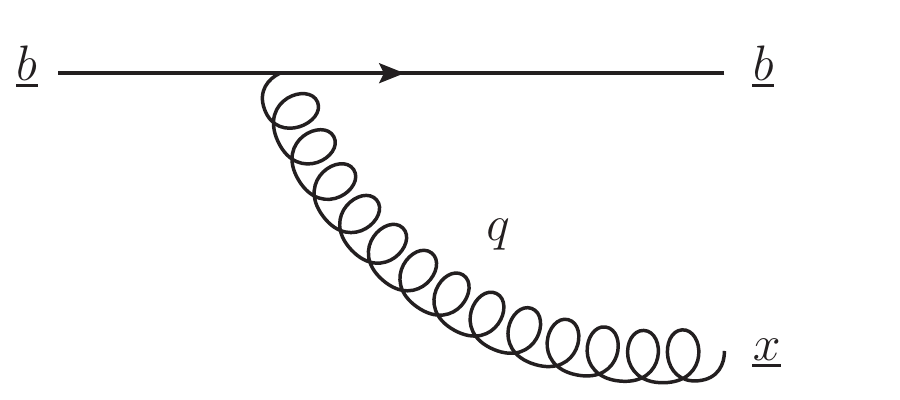}
\caption{The light-front wave function to radiate a soft gluon at mid-rapidity from a valence source, shown here as a quark.} 
\label{f:G_WF}
\end{figure}
%

In comparison with \eqref{e:coord1} for the production amplitude of a soft $q \barq$ pair in coordinate space, the corresponding amplitude to emit a soft gluon is illustrated in Fig.~\ref{f:G_WF} and is given by
\begin{align} \label{e:Gcoord}
\mathcal{A}_{\mathrm{glue}}^a (\ul{x} , \ul{b}) = (V_{\ul b} \, t^b) \, (U_{\ul x})^{a b} \, \phi (\ul{x} - \ul{b}) \: - \:
(t^a \, V_{\ul b}) \, \phi (\ul{x} - \ul{b}) ,
\end{align}
where $\phi$ is the light-front wave function
\begin{subequations} \label{e:GWF}
\begin{align}
\phi(\ul{q}) &= 2 \, \frac{\ul{\epsilon}_\lambda^* \cdot \ul{q} }{q_T^2} \delta_{\sigma_v \, \sigma_v^\prime}
\\
\phi(\ul{x} - \ul{b}) &= \frac{i}{\pi} \frac{\ul{\epsilon}_\lambda^* \cdot (\ul{x} - \ul{b}) }{(\ul{x} - \ul{b})_T^2}
\delta_{\sigma_v \, \sigma_v^\prime}
\end{align}
\end{subequations}
to radiate a soft gluon from a valence quark projectile.  Here $\sigma_v$ and $\sigma_v^\prime$ are the spin states of the valence quark before and after gluon emission, and $\lambda$ is the spin of the emitted gluon.  When we write the trace over the square of these wave functions, we mean the averaging over the quantum numbers of the initial state, together with a sum over the quantum numbers of the final state:
\begin{subequations} \label{e:GWFsq}
\begin{align}
\tr_D [ \phi(q_1) \, \phi^\dagger (q_2) ] & \equiv \half\sum_{\lambda \sigma_v \sigma_v^\prime} 
\phi(q_1) \phi^* (q_2) = 4 \frac{ \ul{q_1} \cdot \ul{q_2}}{q_{1T}^2 \, q_{2T}^2}
\\
\tr_D [ \phi(x) \, \phi^\dagger (y) ] & \equiv \half\sum_{\lambda \sigma_v \sigma_v^\prime}  
\phi(x) \phi^* (y) = \frac{1}{\pi^2} \frac{\ul{x} \cdot \ul{y}}{x_T^2 \, y_T^2} .
\end{align}
\end{subequations}
The first term of \eqref{e:Gcoord} corresponds to the shockwave passing through the gluon, the second term corresponds to the shockwave passing through the valence quark before the gluon is emitted, and we have used the fact that the wave functions for the two time orderings differ by a minus sign (similar to $\psi_1 + \psi_2 + \psi_3 = 0$ for the quark pair case \eqref{e:WFmom}).

As we did in Sec.~\ref{sec:qqbaramp}, we can rewrite the second time ordering so that the valence quark scattering looks the same as the first one, and we can convert to the continuous charge density to write
\begin{align} \label{e:Gamp1}
\mathcal{A}_{\mathrm{glue}}^a (\ul{q}) &= 
\int\frac{d^2 q'}{(2\pi)^2} \, \frac{d^2 k'}{(2\pi)^2} \, \rho^b (\ul{q'}) \:
2 \tr_c \left[  V (\ul{k'}) \, t^b \, V^\dagger (\ul{k'} - \ul{q} + \ul{q'}) \, t^a \right]
\: \left[ \phi(\ul{q'}) - \phi(\ul{q}) \right] .
\end{align}
Comparing the gluon amplitude \eqref{e:Gamp1} with the pair amplitude \eqref{e:ampl1}, we note that the gluon has a specified color $a$ in the final state, resulting in the pair-like structure $V t^b V^\dagger$ being contracted with $2 t^a$ to form a trace.  Squaring the gluon amplitude gives
\begin{align}
\left\langle \left| \mathcal{A}_{\mathrm{glue}}^a (\ul{q}) \right|^2 \right\rangle &=
\int\frac{d^2 q'}{(2\pi)^2} \, \frac{d^2 k'}{(2\pi)^2} \, 
\frac{d^2 q^\pp}{(2\pi)^2} \, \frac{d^2 k^\pp}{(2\pi)^2} \, 
\left\langle \rho^b (\ul{q'}) \, \rho^{c \, *} (\ul{q^\pp}) \right\rangle_{\mathrm{proj}}
\notag \\ & \times
4 \tr_D \left[ \Big(\phi(\ul{q'}) - \phi(\ul{q}) \Big) \, 
\Big(\phi^\dagger(\ul{q^\pp}) - \phi^\dagger(\ul{q}) \Big) \right] 
\notag \\ & \times
\left\langle \tr_c \left[ V (\ul{k'}) \, t^b \, V^\dagger (\ul{k'} - \ul{q} + \ul{q'}) \, t^a \right] \,
\tr_c \left[ t^a \, V (\ul{k^\pp} - \ul{q} + \ul{q^\pp}) \, t^c \, V^\dagger (\ul{k^\pp}) \right] \right\rangle_{\mathrm{tgt}} ,
\end{align}
where we denote the averaging over color fields of the projectile and target by $\langle \cdots \rangle_{\mathrm{proj}}$ and $\langle \cdots \rangle_{\mathrm{tgt}}$, respectively.  By $\tr_c$ we denote a trace over color indices and by $\tr_D$ we denote a trace over the $2 \times 2$ spin states of the wave function which averages over spins in the initial state and sums over spins in the final state.  

If we use the Fierz identity over the repeated color index $a$, we can combine the two traces into one, with the $N_c$-suppressed term in the Fierz identity vanishing exactly by the Wilson line identity \eqref{e:Wcancel3}.  This gives
\begin{align}
\left\langle \left| \mathcal{A}_{\mathrm{glue}}^a (\ul{q}) \right|^2 \right\rangle &=
\int\frac{d^2 q'}{(2\pi)^2} \, \frac{d^2 k'}{(2\pi)^2} \, 
\frac{d^2 q^\pp}{(2\pi)^2} \, \frac{d^2 k^\pp}{(2\pi)^2} \, 
\left\langle \rho^b (\ul{q'}) \, \rho^{c \, *} (\ul{q^\pp}) \right\rangle_{\mathrm{proj}}
\notag \\ & \times
2 \tr_D \left[ \Big(\phi(\ul{q'}) - \phi(\ul{q}) \Big) \, 
\Big(\phi^\dagger(\ul{q^\pp}) - \phi^\dagger(\ul{q}) \Big) \right] 
\notag \\ & \times
\left\langle \tr_c \left[ V (\ul{k'}) \, t^b \, V^\dagger (\ul{k'} - \ul{q} + \ul{q'}) \: \:
V (\ul{k^\pp} - \ul{q} + \ul{q^\pp}) \, t^c \, V^\dagger (\ul{k^\pp}) \right] \right\rangle_{\mathrm{tgt}} ,
\end{align}
which has the same Wilson line structure as we would obtain by squaring the pair amplitude \eqref{e:ampl1}.  Thus we can, without loss of generality, replace the gluon amplitude \eqref{e:Gamp1} with the equivalent expression
\begin{align} \label{e:Gamp2}
\mathcal{A}_{\mathrm{glue}} (\ul{q}) &=
\int\frac{d^2 q'}{(2\pi)^2} \, \frac{d^2 k'}{(2\pi)^2} \, \rho^a (\ul{q'})
\left[ V (\ul{k'}) \, t^a \, V^\dagger (\ul{k'} - \ul{q} + \ul{q'}) \right]
\, \times \, \sqrt{2} \left[ \phi(\ul{q'}) - \phi(\ul{q}) \right] ,
\end{align}
that has the same same form as the pair amplitude \eqref{e:ampl1}.

This similar structure appears to suggest a possible mapping between the pair amplitude \eqref{e:ampl1} and the gluon amplitude \eqref{e:Gamp2}.  Comparing the two expressions, we see that in the limit $k = \bark = \thalf q$, the Wilson line structure of the pair amplitude \eqref{e:ampl1} can be cast in the same form as the gluon amplitude \eqref{e:Gamp2}:
\begin{align}
\mathcal{A} (\thalf q, \thalf q) &= 
\int \frac{d^2 k'}{(2\pi)^2} \, \frac{d^2 \bark'}{(2\pi)^2} \: \rho^a (\ul{k'} + \ul{\bark'}) \: 
\left[ V(\thalf\ul{q} - \ul{k'}) \, t^a \, V^\dagger (\ul{\bark'} - \thalf \ul{q}) \right] \, 
\Psi (\thalf q , \thalf q ; \ul{k'} , \ul{\bark'}) .
\end{align}
The change of variables $\ul{\bark'} = \ul{q'} - \ul{k'}$ followed by $\ul{k'} \rightarrow - \ul{k'} + \thalf \ul{q}$ makes the comparison with \eqref{e:Gamp2} explicit,
\begin{align}
\mathcal{A} (\thalf q, \thalf q) &= 
\int \frac{d^2 q'}{(2\pi)^2} \, \frac{d^2 k'}{(2\pi)^2} \: \rho^a (\ul{q'}) \: 
\left[ V(\ul{k'}) \, t^a \, V^\dagger (\ul{k'} - \ul{q} + \ul{q'}) \right] \, 
\Psi (\thalf q , \thalf q ; -\ul{k'} + \thalf \ul{q}, \ul{q'} + \ul{k'} - \thalf\ul{q}) .
\end{align}
Thus we see that by mapping the pair wave function
\begin{align} \label{e:Gmap}
\Psi( k , \bark ; k' , \bark') \rightarrow \Phi(k + \bark ; k' + \bark') \equiv
\sqrt{2} \left[ \phi(k' + \bark') - \phi(k + \bark) \right] 
\end{align}
and setting the final-state quark and antiquark to have equal momenta, $k = \bark = \thalf q$, we can map the pair amplitude \eqref{e:ampl1} onto the gluon amplitude \eqref{e:Gamp2}:
\begin{align} \label{e:Gmap2}
\mathcal{A}_{\mathrm{glue}} (q) = \left. \mathcal{A} (\thalf q , \thalf q) \right|_{\Psi \rightarrow \Phi}
\end{align}

Thus any cross section calculated for the production of multiple $q \barq$ pairs via \eqref{e:ampl1} can be mapped onto the cross section to instead produce gluons via \eqref{e:Gmap} and \eqref{e:Gmap2}.  The existence of this mapping allows us to efficiently compute multiparticle production at mid rapidity, by first computing the production of various $q \barq$ pairs and then mapping them systematically back to gluons.  Aside from \eqref{e:Gmap2}, the only other modifications to the cross section will be a change in the prefactor of $\frac{1}{2(2\pi)^3}$ to reflect the changed number of final-state particles and the exclusion of quark entanglement in the pair which has been mapped to a gluon.  We will use this strategy in Sec.~\ref{sec:singleG} to obtain the single-inclusive gluon cross section from the pair cross section and in Secs.~\ref{sec:pairgluon} and \ref{sec:dblgluon} to obtain the $q \barq G$ and $G G$ cross sections from the double-pair cross section.

%
\section{Cross Sections for Single Pairs and Gluons}
\label{sec:singlexsec}
%

%
\subsection{Cross Section for Single-Pair Production}
\label{sec:singlepair}
%

%
\begin{figure}
\includegraphics[width=\textwidth]{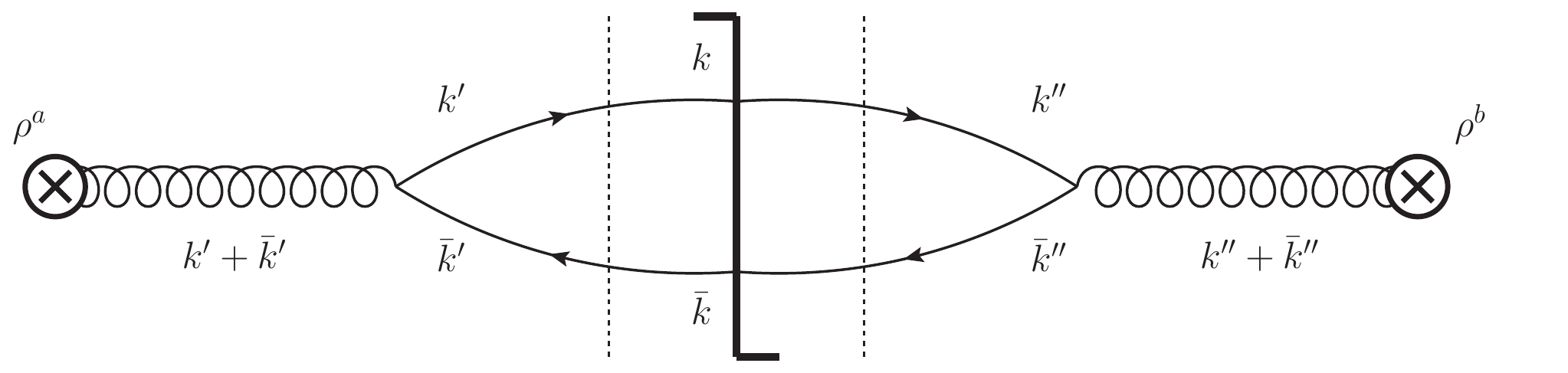}
\caption{The cross section for producing a $q \barq$ pair as written in Eq.~\eqref{e:amplsq1}.  The vertical dotted lines denote the effective positions of the Wilson lines, which shift the quark and antiquark momenta from $k , \bark$ in the final state to $k' , \bark'$ in the amplitude and $k'' , \bark''$ in the complex-conjugate amplitude.  Note that the form of the amplitude \eqref{e:ampl1} makes it possible to write all time orderings as if the $q \barq$ pair passed through the shockwave as illustrated here.} 
\label{f:Pair_Xsec}
\end{figure}
%

The inclusive cross section to produce a single soft $q \barq$ pair with quark (antiquark) transverse momentum $\ul{k}$ ($\ul{\bark}$) and rapidity $y$ ($\bary$) is simply related to the square of the amplitude \eqref{e:ampl1}:
\begin{align} \label{e:sglpairxsec}
\frac{d\sigma^{q \barq}}{d^2 k \, dy \: d^2 \bark \, d\bary} = \left[ \frac{1}{2 (2\pi)^3} \right]^2
\Big\langle \Big| \mathcal{A} (k , \bar{k}) \Big|^2 \Big\rangle ,
\end{align}
where $dy = \frac{dk^+}{k^+}$ and $d\bary = \frac{d\bark^+}{\bark^+}$.
To help keep the notation compact let us introduce the following notation for the differentials:
\begin{subequations}
\begin{align}
\dtwo k &\equiv \frac{d^2 k}{(2\pi)^2}
\\
d^2 \{ x_1 \, x_2 \cdots \, x_n \} &\equiv d^2 x_1 \, d^2 x_2 \cdots d^2 x_n 
\\
\dtwo \{ k_1 \, k_2 \cdots k_n\} &\equiv \dtwo k_1 \, \dtwo k_2 \cdots \dtwo k_n .
\end{align}
\end{subequations}
We will also often exclude the underlines for the many transverse vectors when it is clear from context that they refer to $2$-vectors rather than $4$-vectors.  Then squaring \eqref{e:ampl1} and performing the averaging as in Appendix~\ref{app:Averaging}, we straightforwardly obtain
\begin{align} \label{e:amplsq1}
\left\langle \left| \mathcal{A} (k , \bar{k}) \right|^2 \right\rangle &=
\int \dtwo \{ k' \, \bar{k}' \, k'' \, \bar{k}'' \} \:\:
\left\langle \rho^a (k' + \bar{k}') \, \rho^{b \, *} (k'' + \bar{k}'') \right\rangle_{\mathrm{proj}}
\notag \\ &\times
\tr_D \left[ \Psi (k , \bar{k} ; k' , \bar{k}') \, \Psi^\dagger (k , \bar{k} ; k'' , \bar{k}'') \right]
\notag \\ &\times
\left\langle \tr_c \left[ V(k - k') \, t^a \, V^\dagger (\bark' - \bark) \: 
V(\bark'' - \bark) \, t^b \, V^\dagger (k - k'') \right] \right\rangle_{\mathrm{tgt}} ,
\end{align}
as illustrated in Fig.~\ref{f:Pair_Xsec}.  Using the Gaussian averaging of the projectile with the assumption of Locality from \eqref{e:avgs}, we obtain
\begin{align}
\left\langle \left| \mathcal{A} (k , \bar{k}) \right|^2 \right\rangle &=
\int \dtwo \{ k' \, \bar{k}' \, k'' \, \bar{k}'' \} \:\:
\mu^2 (\ul{k'} + \ul{\bark'} - \ul{k''} - \ul{\bark''} , 0^+)
\notag \\ &\times
\tr_D \left[ \Psi (k , \bar{k} ; k' , \bar{k}') \, \Psi^\dagger (k , \bar{k} ; k'' , \bar{k}'') \right]
\notag \\ &\times
\left\langle \tr_c \left[ V(k - k') \, t^a \, V^\dagger (\bark' - \bark) \: 
V(\bark'' - \bark) \, t^a \, V^\dagger (k - k'') \right] \right\rangle ,
\end{align}
where the second argument of $\mu^2$ vanishes in this case because the eikonal Wilson lines preserve the plus momentum.  This feature will in general be violated when producing multiple pairs.  The color trace is straightforward to simplify using using the Fierz identity, yielding the compact expression 
\begin{align} \label{e:sglpair}
&\Big\langle \Big| \mathcal{A} (k , \bar{k}) \Big|^2 \Big\rangle = \frac{N_c^2}{2}
\int \dtwo \{ k' \, \bar{k}' \, k'' \, \bar{k}'' \} \:\:
\mu^2 (\ul{k'} + \ul{\bark'} - \ul{k''} - \ul{\bark''} , 0^+) \:\:
\tr_D \Big[ \Psi (k , \bar{k} ; k' , \bar{k}') \, \Psi^\dagger (k , \bar{k} ; k'' , \bar{k}'') \Big]
\notag \\ &\times
\Bigg[
\Big\langle \hat{D}_2 (\bark'' - \bark, \bark' - \bark) \: \hat{D}_2 (k - k', k - k'') \Big\rangle
- \frac{1}{N_c^2} \: D_4(k - k', \bark' - \bark, \bark'' - \bark, k - k'') \Bigg] ,
\end{align}
where the fundamental dipole and quadrupole operators are defined in \eqref{e:multipoles}.

Comparing this expression with the corresponding ones (32) and (37) from Ref.~\cite{Martinez:2018ygo} in coordinate space, we see that Eq.~\eqref{e:sglpair} is far more compact.  This is largely because the condensed amplitude \eqref{e:ampl1} in momentum space combines the two time orderings ``1'' and ``2'' into a single form, allowing us to perform a single calculation for all time orderings, rather than requiring a sum over all the distinct time orderings for the pair emission.  The single-pair cross section is then immediately given by combing Eqs.~\eqref{e:sglpair} and \eqref{e:sglpairxsec}.

%
\subsection{Cross Section for Single-Gluon Production}
\label{sec:singleG}
%

As a cross-check and to illustrate the ``gluonic mapping'' of a $q \barq$ pair \eqref{e:Gmap2}, let us map \eqref{e:sglpair} onto the cross section for single-inclusive gluon production.  Applying \eqref{e:Gmap2} to \eqref{e:sglpair} gives 
\begin{align} 
\frac{d\sigma^G}{d^2 q \, dy} &= \frac{N_c^2}{4 (2\pi)^3}
\int \dtwo \{ k' \, \bar{k}' \, k'' \, \bar{k}'' \} \:\:
\mu^2 (\ul{k'} + \ul{\bark'} - \ul{k''} - \ul{\bark''} , 0^+) \:\:
\tr_D \Big[ \Phi (q ; k' + \bar{k}') \, \Phi^\dagger (q ; k'' + \bar{k}'') \Big]
\notag \\ &\times
\Bigg[
\Big\langle \hat{D}_2 (\bark'' - \thalf q , \bark' - \thalf q) \: 
\hat{D}_2(\thalf q - k' , \thalf q - k'') \Big\rangle
\notag \\ & \hspace{1cm}
- \frac{1}{N_c^2} \: D_4 (\thalf q - k' , \bark' - \thalf q , \bark'' - \thalf q , \thalf q - k'') \Bigg] ,
\end{align}
where we changed the prefactor from \eqref{e:sglpairxsec} to $\frac{1}{2(2\pi)^3}$ to reflect the fact that there is now only one particle tagged in the final state.  Changing variables to $q^{ (\prime \, , \, \prime \prime) } \equiv k^{ (\prime \, , \, \prime \prime) } + \bark^{ (\prime \, , \, \prime \prime) }$ and $\delta k^{ (\prime \, , \, \prime \prime) } \equiv \thalf \left( k^{ (\prime \, , \, \prime \prime) } - \bark^{ (\prime \, , \, \prime \prime) } \right)$ gives
\begin{align} \label{e:Gcheck1}
\frac{d\sigma^G}{d^2 q \, dy} &= \frac{N_c^2}{4 (2\pi)^3}
\int\dtwo \{ q' \, q'' \, \delta k' \, \delta k'' \} \:\:
\mu^2 (\ul{q'} - \ul{q''}, 0^+) \:\:
\tr_D \Big[ \Phi (q ; q') \, \Phi^\dagger (q ; q'') \Big]
\notag \\ &\times
\Bigg[
\Big\langle \hat{D}_2
(\thalf q'' - \thalf q - \delta k'' , 
\thalf q' - \thalf q - \delta k') \: 
\hat{D}_2
(\thalf q - \thalf q' - \delta k' , 
\thalf q - \thalf q'' - \delta k'' ) \Big\rangle
\notag \\ & \hspace{0.5cm}
- \frac{1}{N_c^2} \: D_4
(\thalf q - \thalf q' - \delta k' , 
\thalf q' - \thalf q - \delta k' ,
\thalf q'' - \thalf q - \delta k'' ,
\thalf q - \thalf q'' - \delta k'' ) \Bigg] .
\end{align}
At first glance, the mapping \eqref{e:Gmap2} doesn't seem to have accomplished very much.  Even more alarmingly, this expression contains a quadrupole contribution, whereas the explicit gluon production cross section has only double-dipoles.  The resolution is that there are hidden cancellations among the Wilson lines in momentum space, as described in Appendix~\ref{app:WLines}.  These cancellations in momentum space occur after integration over the dummy variables $\delta k'$ and $\delta k''$, which do not couple to the wave functions or to $\mu^2$.  

Consider first the problematic quadrupole term:
\begin{align} \label{e:Wcancel4}
\int \dtwo \{ \delta k' \, \delta k'' \} &\: 
D_4
(\thalf q - \thalf q' - \delta k' ,
\thalf q' - \thalf q - \delta k' ,
\thalf q'' - \thalf q - \delta k'' ,
\thalf q - \thalf q'' - \delta k'' )
\notag \\ &=
(2\pi)^4 \, \delta^2 ( \ul{q}' -  \ul{q} ) \, \delta^2 ( \ul{q} - \ul{q}'' ) ,
\end{align}
where we've used \eqref{e:Wcancel3} to integrate over $\delta k'$ in the first two arguments and $\delta k''$ in the last two.  The quadrupole term therefore sets $\ul{q}' = \ul{q}'' = \ul{q}$, but we immediately see from \eqref{e:Gmap} that
\begin{align} \label{e:Wcancel5}
\Phi( q ; q ) = 0 ,
\end{align}
so the quadrupole term vanishes identically in the ``gluonic limit,'' as it must.  This cancellation in the wave function is just a reflection of the fact that the sum of time orderings must vanish by the definition of the $T$-matrix.  For the double-dipole term of \eqref{e:Gcheck1}, the situation is more subtle, since the shared momenta $\delta k' , \delta k ''$ live in opposite traces and cannot simply cancel each other.  To see clearly what is going on, it is useful to transform momentarily back to coordinate space:
\begin{align}  \label{e:Wcancel6}
\int \dtwo \{ \delta k' \, \delta k'' \} &\: 
\Big\langle \hat{D}_2
(\thalf q'' - \thalf q - \delta k'' ,
\thalf q' - \thalf q - \delta k' ) \: 
\hat{D}_2
( \thalf q - \thalf q' - \delta k' ,
\thalf q - \thalf q'' - \delta k'' ) \Big\rangle
\notag \\ \notag \\ &=
\int d^2 \{ x \, y \, z \, w \} \,
\int \dtwo \{ \delta k' \, \delta k'' \} \,
e^{-i \left( \thalf \ul{q''} - \thalf \ul{q} - \ul{\delta k''} \right) \cdot \ul{x} } \, 
e^{i \left( \thalf \ul{q'} - \thalf \ul{q} - \ul{\delta k'} \right) \cdot \ul{y} } \, 
\notag \\ &\hspace{1cm} \times
e^{-i \left( \thalf \ul{q} - \thalf \ul{q'} - \ul{\delta k'} \right) \cdot \ul{z} } \, 
e^{i \left( \thalf \ul{q} - \thalf \ul{q''} - \ul{\delta k''} \right) \cdot \ul{w} }
\left\langle \Dtwohat{x}{y} \, \Dtwohat{z}{w} \right\rangle
\notag \\ \notag \\ &=
\int d^2 x \, d^2 y \, e^{-i (  \ul{q}'' -  \ul{q} ) \cdot \ul{x} } \, e^{i ( \ul{q}' - \ul{q}) \cdot \ul{y} } \, 
\left\langle \left| \Dtwohat{x}{y} \right|^2 \right\rangle
\notag \\ \notag \\ &\equiv
\left| D_2 \right|^2  ( \ul{q}'' -  \ul{q} , \ul{q}' - \ul{q} ) .
\end{align}
This quantity, after integration over $\delta k' , \delta k''$, is just the Fourier transform to momentum space of a squared dipole amplitude.  (Note that this is NOT the same thing as the square of the momentum-space dipole amplitude!)  Using these results, the single-inclusive gluon cross section \eqref{e:Gcheck1} becomes
\begin{align} \label{e:Gcheck2}
\frac{d\sigma^G}{d^2 q \, dy} &= \frac{N_c^2}{4 (2\pi)^3}
\int \dtwo \{ q' \, q'' \} \, \mu^2 (\ul{q'} - \ul{q''}, 0^+) \:\:
\tr_D \Big[ \Phi (q ; q') \, \Phi^\dagger (q ; q'') \Big]
\notag \\ &\times
\left| D_2 \right|^2  ( \ul{q}'' -  \ul{q} , \ul{q}' - \ul{q} ) ,
\end{align}
which is again an incredibly compact expression in momentum space.  

To complete the cross-check, let us insert the Fourier transform back to coordinate space, obtaining
\begin{align} 
\frac{d\sigma^G}{d^2 q \, dy} &= \frac{N_c^2}{4 (2\pi)^3}
\int d^2 \{ x \, y \, x' \, y' \, b \} \, e^{- i \ul{q}  \cdot (\ul{x}' - \ul{y}') } \, 
e^{-i \ul{q} \cdot (\ul{x} - \ul{b})} \, e^{i \ul{q} \cdot (\ul{y} - \ul{b})} \, 
\mu^2 ( \ul{b} , 0^+) 
\notag \\ &\hspace{1cm} \times
\tr_D \Big[ \Phi (x - b ; x' - b) \, \Phi^\dagger (y - b ; y' - b) \Big] \:
\left\langle \left| \hat{D}_2 (x', y') \right|^2 \right\rangle .
\end{align}
From \eqref{e:Gmap}, it's straightforward to see that the Fourier transform of the wave function is
\begin{align} \label{e:GWF2}
\Phi( x - b ; x' - b) = \sqrt{2} \left[ \delta^2 ( \ul{x} - \ul{b} ) \, \phi (\ul{x}' - \ul{b}) -
\delta^2 ( \ul{x}' - \ul{b} ) \, \phi (\ul{x} - \ul{b}) \right] ,
\end{align}
giving
\begin{align}
\frac{d\sigma^G}{d^2 q \, dy} &= \frac{N_c^2}{2 (2\pi)^3}
\int d^2 \{ x \, y \, b \} \, e^{- i \ul{q}  \cdot (\ul{x} - \ul{y}) } \, \mu^2 ( \ul{b} , 0^+) \:
\tr\left[ \phi(\ul{x} - \ul{b}) \, \phi^\dagger (\ul{y} - \ul{b}) \right]
\notag \\ &\hspace{1cm} \times
\Bigg\{
\left\langle \left| \Dtwohat{x}{y} \right|^2 \right\rangle
- \left\langle \left| \Dtwohat{x}{b} \right|^2 \right\rangle
- \left\langle \left| \Dtwohat{b}{y} \right|^2 \right\rangle
+ 1
\Bigg\} ,
\end{align}
where we have dropped the primes on any remaining integration variables in a given term.  The last step is to insert the explicit gluon-emission wave functions \eqref{e:GWFsq}, obtaining
\begin{align} \label{e:Gcheck3}
\frac{d\sigma^G}{d^2 q \, dy} &= \frac{a}{(2\pi)^2} \left( \frac{\alpha_s N_c}{2\pi^2} \right) 
\int d^2 \{ x \, y \, b \} \, e^{- i \ul{q}  \cdot (\ul{x} - \ul{y}) } \, 
\left[ \frac{2 N_c}{g^2 a} \mu^2 ( \ul{b} , 0^+) \right] \:
\frac{ (\ul{x} - \ul{b}) \cdot (\ul{y} - \ul{b}) }{ (\ul{x} - \ul{b})_T^2 \, (\ul{y} - \ul{b})_T^2 }
\notag \\ &\hspace{1cm} \times
\Bigg\{ 1 - \left\langle \left| \Dtwohat{x}{b} \right|^2 \right\rangle
- \left\langle \left| \Dtwohat{b}{y} \right|^2 \right\rangle
+ \left\langle \left| \Dtwohat{x}{y} \right|^2 \right\rangle \Bigg\} ,
\end{align}
which agrees perfectly with Eq. (38) of \cite{Martinez:2018ygo} and, after employing \eqref{e:Wconvert1}, with Eq.~(8.18) of \cite{Kovchegov:2012mbw}.  This confirmation validates the cross-check of the  ``gluonic limit'' \eqref{e:Gmap2} of a $q \barq$ pair, as well as the dictionary \eqref{e:mumap} between the langauge of continous color charge densities and discrete valence quarks, since
\begin{align}
\frac{2 N_c}{g^2 a} \mu^2 (\ul{b} , 0^+) \rightarrow \frac{1}{a} \int d^2 B \, 
T_{\mathrm{proj}} (\ul{b} - \ul{B}) = 1 ,
\end{align}
where $a = \int d^2 B \, T_{\mathrm{proj}} (\ul{B})$ is the number of nucleons in the projectile.  In the same way, we can next calculate the double-pair cross section and then use the mapping \eqref{e:Gmap2} to convert it into the $q \barq G$ cross section and then into the $G G$ cross section.

%
\section{Double-Inclusive Cross Sections for Pairs and Gluons}
\label{sec:dblxsec}
%

%
\subsection{Cross Section for Double-Pair Production}
\label{sec:dblpair}
%

For double-pair production, we consider a final state with two quarks $k_1 , k_2$ and two antiquarks $\bark_1 , \bark_2$.  As emphasized in Ref.~\cite{Altinoluk:2016vax}, if the produced (anti)quarks have the same flavor, then we need to take care to explicitly antisymmetrize the full amplitude $\mathcal{A}_{\mathrm{full}}$ under the interchange of identical quarks and antiquarks to satisfy Fermi-Dirac statistics:
\begin{align}
\mathcal{A}_{\mathrm{full}} (k_1 , \bark_1 , k_2 , \bark_2) = 
- \mathcal{A}_{\mathrm{full}} (k_1 , \bark_2 , k_2 , \bark_1) =
- \mathcal{A}_{\mathrm{full}} (k_2 , \bark_1 , k_1 , \bark_2) .
\end{align}
Moreover, since in our case the two pairs are radiated independently, the amplitude is automatically symmetric under the interchange of both pairs:
\begin{align}
\mathcal{A}_{\mathrm{full}} (k_1 , \bark_1 , k_2 , \bark_2) = 
+ \mathcal{A}_{\mathrm{full}} (k_2 , \bark_2 , k_1 , \bark_1) ,
\end{align}
which can be seen because of the sum over color sources $a$ in \eqref{e:ampl1}.  As such, it is sufficient to only antisymmetrize the amplitude under the exchange of the antiquarks:
\begin{align} \label{e:symmamp}
\mathcal{A}_{\mathrm{full}} (k_1 , \bark_1 , k_2 , \bark_2) &= 
\mathcal{A} (k_1 , \bark_1 , k_2 , \bark_2) - (\bark_1 \leftrightarrow \bark_2)
\notag \\ &=
\mathcal{A} (k_1 , \bark_1) \otimes \mathcal{A}(k_2 , \bark_2) - (\bark_1 \leftrightarrow \bark_2)
\end{align}
with the elementary single-pair amplitudes being given by \eqref{e:ampl1}.  Written explicitly, the (unsymmetrized) double-pair amplitude $\mathcal{A} (k_1 , \bark_1 , k_2 , \bark_2) = \mathcal{A} (k_1 , \bark_1) \otimes \mathcal{A}(k_2 , \bark_2)$ is
\begin{align} \label{e:ampl2}
\mathcal{A} (k_1 , \bark_1 , k_2 , \bark_2) &= \int \dtwo \{ k_1^\prime \, \bark_1^\prime \, k_2^\prime \, \bark_2^\prime \} \: \rho^a (k_1^\prime + \bark_1^\prime) \, \rho^b (k_2^\prime + \bark_2^\prime)
\notag \\ &\times
\Bigg\{
\Psi(k_1 , \bark_1 ; k_1^\prime , \bark_1^\prime) \:
\left[ V(k_1 - k_1^\prime) \, t^a \, V^\dagger (\bark_1^\prime - \bark_1) \right]
\notag \\ & \hspace{1cm}\otimes
\Psi(k_2 , \bark_2 ; k_2^\prime , \bark_2^\prime) \:
\left[ V(k_2 - k_2^\prime) \, t^b \, V^\dagger (\bark_2^\prime - \bark_2) \right] \Bigg\} .
\end{align}
Note that, as claimed, the amplitude is automatically symmetric under the exchange of the pairs $(k_1 , \bark_1) \leftrightarrow (k_2 , \bark_2)$ if one also relabels the dummy momenta $(k_1^\prime , \bark_1^\prime) \leftrightarrow (k_2^\prime , \bark_2^\prime)$ and dummy color indices $a \leftrightarrow b$.  Squaring the full symmetrized amplitude \eqref{e:symmamp} and converting to the cross section yields
\begin{align} \label{e:dblxsec}
&\frac{d\sigma^{(q \barq) \, (q \barq)}}{d^2 k_1 \, dy_1 \: d^2 \bark_1 \, d\bary_1 \: d^2 k_2 \, dy_2 \: d^2 \bark_2 \, d\bary_2} =
\left[ \frac{1}{2 (2\pi)^3} \right]^4
\notag \\ & \hspace{1cm} \times
\Bigg[
\Big\langle \Big| \mathcal{A} (k_1 , \bark_1 , k_2 , \bark_2) \Big|^2 \Big\rangle 
- \Big\langle \mathcal{A} (k_1 , \bark_1 , k_2 , \bark_2) \, \mathcal{A}^\dagger (k_1 , \bark_2 , k_2 , \bark_1) \Big\rangle 
+ (\bark_1 \leftrightarrow \bark_2) \Bigg] ,
\end{align}
where the notation $+ (\bark_1 \leftrightarrow \bark_2)$ applies to both preceding terms.  The task has now been reduced to calculating the two contributions in brackets: the case without fermion entanglement in the first term and the case with fermion entanglement in the second term.  Both exercises are straightforward, and we calculate them in the following subsections.  For maximum generality, we have considered here the case in which both produced pairs have the same flavor; if the flavor of the quark pairs is different, then all particles are distinguishable and only the first term of Eq.~\eqref{e:dblxsec} contributes.

%
\subsubsection{Case 1: No Fermion Entanglement}
\label{sec:sausage} 
%

%
\begin{figure}
\includegraphics[width=0.8\textwidth]{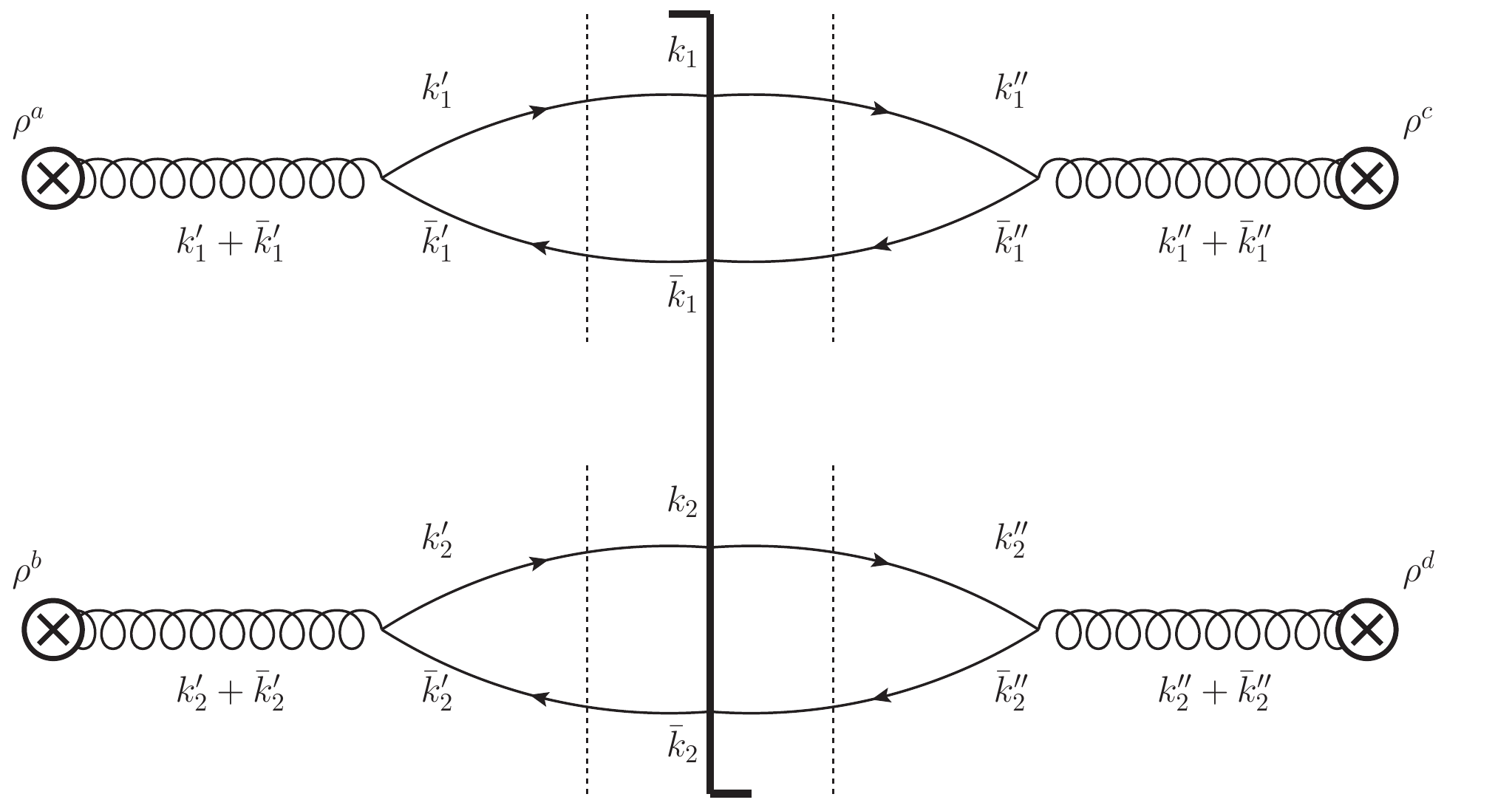}
\caption{Double-pair production topologies without fermion entanglement, as calculated in Eq.~\eqref{e:saus1}.} 
\label{f:Sausage}
\end{figure}
%

Squaring \eqref{e:ampl2} for topologies with no fermion entanglement, as in Fig.~\ref{f:Sausage}, leads directly to
\begin{align} \label{e:saus1}
\Big\langle \Big| \mathcal{A} (k_1 , \bark_1 &, k_2 , \bark_2) \Big|^2 \Big\rangle = \int \dtwo 
\{ k_1^\prime \, \bark_1^\prime \, k_2^\prime \, \bark_2^\prime \, k_1^\pp \, \bark_1^\pp \, k_2^\pp \, \bark_2^\pp \} \: 
\notag \\ &\times
\left\langle
\rho^a (k_1^\prime + \bark_1^\prime) \, \rho^b (k_2^\prime + \bark_2^\prime) \,
\rho^{c \, *} (k_1^\pp + \bark_1^\pp) \, \rho^{d \, *} (k_2^\pp + \bark_2^\pp)
\right\rangle_{\mathrm{proj}}
\notag \\ &\times
\tr_D \left[ \Psi(k_1 , \bark_1 ; k_1^\prime , \bark_1^\prime) \: 
\Psi^\dagger(k_1 , \bark_1 ; k_1^\pp , \bark_1^\pp) \right] \:
\tr_D \left[ \Psi(k_2 , \bark_2 ; k_2^\prime , \bark_2^\prime) \:
\Psi^\dagger(k_2 , \bark_2 ; k_2^\pp , \bark_2^\pp) \right]
\notag \\ &\times
\Big\langle
\tr_c
\left[ V(k_1 - k_1^\prime) \, t^a \, V^\dagger (\bark_1^\prime - \bark_1) \:
V (\bark_1^\pp - \bark_1) \, t^c \, V^\dagger (k_1 - k_1^\pp)  \right] \:
\notag \\ & \hspace{1cm} \times
\tr_c
\left[ V(k_2 - k_2^\prime) \, t^b \, V^\dagger (\bark_2^\prime - \bark_2) \:
V (\bark_2^\pp - \bark_2) \, t^b \, V^\dagger (k_2 - k_2^\pp) \right] 
\Big\rangle_{\mathrm{tgt}} . 
\end{align}
There are now 3 possible ``contractions'' of the source colors, obtained in terms of \eqref{e:avgs}:
\begin{align} \label{e:sausctr}
\Big\langle \rho^a (k_1^\prime + \bark_1^\prime) &\, \rho^b (k_2^\prime + \bark_2^\prime) \,
\rho^{c \, *} (k_1^\pp + \bark_1^\pp) \, \rho^{d \, *} (k_2^\pp + \bark_2^\pp) \Big\rangle_{\mathrm{proj}} =
\notag \\ \notag \\ &=
\delta^{a b} \delta^{c d} \: 
\mu^2 ( \ul{k_1^\prime} + \ul{\bark_1^\prime} + \ul{k_2^\prime} + \ul{\bark_2^\prime} \: , \: 
k_1^+ + \bark_1^+ + k_2^+ + \bark_2^+ ) 
\notag \\ &\hspace{1cm}\times
\mu^2 ( -\ul{k_1^\pp} - \ul{\bark_1^\pp} - \ul{k_2^\pp} - \ul{\bark_2^\pp} \: , \:
-k_1^+ - \bark_1^+ - k_2^+ - \bark_2^+ ) 
\notag \\ \notag \\ &+
\delta^{a c} \delta^{b d} \: 
\mu^2 ( \ul{k_1^\prime} + \ul{\bark_1^\prime} - \ul{k_1^\pp} - \ul{\bark_1^\pp} \: , \: 0^+ ) \:
\mu^2 ( \ul{k_2^\prime} + \ul{\bark_2^\prime} - \ul{k_2^\pp} - \ul{\bark_2^\pp} \: , \: 0^+ )
\notag \\ \notag \\ &+
\delta^{a d} \delta^{b c} \: 
\mu^2 ( \ul{k_1^\prime} + \ul{\bark_1^\prime} - \ul{k_2^\pp} - \ul{\bark_2^\pp} \: , \: 
k_1^+ + \bark_1^+ - k_2^+ - \bark_2^+) \:
\notag \\ &\hspace{1cm} \times
\mu^2 ( \ul{k_2^\prime} + \ul{\bark_2^\prime} - \ul{k_1^\pp} - \ul{\bark_1^\pp} \: , \: 
k_2^+ + \bark_2^+ - k_1^+ - \bark_1^+ ) .
\end{align}
Note that only in the second term do the plus momenta combine to give $0^+$, even for this topology with no fermion entanglement.  For a given set of color contractions, we will need to Fierz reduce the Wilson line traces of \eqref{e:saus1} twice.  The algebra is straightforward, but it is convenient to define the following abbreviated notation:
\begin{align} \label{e:abbrev}
V_1 &\equiv V(k_1 - k_1^\prime) &
V_5 &\equiv V(k_2 - k_2^\prime)
\notag \\
V_2^\dagger &\equiv V^\dagger (\bark_1^\prime - \bark_1) &
V_6^\dagger &\equiv V^\dagger (\bark_2^\prime - \bark_2)
\notag \\
V_3 &\equiv V (\bark_1^\pp - \bark_1) &
V_7 &\equiv V (\bark_2^\pp - \bark_2)
\notag \\
V_4^\dagger &\equiv V^\dagger (k_1 - k_1^\pp) &
V_8^\dagger &\equiv V^\dagger (k_2 - k_2^\pp) .
\end{align}
With this shorthand, the Wilson line tensor entering \eqref{e:saus1} is
\begin{align}
\Omega^{a b c d}_1 = \tr_c [ V_1 \, t^a \, V_2^\dagger \, V_3 \, t^c \, V_4^\dagger] \:
\tr [ V_5 \, t^b \, V_6^\dagger \, V_7 \, t^d \, V_8^\dagger] ,
\end{align}
and we can straightforwardly compute the various contraction of \eqref{e:sausctr}:  
\begin{subequations}
\begin{align}
\delta^{a b} \delta^{c d} \, \Omega_1^{a b c d} &=
\frac{N_c^2}{4} \left\langle \hat{D}_4 (1 6 7 4) \, \hat{D}_4 (5 2 3 8) \right\rangle
- \frac{1}{4} D_8 (1 6 7 8 5 2 3 4)
\notag \\ & \hspace{1cm}
- \frac{1}{4} D_8 (1 2 3 8 5 6 7 4)
+ \frac{1}{4} \left\langle \hat{D}_4 (1 2 3 4) \, \hat{D}_4 (5 6 7 8) \right\rangle ,
\\ \notag \\
\delta^{a c} \delta^{b d} \, \Omega_1^{a b c d} &= \frac{N_c^4}{4}
\left\langle \hat{D}_2 (3 2) \, \hat{D}_2 (1 4) \, \hat{D}_2 (7 6) \, \hat{D}_2 (5 8) \right\rangle
- \frac{N_c^2}{4}
\left\langle \hat{D}_2 (3 2) \, \hat{D}_2 (1 4) \, \hat{D}_4 (5 6 7 8) \right\rangle
\notag \\ &\hspace{1cm}
- \frac{N_c^2}{4}
\left\langle \hat{D}_4 (1 2 3 4) \, \hat{D}_2 (7 6) \, \hat{D}_2 (5 8) \right\rangle 
+ \frac{1}{4}
\left\langle \hat{D}_4 (1 2 3 4) \, \hat{D}_4 (5 6 7 8) \right\rangle ,
\\ \notag \\
\delta^{a d} \delta^{b c} \, \Omega_1^{a b c d} &= \frac{N_c^2}{4} 
\left\langle \hat{D}_4 (1 8 5 4) \, \hat{D}_4 (7 2 3 6) \right\rangle
- \frac{1}{4} D_8 (3 4 1 8 5 6 7 2)
\notag \\ &\hspace{1cm}
- \frac{1}{4} D_8 (1 2 3 6 7 8 5 4)
+ \frac{1}{4} \left\langle \hat{D}_4 (1 2 3 4) \, \hat{D}_4 (5 6 7 8) \right\rangle ,
\end{align}
\end{subequations}
with the numbers in parentheses denoting the arguments of the corresponding Wilson lines in \eqref{e:abbrev}.  These various traces are ilustrated in Fig.~\ref{f:Traces_1}.  Fantastically, we only have to do one calculation {\it{per topology}} (i.e., source contraction) because all of the different time orderings enter on the same footing.  Combining all these terms back into \eqref{e:saus1} yields the complete result for topologies with no fermion entanglement:
%
%
\begin{figure}
\includegraphics[width=\textwidth]{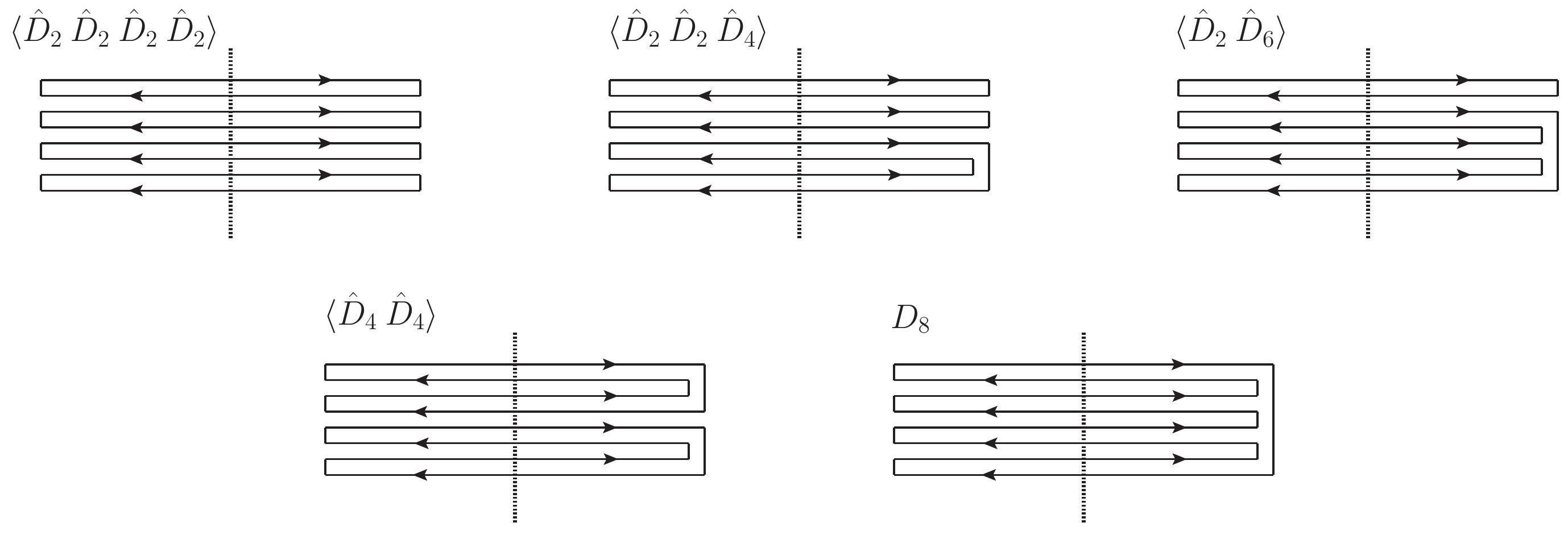}
\caption{Illustration of the Wilson line traces $\langle \hat{D}_2 \: \hat{D}_2 \:  \hat{D}_2 \:  \hat{D}_2 \rangle$, $\langle \hat{D}_2 \:  \hat{D}_2 \:  \hat{D}_4 \rangle$, $\langle \hat{D}_2 \:  \hat{D}_6 \rangle$, $\langle \hat{D}_4 \:  \hat{D}_4 \rangle$, and $D_8$ contributing to double-pair production in Eqs.~\eqref{e:saus2} and \eqref{e:Pac2}.} 
\label{f:Traces_1}
\end{figure}
%
%
\newpage
\begin{align} \label{e:saus2}
\Big\langle \Big| & \mathcal{A} (k_1 , \bark_1 , k_2 , \bark_2) \Big|^2 \Big\rangle = 
\frac{N_c^4}{4}
\int \dtwo \{ k_1^\prime \, \bark_1^\prime \, k_2^\prime \, \bark_2^\prime \, k_1^\pp \, \bark_1^\pp \, k_2^\pp \, \bark_2^\pp \} \: 
\notag \\ &\times
\tr_D \left[ \Psi(k_1 , \bark_1 ; k_1^\prime , \bark_1^\prime) \: 
\Psi^\dagger(k_1 , \bark_1 ; k_1^\pp , \bark_1^\pp) \right] \:
\tr_D \left[ \Psi(k_2 , \bark_2 ; k_2^\prime , \bark_2^\prime) \:
\Psi^\dagger(k_2 , \bark_2 ; k_2^\pp , \bark_2^\pp) \right]
\notag \\ &\times
\Bigg\{
\mu^2 ( \ul{k_1^\prime} + \ul{\bark_1^\prime} + \ul{k_2^\prime} + \ul{\bark_2^\prime} \: , \: 
k_1^+ + \bark_1^+ + k_2^+ + \bark_2^+ ) \:
\mu^2 ( -\ul{k_1^\pp} - \ul{\bark_1^\pp} - \ul{k_2^\pp} - \ul{\bark_2^\pp} \: , \:
-k_1^+ - \bark_1^+ - k_2^+ - \bark_2^+ ) 
\notag \\ & \hspace{0.5cm} \times
\Bigg[ \frac{1}{N_c^2} \left\langle 
\hat{D}_4 (k_1 - k_1^\prime ,  \bark_2^\prime - \bark_2 , \bark_2^\pp - \bark_2 , k_1 - k_1^\pp) 
\, 
\hat{D}_4 (k_2 - k_2^\prime , \bark_1^\prime - \bark_1 , \bark_1^\pp - \bark_1 , k_2 - k_2^\pp) 
\right\rangle
\notag \\ & \hspace{0.5cm}
- \frac{1}{N_c^4 } 
D_8 (k_1 - k_1^\prime ,  \bark_2^\prime - \bark_2 , \bark_2^\pp - \bark_2 , k_2 - k_2^\pp , k_2 - k_2^\prime , \bark_1^\prime - \bark_1 , \bark_1^\pp - \bark_1 , k_1 - k_1^\pp)
\notag \\ & \hspace{0.5cm}
- \frac{1}{N_c^4} 
D_8 (k_1 - k_1^\prime ,  \bark_1^\prime - \bark_1 , \bark_1^\pp - \bark_1 , k_2 - k_2^\pp , k_2 - k_2^\prime , \bark_2^\prime - \bark_2 , \bark_2^\pp - \bark_2 , k_1 - k_1^\pp)
\notag \\ & \hspace{0.5cm}
+ \frac{1}{N_c^4} 
\left\langle \hat{D}_4 (k_1 - k_1^\prime ,  \bark_1^\prime - \bark_1 , \bark_1^\pp - \bark_1 , k_1 - k_1^\pp) 
\, 
\hat{D}_4 (k_2 - k_2^\prime , \bark_2^\prime - \bark_2 , \bark_2^\pp - \bark_2 , k_2 - k_2^\pp) 
\right\rangle \Bigg]
\notag \\ \notag \\ &+
\mu^2 ( \ul{k_1^\prime} + \ul{\bark_1^\prime} - \ul{k_1^\pp} - \ul{\bark_1^\pp} \: , \: 0^+ ) \:
\mu^2 ( \ul{k_2^\prime} + \ul{\bark_2^\prime} - \ul{k_2^\pp} - \ul{\bark_2^\pp} \: , \: 0^+ )
\notag \\ & \hspace{0.5cm} \times
\Bigg[ \left\langle 
\hat{D}_2 (\bark_1^\pp - \bark_1 , \bark_1^\prime - \bark_1 ) \, 
\hat{D}_2 (k_1 - k_1^\prime , k_1 - k_1^\pp ) \, 
\hat{D}_2 (\bark_2^\pp - \bark_2 , \bark_2^\prime - \bark_2 ) \, 
\hat{D}_2 (k_2 - k_2^\prime , k_2 - k_2^\pp ) \right\rangle
\notag \\ & \hspace{0.5cm}
- \frac{1}{N_c^2} \left\langle 
\hat{D}_2 (\bark_1^\pp - \bark_1 , \bark_1^\prime - \bark_1 ) \, 
\hat{D}_2 (k_1 - k_1^\prime , k_1 - k_1^\pp ) \, 
\hat{D}_4 (k_2 - k_2^\prime , \bark_2^\prime - \bark_2 , \bark_2^\pp - \bark_2 , k_2 - k_2^\pp ) 
\right\rangle
\notag \\ & \hspace{0.5cm}
- \frac{1}{N_c^2} \left\langle 
\hat{D}_4 (k_1 - k_1^\prime , \bark_1^\prime - \bark_1 , \bark_1^\pp - \bark_1 , k_1 - k_1^\pp) \, 
\hat{D}_2 (\bark_2^\pp - \bark_2 , \bark_2^\prime - \bark_2 ) \, 
\hat{D}_2 (k_2 - k_2^\prime , k_2 - k_2^\pp ) 
\right\rangle 
\notag \\ & \hspace{0.5cm}
+ \frac{1}{N_c^4} \left\langle 
\hat{D}_4 (k_1 - k_1^\prime , \bark_1^\prime - \bark_1 , \bark_1^\pp - \bark_1 , k_1 - k_1^\pp ) \, 
\hat{D}_4 (k_2 - k_2^\prime , \bark_2^\prime - \bark_2 , \bark_2^\pp - \bark_2 , k_2 - k_2^\pp ) 
\right\rangle \Bigg]
\notag \\ \notag \\ &+
\mu^2 ( \ul{k_1^\prime} + \ul{\bark_1^\prime} - \ul{k_2^\pp} - \ul{\bark_2^\pp} \: , \: 
k_1^+ + \bark_1^+ - k_2^+ - \bark_2^+) \:
\mu^2 ( \ul{k_2^\prime} + \ul{\bark_2^\prime} - \ul{k_1^\pp} - \ul{\bark_1^\pp} \: , \: 
k_2^+ + \bark_2^+ - k_1^+ - \bark_1^+ )
\notag \\ & \hspace{0.5cm} \times
\Bigg[
\frac{1}{N_c^2} \left\langle 
\hat{D}_4 (k_1 - k_1^\prime , k_2 - k_2^\pp , k_2 - k_2^\prime , k_1 - k_1^\pp ) \, 
\hat{D}_4 (\bark_2^\pp - \bark_2 , \bark_1^\prime - \bark_1 , \bark_1^\pp - \bark_1 , \bark_2^\prime - \bark_2 ) \right\rangle
\notag \\ & \hspace{0.5cm}
- \frac{1}{N_c^4} 
D_8 (\bark_1^\pp - \bark_1 , k_1 - k_1^\pp , k_1 - k_1^\prime , k_2 - k_2^\pp , k_2 - k_2^\prime , \bark_2^\prime - \bark_2 , \bark_2^\pp - \bark_2 , \bark_1^\prime - \bark_1 )
\notag \\ & \hspace{0.5cm}
- \frac{1}{N_c^4} 
D_8 (k_1 - k_1^\prime , \bark_1^\prime - \bark_1 , \bark_1^\pp - \bark_1 , \bark_2^\prime - \bark_2 , \bark_2^\pp - \bark_2 , k_2 - k_2^\pp , k_2 - k_2^\prime , k_1 - k_1^\pp )
\notag \\ & \hspace{0.5cm}
+ \frac{1}{N_c^4} \left\langle 
\hat{D}_4 (k_1 - k_1^\prime , \bark_1^\prime - \bark_1 , \bark_1^\pp - \bark_1 , k_1 - k_1^\pp ) \, 
\hat{D}_4 (k_2 - k_2^\prime , \bark_2^\prime - \bark_2 , \bark_2^\pp - \bark_2 , k_2 - k_2^\pp ) 
\right\rangle \Bigg] \Bigg\}
\end{align}
%

%
\subsubsection{Case 2: Fermion Entanglement}
\label{sec:PacMan} 
%

%
\begin{figure}
\includegraphics[width=0.8\textwidth]{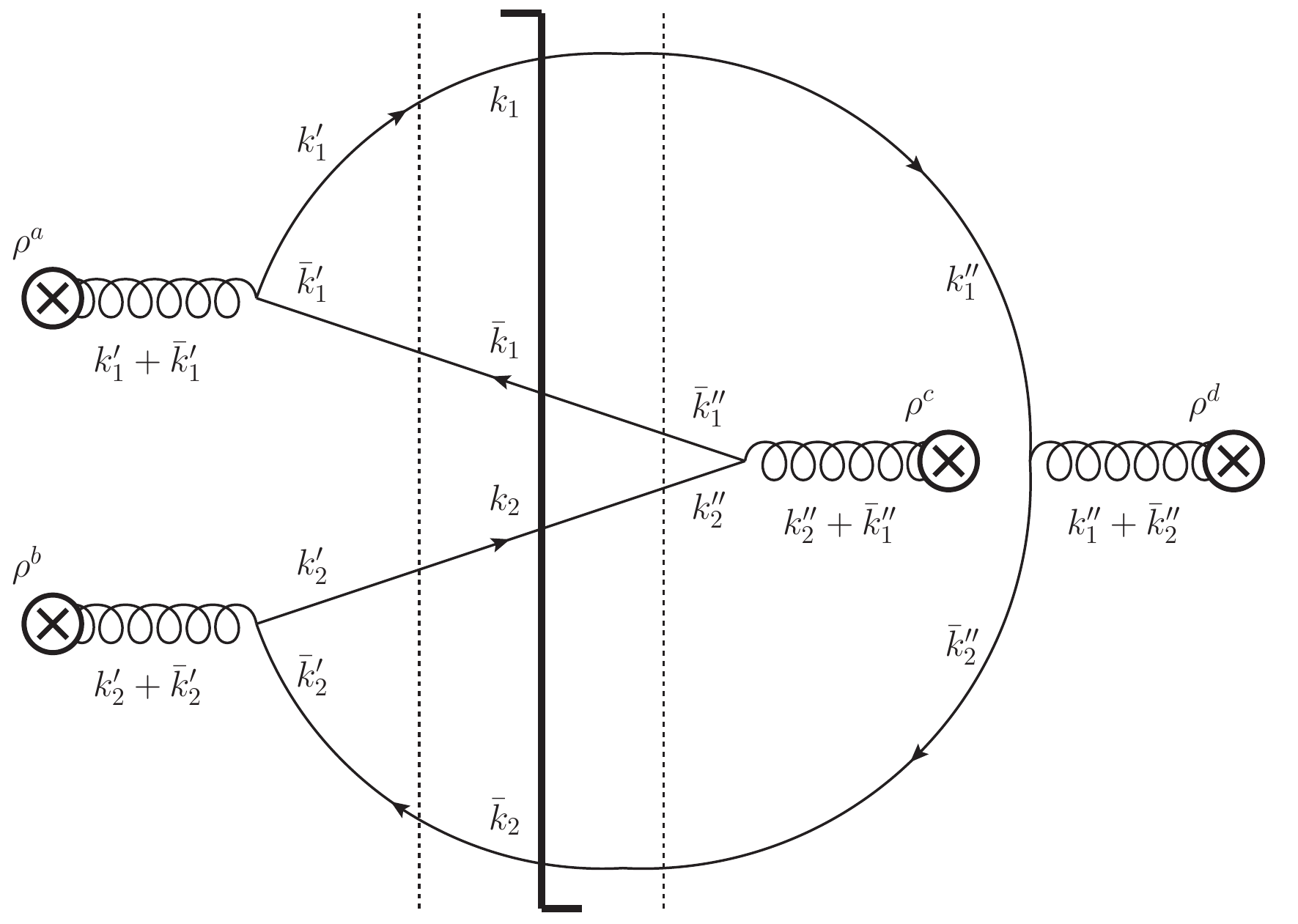}
\caption{Double-pair production ``Pac Man'' topologies with fermion entanglement, as calculated in Eq.~\eqref{e:Pac1}.} 
\label{f:PacMan}
\end{figure}
%

The interference term in \eqref{e:dblxsec} contains the topologies with the fermions being entangled, such that the pairs in the amplitude ``swap ownership'' of one of the fermions in going to the complex-conjugate amplitude.  This contribution arises from the Fermi-Dirac statistics of identical particles and therefore does not contribute if the pairs have different flavors.  Taking the interference of \eqref{e:ampl2} as shown in the ``Pac Man'' type diagram of Fig.~\ref{f:PacMan} we have directly
\begin{align} \label{e:Pac1}
\Big\langle \mathcal{A} &(k_1 , \bark_1 , k_2 , \bark_2) \, 
\mathcal{A}^\dagger (k_1 , \bark_2 , k_2 , \bark_1) \Big\rangle = \int \dtwo
\{ k_1^\prime \, \bark_1^\prime \, k_2^\prime \, \bark_2^\prime \, k_1^\pp \, \bark_1^\pp \, k_2^\pp \, \bark_2^\pp \} 
\notag \\ &\times
\left\langle
\rho^a (k_1^\prime + \bark_1^\prime) \, \rho^b (k_2^\prime + \bark_2^\prime) \,
\rho^{c \, *} (k_2^\pp + \bark_1^\pp) \, \rho^{d \, *} (k_1^\pp + \bark_2^\pp)
\right\rangle_{\mathrm{proj}}
\notag \\ &\times
\tr_D \left[
\Psi (k_1 , \bark_1 ; k_1^\prime , \bark_1^\prime) \,
\Psi^\dagger (k_2 , \bark_1 ; k_2^\pp , \bark_1^\pp) \,
\Psi (k_2 , \bark_2 ; k_2^\prime , \bark_2^\prime)
\Psi^\dagger (k_1 , \bark_2 ; k_1^\pp , \bark_2^\pp)
\right]
\notag \\ &\times
\Big\langle \tr_c \Big[ 
V(k_1 - k_1^\prime) \, t^a \, V^\dagger (\bark_1^\prime - \bark_1) \:
V (\bark_1^\pp - \bark_1) \, t^c \, V^\dagger (k_2 - k_2^\pp) \:
\notag \\ & \hspace{1cm} \times
V(k_2 - k_2^\prime) \, t^b \, V^\dagger (\bark_2^\prime - \bark_2) \:
V (\bark_2^\pp - \bark_2) \, t^d \, V^\dagger (k_1 - k_1^\pp) \Big] \Big\rangle_{\mathrm{tgt}} .
\end{align}
In the same way as \eqref{e:sausctr}, we form the 3 contractions of the projectile sources:
\begin{align} \label{e:Pacctr}
\Big\langle 
\rho^a (k_1^\prime + \bark_1^\prime) &\, 
\rho^b (k_2^\prime + \bark_2^\prime) \,
\rho^{c \, *} (k_2^\pp + \bark_1^\pp) \, 
\rho^{d \, *} (k_1^\pp + \bark_2^\pp) 
\Big\rangle_{\mathrm{proj}} =
\notag \\ \notag \\ &=
\delta^{a b} \delta^{c d} \:
\mu^2 (\ul{k_1^\prime} + \ul{\bark_1^\prime} + \ul{k_2^\prime} + \ul{\bark_2^\prime} \: , \:
k_1^+ + \bark_1^+ + k_2^+ + \bark_2^+) \:
\notag \\ &\hspace{1cm}\times
\mu^2 (-\ul{k_2^\pp} - \ul{\bark_1^\pp} - \ul{k_1^\pp} - \ul{\bark_2^\pp} \: , \:
-k_2^+ - \bark_1^+ - k_1^+ - \bark_2^+)
\notag \\ \notag \\ &+
\delta^{a c} \delta^{b d}
\mu^2 (\ul{k_1^\prime} + \ul{\bark_1^\prime} - \ul{k_2^\pp} - \ul{\bark_1^\pp} \: , \:
k_1^+ - k_2^+) \:
\mu^2 (\ul{k_2^\prime} + \ul{\bark_2^\prime} - \ul{k_1^\pp} - \ul{\bark_2^\pp} \: , \:
k_2^+ - k_1^+)
\notag \\ \notag \\ &+
\delta^{a d} \delta^{b c}
\mu^2 (\ul{k_1^\prime} + \ul{\bark_1^\prime} - \ul{k_1^\pp} - \ul{\bark_2^\pp} \: , \:
\bark_1^+ - \bark_2^+) \:
\mu^2 (\ul{k_2^\prime} + \ul{\bark_2^\prime} - \ul{k_2^\pp} - \ul{\bark_1^\pp} \: , \:
\bark_2^+ - \bark_1^+) .
\end{align}
Using the same shorthand notation as \eqref{e:abbrev}, the Wilson line tensor entering \eqref{e:Pac1} is
\begin{align}
\Omega_2^{a b c d} = 
\tr_c[ V_1 \, t^a \, V_2^\dagger \, V_3 \, t^c \, V_8^\dagger V_5 \, t^b \, V_6^\dagger \, V_7 \, t^d \, V_4^\dagger] ,
\end{align}
and we can compute the various color contractions in the same way:
\begin{subequations}
\begin{align}
\delta^{a b} \delta^{c d} \, \Omega_2^{a b c d} &= \frac{N_c}{4} D_8 (1 6 7 8 5 2 3 4)
- \frac{N_c}{4}\left\langle \hat{D}_4 (3 8 5 2) \, \hat{D}_4 (1 6 7 4) \right\rangle
\notag \\ &\hspace{1cm}
- \frac{N_c}{4}
\left\langle \hat{D}_4 (5 6 7 8) \, \hat{D}_4 (1 2 3 4) \right\rangle
+ \frac{1}{4 N_c} D_8 (1 2 3 8 5 6 7 4) .
\\ \notag \\
\delta^{a c} \delta^{b d} \, \Omega_2^{a b c d} &= \frac{N_c^3}{4}
\left\langle \hat{D}_2 (3 2) \, \hat{D}_2 (7 6) \, \hat{D}_4 (1 8 5 4) \right\rangle
-\frac{N_c}{4}
\left \langle \hat{D}_2 (3 2) \, \hat{D}_6 (1 8 5 6 7 4) \right\rangle
\notag \\ & \hspace{1cm}
- \frac{N_c}{4}
\left\langle \hat{D}_2 (7 6) \, \hat{D}_6 (1 2 3 8 5 4) \right\rangle
+ \frac{1}{4 N_c} D_8 (1 2 3 8 5 6 7 4) .
\\ \notag \\
\delta^{a d} \delta^{b c} \, \Omega_2^{a b c d} &= 
\frac{N_c^3}{4}
\left\langle \hat{D}_2 (5 8) \, \hat{D}_2 (1 4) \, \hat{D}_4 (7 2 3 6) \right\rangle
- \frac{N_c}{4}
\left\langle \hat{D}_2 (1 4) \, \hat{D}_6 (3 8 5 6 7 2) \right\rangle
\notag \\ & \hspace{1cm}
- \frac{N_c}{4}
\left\langle \hat{D}_2 (5 8) \, \hat{D}_6 (1 2 3 6 7 4) \right\rangle
+ \frac{1}{4 N_c} D_8 (1 2 3 8 5 6 7 4) .
\end{align}
\end{subequations}
Combining these back into \eqref{e:Pac1} yields the complete result for topologies with fermion entanglement:
\newpage
\begin{align} \label{e:Pac2}
\Big\langle &\mathcal{A} (k_1 , \bark_1 , k_2 , \bark_2) \, 
\mathcal{A}^\dagger (k_1 , \bark_2 , k_2 , \bark_1) \Big\rangle = \frac{N_c^3}{4}
\int \dtwo \{ k_1^\prime \, \bark_1^\prime \, k_2^\prime \, \bark_2^\prime \, k_1^\pp \, \bark_1^\pp \, k_2^\pp \, \bark_2^\pp \} 
\notag \\ &\times
\tr_D \left[
\Psi (k_1 , \bark_1 ; k_1^\prime , \bark_1^\prime) \,
\Psi^\dagger (k_2 , \bark_1 ; k_2^\pp , \bark_1^\pp) \,
\Psi (k_2 , \bark_2 ; k_2^\prime , \bark_2^\prime)
\Psi^\dagger (k_1 , \bark_2 ; k_1^\pp , \bark_2^\pp)
\right]
\notag \\ &\times
\Bigg\{
\mu^2 (\ul{k_1^\prime} + \ul{\bark_1^\prime} + \ul{k_2^\prime} + \ul{\bark_2^\prime} \: , \:
k_1^+ + \bark_1^+ + k_2^+ + \bark_2^+) \:
\mu^2 (-\ul{k_2^\pp} - \ul{\bark_1^\pp} - \ul{k_1^\pp} - \ul{\bark_2^\pp} \: , \:
- k_2^+ - \bark_1^+ - k_1^+ - \bark_2^+)
\notag \\ & \hspace{0.5cm} \times
\Bigg[
\frac{1}{N_c^2} 
D_8 (k_1 - k_1^\prime , \bark_2^\prime - \bark_2 , \bark_2^\pp - \bark_2 , k_2 - k_2^\pp , k_2 - k_2^\prime , \bark_1^\prime - \bark_1 , \bark_1^\pp - \bark_1 , k_1 - k_1^\pp )
\notag \\ & \hspace{0.5cm}
- \frac{1}{N_c^2}\left\langle 
\hat{D}_4 (\bark_1^\pp - \bark_1 , k_2 - k_2^\pp , k_2 - k_2^\prime , \bark_1^\prime - \bark_1 ) \, 
\hat{D}_4 (k_1 - k_1^\prime , \bark_2^\prime - \bark_2 , \bark_2^\pp - \bark_2 , k_1 - k_1^\pp ) 
\right\rangle
\notag \\ & \hspace{0.5cm}
- \frac{1}{N_c^2} \left\langle 
\hat{D}_4 (k_2 - k_2^\prime , \bark_2^\prime - \bark_2 , \bark_2^\pp - \bark_2 , k_2 - k_2^\pp ) \, 
\hat{D}_4 (k_1 - k_1^\prime , \bark_1^\prime - \bark_1 , \bark_1^\pp - \bark_1 , k_1 - k_1^\pp ) 
\right\rangle
\notag \\ & \hspace{0.5cm}
+ \frac{1}{N_c^4} 
D_8 (k_1 - k_1^\prime , \bark_1^\prime - \bark_1 , \bark_1^\pp - \bark_1 , k_2 - k_2^\pp , k_2 - k_2^\prime , \bark_2^\prime - \bark_2 , \bark_2^\pp - \bark_2 , k_1 - k_1^\pp )
\Bigg]
\notag \\\notag \\ &+
\mu^2 (\ul{k_1^\prime} + \ul{\bark_1^\prime} - \ul{k_2^\pp} - \ul{\bark_1^\pp} \: , \: k_1^+ - k_2^+) \:
\mu^2 (\ul{k_2^\prime} + \ul{\bark_2^\prime} - \ul{k_1^\pp} - \ul{\bark_2^\pp} \: , \: k_2^+ - k_1^+)
\notag \\ & \hspace{0.5cm} \times
\Bigg[
\left\langle 
\hat{D}_2 (\bark_1^\pp - \bark_1 , \bark_1^\prime - \bark_1 ) \, 
\hat{D}_2 (\bark_2^\pp - \bark_2 , \bark_2^\prime - \bark_2 ) \, 
\hat{D}_4 (k_1 - k_1^\prime , k_2 - k_2^\pp , k_2 - k_2^\prime , k_1 - k_1^\pp ) \right\rangle
\notag \\ & \hspace{0.5cm}
-\frac{1}{N_c^2} \left \langle 
\hat{D}_2 (\bark_1^\pp - \bark_1 , \bark_1^\prime - \bark_1 ) \, 
\hat{D}_6 (k_1 - k_1^\prime , k_2 - k_2^\pp , k_2 - k_2^\prime , \bark_2^\prime - \bark_2 , \bark_2^\pp - \bark_2 , k_1 - k_1^\pp ) 
\right\rangle
\notag \\ & \hspace{0.5cm}
- \frac{1}{N_c^2} \left\langle 
\hat{D}_2 (\bark_2^\pp - \bark_2 , \bark_2^\prime - \bark_2 ) \, 
\hat{D}_6 (k_1 - k_1^\prime , \bark_1^\prime - \bark_1 , \bark_1^\pp - \bark_1 , k_2 - k_2^\pp , k_2 - k_2^\prime , k_1 - k_1^\pp ) 
\right\rangle
\notag \\ & \hspace{0.5cm}
+ \frac{1}{N_c^4} 
D_8 (k_1 - k_1^\prime , \bark_1^\prime - \bark_1 , \bark_1^\pp - \bark_1 , k_2 - k_2^\pp , k_2 - k_2^\prime , \bark_2^\prime - \bark_2 , \bark_2^\pp - \bark_2 , k_1 - k_1^\pp )
\Bigg]
\notag \\\notag \\ &+
\mu^2 (\ul{k_1^\prime} + \ul{\bark_1^\prime} - \ul{k_1^\pp} - \ul{\bark_2^\pp} \: , \: \bark_1^+ - \bark_2^+) \:
\mu^2 (\ul{k_2^\prime} + \ul{\bark_2^\prime} - \ul{k_2^\pp} - \ul{\bark_1^\pp} \: , \: \bark_2^+ - \bark_1^+)
\notag \\ & \hspace{0.5cm} \times
\Bigg[
\left\langle 
\hat{D}_2 (k_2 - k_2^\prime , k_2 - k_2^\pp ) \, 
\hat{D}_2 (k_1 - k_1^\prime , k_1 - k_1^\pp ) \, 
\hat{D}_4 (\bark_2^\pp - \bark_2 , \bark_1^\prime - \bark_1 , \bark_1^\pp - \bark_1 , \bark_2^\prime - \bark_2 ) 
\right\rangle
\notag \\ & \hspace{0.5cm}
- \frac{1}{N_c^2} \left\langle 
\hat{D}_2 (k_1 - k_1^\prime , k_1 - k_1^\pp ) \, 
\hat{D}_6 (\bark_1^\pp - \bark_1 , k_2 - k_2^\pp , k_2 - k_2^\prime , \bark_2^\prime - \bark_2 , \bark_2^\pp - \bark_2 , \bark_1^\prime - \bark_1 ) 
\right\rangle
\notag \\ & \hspace{0.5cm}
- \frac{1}{N_c^2} \left\langle 
\hat{D}_2 (k_2 - k_2^\prime , k_2 - k_2^\pp ) \, 
\hat{D}_6 (k_1 - k_1^\prime , \bark_1^\prime - \bark_1 , \bark_1^\pp - \bark_1 , \bark_2^\prime - \bark_2 , \bark_2^\pp - \bark_2 , k_1 - k_1^\pp ) 
\right\rangle
\notag \\ & \hspace{0.5cm}
+ \frac{1}{N_c^4} 
D_8 (k_1 - k_1^\prime , \bark_1^\prime - \bark_1 , \bark_1^\pp - \bark_1 , k_2 - k_2^\pp , k_2 - k_2^\prime , \bark_2^\prime - \bark_2 , \bark_2^\pp - \bark_2 , k_1 - k_1^\pp )
\Bigg] \Bigg\}
\end{align}
The expression for the cross section \eqref{e:dblxsec}, together with the two classes of topologies \eqref{e:saus2} and \eqref{e:Pac2} constitute the first complete and exact solution to the 4-particle inclusive $(q \barq) \, (q \barq)$ cross section at this order.  These expressions are one of the primary results of this paper.

%
\subsection{Cross Section for Quark + Antiquark + Gluon Production}
\label{sec:pairgluon} 
%

In the same way as in Sec.~\ref{sec:singleG}, we can now take the ``gluonic limit'' of the $q \bar q$ pair $k_2 , \bark_2 \rightarrow \half q_2$ in the double-pair expression \eqref{e:saus2}.  Note that, once we replace one of the pairs with a gluon, there are no longer any possible fermion entanglement topologies, since all three final state particles $q \barq G$ are now distinguishable. Thus we need only take the limit of the topologies in \eqref{e:saus2}, immediately obtaining:
\begin{align}
& \frac{d\sigma^{(q \barq) \, G}}{d^2 k_1 \, dy_1 \: d^2 \bark_1 \, d\bary_1 \: d^2 q_2 \, dy_2} =
\left(\frac{1}{2(2\pi)^3}\right)^3 \frac{N_c^4}{4}
\int \dtwo \{ k_1^\prime \, \bark_1^\prime \, k_1^\pp \, \bark_1^\pp \, 
q_2^\prime \, q_2^\pp \, \delta k_2^\prime \, \delta k_2^\pp \} \: 
\notag \\ &\times
\tr_D \left[ \Psi(k_1 , \bark_1 ; k_1^\prime , \bark_1^\prime) \: 
\Psi^\dagger(k_1 , \bark_1 ; k_1^\pp , \bark_1^\pp) \right] \:
\tr_D \left[ \Phi(q_2 ; q_2^\prime) \: \Phi^\dagger(q_2 ; q_2^\pp) \right]
\notag \\ &\times
\Bigg\{
\mu^2 ( \ul{k_1^\prime} + \ul{\bark_1^\prime} + \ul{q_2^\prime} \: , \: k_1^+ + \bark_1^+ + q_2^+) \:
\mu^2 ( -\ul{k_1^\pp} - \ul{\bark_1^\pp} - \ul{q_2^\pp} \: , \: -k_1^+ - \bark_1^+ - q_2^+) 
\notag \\ & \hspace{0.5cm} \times
\Bigg[ \frac{1}{N_c^2} \Big\langle 
\hat{D}_4 (k_1 - k_1^\prime ,  \thalf q_2^\prime - \thalf q_2 - \delta k_2^\prime , 
\thalf q_2^\pp - \thalf q_2 - \delta k_2^\pp , k_1 - k_1^\pp) 
\notag \\ &\hspace{2cm} \times
\hat{D}_4 (\thalf q_2 - \thalf q_2^\prime - \delta k_2^\prime , \bark_1^\prime - \bark_1 , 
\bark_1^\pp - \bark_1 , \thalf q_2 - \thalf q_2^\pp - \delta k_2^\pp) 
\Big\rangle
\notag \\ & \hspace{0.5cm}
- \frac{1}{N_c^4 } 
D_8 (k_1 - k_1^\prime ,  \thalf q_2^\prime - \thalf q_2 - \delta k_2^\prime , 
\thalf q_2^\pp - \thalf q_2 - \delta k_2^\pp , \thalf q_2 - \thalf q_2^\pp - \delta k_2^\pp , 
\notag \\ &\hspace{2.5cm} 
\thalf q_2 - \thalf q_2^\prime - \delta k_2^\prime , \bark_1^\prime - \bark_1 , \bark_1^\pp - \bark_1 , k_1 - k_1^\pp)
\notag \\ & \hspace{0.5cm}
- \frac{1}{N_c^4} 
D_8 (k_1 - k_1^\prime ,  \bark_1^\prime - \bark_1 , \bark_1^\pp - \bark_1 , \thalf q_2 - \thalf q_2^\pp - \delta k_2^\pp , 
\notag \\ &\hspace{2.5cm} 
\thalf q_2 - \thalf q_2^\prime - \delta k_2^\prime , \thalf q_2^\prime - \thalf q_2 - \delta k_2^\prime , \thalf q_2^\pp - \thalf q_2 - \delta k_2^\pp , k_1 - k_1^\pp)
\notag \\ & \hspace{0.5cm}
+ \frac{1}{N_c^4} 
\Big\langle \hat{D}_4 (k_1 - k_1^\prime ,  \bark_1^\prime - \bark_1 , \bark_1^\pp - \bark_1 , k_1 - k_1^\pp) 
\notag \\ &\hspace{2cm} \times
\hat{D}_4 (\thalf q_2 - \thalf q_2^\prime - \delta k_2^\prime , 
\thalf q_2^\prime - \thalf q_2 - \delta k_2^\prime, \thalf q_2^\pp - \thalf q_2 - \delta k_2^\pp , 
\thalf q_2 - \thalf q_2^\pp - \delta k_2^\pp) 
\Big\rangle \Bigg]
\notag \\ \notag \\ &+
\mu^2 ( \ul{k_1^\prime} + \ul{\bark_1^\prime} - \ul{k_1^\pp} - \ul{\bark_1^\pp} \: , \: 0^+ ) \:
\mu^2 ( \ul{q_2^\prime} - \ul{q_2^\pp} \: , \: 0^+ )
\notag \\ & \hspace{0.5cm} \times
\Bigg[ \Big\langle 
\hat{D}_2 (\bark_1^\pp - \bark_1 , \bark_1^\prime - \bark_1 ) \, 
\hat{D}_2 (k_1 - k_1^\prime , k_1 - k_1^\pp ) \, 
\hat{D}_2 (\thalf q_2^\pp - \thalf q_2 - \delta k_2^\pp , \thalf q_2^\prime - \thalf q - \delta k_2^\prime ) 
\notag \\ &\hspace{2cm} \times
\hat{D}_2 (\thalf q_2 - \thalf q_2^\prime - \delta k_2^\prime , \thalf q_2 - \thalf q_2^\pp - \delta k_2^\pp ) \Big\rangle
\notag \\ & \hspace{0.5cm}
- \frac{1}{N_c^2} \Big\langle 
\hat{D}_2 (\bark_1^\pp - \bark_1 , \bark_1^\prime - \bark_1 ) \, 
\hat{D}_2 (k_1 - k_1^\prime , k_1 - k_1^\pp ) 
\notag \\ &\hspace{2cm} \times
\hat{D}_4 (\thalf q_2 - \thalf q_2^\prime - \delta k_2^\prime , \thalf q_2^\prime - \thalf q_2 - \delta k_2^\prime , \thalf q_2^\pp - \thalf q_2 - \delta k_2^\pp , \thalf q_2 - \thalf q_2^\pp - \delta k_2^\pp ) 
\Big\rangle
\notag \\ & \hspace{0.5cm}
- \frac{1}{N_c^2} \Big\langle 
\hat{D}_4 (k_1 - k_1^\prime , \bark_1^\prime - \bark_1 , \bark_1^\pp - \bark_1 , k_1 - k_1^\pp) \, 
\hat{D}_2 (\thalf q_2^\pp - \thalf q_2 - \delta k_2^\pp , \thalf q_2^\prime - \thalf q_2 - \delta k_2^\prime ) 
\notag \\ &\hspace{2cm} \times
\hat{D}_2 (\thalf q_2 - \thalf q_2^\prime - \delta k_2^\prime , \thalf q_2 - \thalf q_2^\pp - \delta k_2^\pp ) 
\Big\rangle 
\notag \\ & \hspace{0.5cm}
+ \frac{1}{N_c^4} \Big\langle 
\hat{D}_4 (k_1 - k_1^\prime , \bark_1^\prime - \bark_1 , \bark_1^\pp - \bark_1 , k_1 - k_1^\pp ) 
\notag \\ &\hspace{2cm} \times
\hat{D}_4 (\thalf q_2 - \thalf q_2^\prime - \delta k_2^\prime , \thalf q_2^\prime - \thalf q_2 - \delta k_2^\prime , \thalf q_2^\pp - \thalf q_2 - \delta k_2^\pp , \thalf q_2 - \thalf q_2^\pp - \delta k_2^\pp) \Big\rangle \Bigg]
\notag \\ \notag \\ &+
\mu^2 ( \ul{k_1^\prime} + \ul{\bark_1^\prime} - \ul{q_2^\pp} \: , \: k_1^+ + \bark_1^+ - q_2^+) \:
\mu^2 ( \ul{q_2^\prime} - \ul{k_1^\pp} - \ul{\bark_1^\pp} \: , \: q_2^+ - k_1^+ - \bark_1^+ )
\notag \\ & \hspace{0.5cm} \times
\Bigg[
\frac{1}{N_c^2} \Big\langle 
\hat{D}_4 (k_1 - k_1^\prime , \thalf q_2 - \thalf q_2^\pp - \delta k_2^\pp, \thalf q_2 - \thalf q_2^\prime - \delta k_2^\prime , k_1 - k_1^\pp ) 
\notag \\ & \hspace{2cm} \times
\hat{D}_4 (\thalf q_2^\pp - \thalf q_2 - \delta k_2^\pp , \bark_1^\prime - \bark_1 , \bark_1^\pp - \bark_1 , \thalf q_2^\prime - \thalf q_2 - \delta k_2^\prime ) \Big\rangle
\notag \\ & \hspace{0.5cm}
- \frac{1}{N_c^4} 
D_8 (\bark_1^\pp - \bark_1 , k_1 - k_1^\pp , k_1 - k_1^\prime , \thalf q_2 - \thalf q_2^\pp - \delta k_2^\pp ,
\notag \\ & \hspace{2.5cm}
\thalf q_2 - \thalf q_2^\prime - \delta k_2^\prime , \thalf q_2^\prime - \thalf q_2 - \delta k_2^\prime, \thalf q_2^\pp - \thalf q_2 - \delta k_2^\pp , \bark_1^\prime - \bark_1 )
\notag \\ & \hspace{0.5cm}
- \frac{1}{N_c^4} 
D_8 (k_1 - k_1^\prime , \bark_1^\prime - \bark_1 , \bark_1^\pp - \bark_1 , \thalf q_2^\prime - \thalf q_2 - \delta k_2^\prime , 
\notag \\ & \hspace{2.5cm}
\thalf q_2^\pp - \thalf q_2 - \delta k_2^\pp , \thalf q_2 - \thalf q_2^\pp - \delta k_2^\pp , \thalf q_2 - \thalf q_2^\prime - \delta k_2^\prime , k_1 - k_1^\pp )
\notag \\ & \hspace{0.5cm}
+ \frac{1}{N_c^4} \Big\langle 
\hat{D}_4 (k_1 - k_1^\prime , \bark_1^\prime - \bark_1 , \bark_1^\pp - \bark_1 , k_1 - k_1^\pp )
\notag \\ & \hspace{2cm} \times
\hat{D}_4 (\thalf q_2 - \thalf q_2^\prime - \delta k_2^\prime , \thalf q_2^\prime - \thalf q_2 - \delta k_2^\prime , \thalf q_2^\pp - \thalf q_2 - \delta k_2^\pp , \thalf q_2 - \thalf q_2^\pp - \delta k_2^\pp ) 
\Big\rangle \Bigg] \Bigg\} ,
\end{align}
where as before we have changed variables into $\ul{q_2^\prime} \equiv k_2^\prime + \bark_2^\prime$ and $\ul{\delta k_2^\prime} \equiv \thalf ( \ul{k_2^\prime} - \ul{\bark_2^\prime} )$, and similarly for $\ul{q_2^\pp}$ and $\ul{\delta k_2^\pp}$.

After integration over $\delta k_2^\prime , \delta k_2^\pp$, most of these terms vanish.  These integrals act only on the interaction terms, and as we saw in Eqs.~\eqref{e:Wcancel4} and \eqref{e:Wcancel5}, any time the same momentum $\delta k_2^{(\prime \: , \: \pp)}$ appears in adjacent arguments of the same trace, those Wilson lines will cancel.  In canceling, they lead to delta functions of the momentum difference following \eqref{e:Wcancel3}, which is always either $\delta^2 (\ul{q_2^\prime} - \ul{q_2})$ or $\delta^2 (\ul{q_2^\pp} - \ul{q_2})$, and these terms drop out due to the vanishing of the wave function as in \eqref{e:Wcancel5}.  The only interaction terms which do not vanish correspond to lines 1, 5, 7, and 9 out of the 12 terms in braces.  As in \eqref{e:Wcancel6}, these terms instead simplify the Wilson line traces by causing some of the coordinate-space arguments to be repeated, with the composite object being Fourier transformed to momentum space.  
One of the two surviving operators was already calculated in \eqref{e:Wcancel6}: the square of the dipole amplitude.  The other nonvanishing operator is a partial reduction of the double quadrupole (see Fig.~\ref{f:Traces_2}):
\begin{align} \label{e:dblquad1}
&\int \dtwo \{ \delta k_2^\prime \, \delta k_2^\pp \} \:
\Big\langle 
\hat{D}_4 (p_1 ,  \thalf q_2^\prime - \thalf q_2 - \delta k_2^\prime , 
\thalf q_2^\pp - \thalf q_2 - \delta k_2^\pp , p_2) 
\notag \\ &\hspace{1cm} \times
\hat{D}_4 (\thalf q_2 - \thalf q_2^\prime - \delta k_2^\prime , p_3 , p_4 , 
\thalf q_2 - \thalf q_2^\pp - \delta k_2^\pp) 
\Big\rangle
\notag \\ \notag \\ &=
\int d^2 \{ x_1 \, y_1 \, x_2 \, y_2 \, z_1 \, w_1 \, z_2 \, w_2 \} \,
\int \dtwo \{ \delta k_2^\prime \, \delta k_2^\pp \}
\notag \\ &\hspace{1cm} \times
e^{-i \ul{p_1} \cdot \ul{x_1}} \, e^{i \left( \thalf \ul{q_2^\prime} - \thalf \ul{q_2} - \ul{\delta k_2^\prime} \right) \cdot \ul{y_1}} \,
e^{-i \left( \thalf \ul{q_2^\pp} - \thalf \ul{q_2} - \ul{\delta k_2^\pp} \right) \cdot \ul{x_2}} \, e^{i \ul{p_2} \cdot \ul{y_2}}
\notag \\ &\hspace{1cm} \times
e^{-i \left( \thalf \ul{q_2} - \thalf \ul{q_2^\prime} - \ul{\delta k_2^\prime} \right) \cdot \ul{z_1}} \, 
e^{i \ul{p_3} \cdot \ul{w_1}} \, e^{-i \ul{p_4} \cdot \ul{z_2}} \, 
e^{i \left( \thalf \ul{q_2} - \thalf \ul{q_2^\pp} - \ul{\delta k_2^\pp} \right) \cdot \ul{w_2}} \,
\notag \\ &\hspace{1cm} \times
\left\langle \Dfourhat{x_1}{y_1}{x_2}{y_2} \, \Dfourhat{z_1}{w_1}{z_2}{w_2} \right\rangle
\notag \\ \notag \\ &=
\int d^2 \{ x_1 \, y_1 \, x_2 \, y_2 \, w_1 \, z_2 \} \,
e^{-i \ul{p_1} \cdot \ul{x_1}} \, e^{i ( \ul{q_2^\prime} - \ul{q_2} ) \cdot \ul{y_1}} \, e^{-i ( \ul{q_2^\pp} - \ul{q_2} ) \cdot \ul{x_2}} \, 
e^{i \ul{p_2} \cdot \ul{y_2}} \, e^{i \ul{p_3} \cdot \ul{w_1}} \, e^{-i \ul{p_4} \cdot \ul{z_2}} \, 
\notag \\ &\hspace{1cm} \times
\left\langle \Dfourhat{x_1}{y_1}{x_2}{y_2} \, \Dfourhat{y_1}{w_1}{z_2}{x_2} \right\rangle
\notag \\ \notag \\ &\equiv
D_{4 \, , \, 4} (\ul{p_1} , \ul{q_2^\prime} - \ul{q_2} , \ul{q_2^\pp} - \ul{q_2} , \ul{p_2} \: ; \: \ul{p_3} \, \ul{p_4}) .
\end{align}
The result is the complete expression for the $q \barq G$ cross section in momentum space, with the corresponding operators illustrated in Fig.~\ref{f:Traces_2}:
%
%
\begin{figure}
\includegraphics[width=\textwidth]{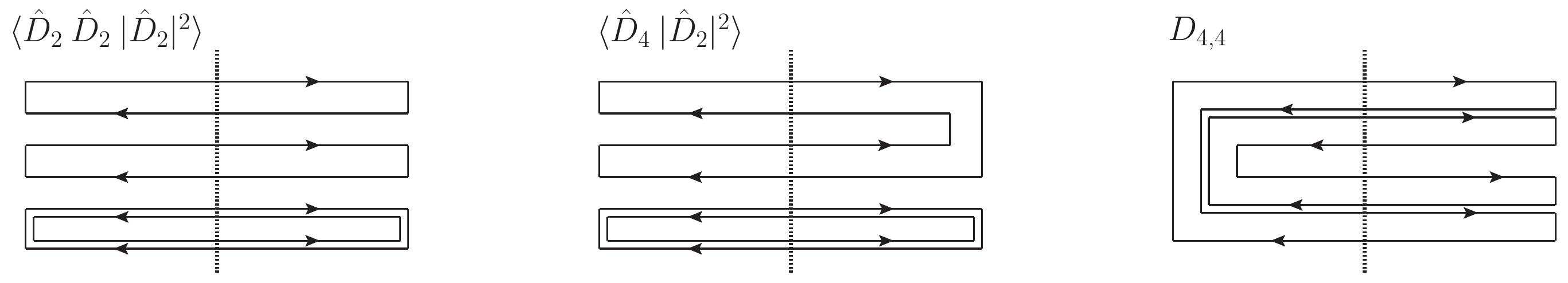}
\caption{Illustration of the Wilson line traces $\langle \hat{D}_2 \: \hat{D}_2 \: | \hat{D}_2 |^2 \rangle$, $\langle \hat{D}_4 \: |\hat{D}_2|^2 \rangle$ and $D_{4 , 4}$ contributing to $q \barq G$ production in Eq.~\eqref{e:mixedxsec}.} 
\label{f:Traces_2}
\end{figure}
%
%
\newpage
\begin{align} \label{e:mixedxsec}
& \frac{d\sigma^{(q \barq) \, G}}{d^2 k_1 \, dy_1 \: d^2 \bark_1 \, d\bary_1 \: d^2 q_2 \, dy_2} =
\left(\frac{1}{2(2\pi)^3}\right)^3 \frac{N_c^4}{4}
\int \dtwo \{ k_1^\prime \, \bark_1^\prime \, k_1^\pp \, \bark_1^\pp \, 
q_2^\prime \, q_2^\pp \} \: 
\notag \\ &\times
\tr_D \left[ \Psi(k_1 , \bark_1 ; k_1^\prime , \bark_1^\prime) \: 
\Psi^\dagger(k_1 , \bark_1 ; k_1^\pp , \bark_1^\pp) \right] \:
\tr_D \left[ \Phi(q_2 ; q_2^\prime) \: \Phi^\dagger(q_2 ; q_2^\pp) \right]
\notag \\ &\times
\Bigg\{
\mu^2 ( \ul{k_1^\prime} + \ul{\bark_1^\prime} + \ul{q_2^\prime} \: , \: k_1^+ + \bark_1^+ + q_2^+) \:
\mu^2 ( -\ul{k_1^\pp} - \ul{\bark_1^\pp} - \ul{q_2^\pp} \: , \: -k_1^+ - \bark_1^+ - q_2^+) 
\notag \\ & \hspace{0.5cm} \times
\Bigg[ \frac{1}{N_c^2} \, D_{4 \, , \, 4} (k_1 - k_1^\prime , q_2^\prime - q_2 , q_2^\pp - q_2 , k_1 - k_1^\pp \: ; \: 
\bark_1^\prime - \bark_1 \, \bark_1^\pp - \bark_1) \Bigg]
\notag \\ \notag \\ &+
\mu^2 ( \ul{k_1^\prime} + \ul{\bark_1^\prime} - \ul{k_1^\pp} - \ul{\bark_1^\pp} \: , \: 0^+ ) \:
\mu^2 ( \ul{q_2^\prime} - \ul{q_2^\pp} \: , \: 0^+ )
\notag \\ & \hspace{0.5cm} \times
\Bigg[ \Big\langle \hat{D}_2 (\bark_1^\pp - \bark_1 , \bark_1^\prime - \bark_1 ) \, 
\hat{D}_2 (k_1 - k_1^\prime , k_1 - k_1^\pp ) \,
\left| \hat{D}_2 \right|^2 (q_2^\pp - q_2 , q_2^\prime - q_2) \Big\rangle
\notag \\ & \hspace{0.5cm}
- \frac{1}{N_c^2} \Big\langle 
\hat{D}_4 (k_1 - k_1^\prime , \bark_1^\prime - \bark_1 , \bark_1^\pp - \bark_1 , k_1 - k_1^\pp) \, 
\left| \hat{D}_2 \right|^2 (q_2^\pp - q_2 , q_2^\prime - q_2) \Big\rangle \Bigg]
\notag \\ \notag \\ &+
\mu^2 ( \ul{k_1^\prime} + \ul{\bark_1^\prime} - \ul{q_2^\pp} \: , \: k_1^+ + \bark_1^+ - q_2^+) \:
\mu^2 ( \ul{q_2^\prime} - \ul{k_1^\pp} - \ul{\bark_1^\pp} \: , \: q_2^+ - k_1^+ - \bark_1^+ )
\notag \\ & \hspace{0.5cm} \times
\Bigg[
\frac{1}{N_c^2} \, 
D_{4 \, , \, 4} (k_1 - k_1^\prime, q_2 - q_2^\pp , q_2 - q_2^\prime , k_1 - k_1^\pp \: ; \: 
\bark_1^\prime - \bark_1  \, \bark_1^\pp - \bark_1 )
\Bigg] \Bigg\} .
\end{align}
To our knowledge, Eq.~\eqref{e:mixedxsec} represents the first complete calculation of $q \barq G$ production in the CGC framework, and this compact form in momentum space comprises an exact solution at this order, at finite $N_c$.  We emphasize, however, that this cross section applies only for the production of a quark and antiquark of the same flavor.  This new expression is the second primary result of this paper.

%
\subsection{Double-Gluon Production}
\label{sec:dblgluon} 
%

As a final cross-check of the preceding calculations, let us use the mapping \eqref{e:Gmap2} on the remaining $q \barq$ pair in \eqref{e:mixedxsec} to obtain the double-gluon production cross section, which we can compare with explicit results in the literature.  As before, we take the limit $k_1 , \bark_1 \rightarrow \thalf q_1$ in \eqref{e:mixedxsec}, adjust the count of $\frac{1}{2(2\pi)^3}$ in the prefactor, and change integration variables to $\ul{q_1^\prime} \equiv k_1^\prime + \bark_1^\prime$ and $\ul{\delta k_1^\prime} \equiv \thalf ( \ul{k_1^\prime} - \ul{\bark_1^\prime} )$, and similarly for $\ul{q_1^\pp}$ and $\ul{\delta k_1^\pp}$.  This gives
\begin{align}
& \frac{d\sigma^{G G}}{d^2 q_1 \, dy_1 \: d^2 q_2 \, dy_2} =
\left( \frac{1}{2(2\pi)^3}\right)^2 \frac{N_c^4}{4}
\int \dtwo \{ q_1^\prime \, q_1^\pp \, q_2^\prime \, q_2^\pp \, \delta k_1^\prime \, \delta k_1^\pp \} %
\notag \\ &\times
\tr_D \left[ \Phi(q_1 ; q_1^\prime) \: \Phi^\dagger(q_1 ; q_1^\pp) \right] \:
\tr_D \left[ \Phi(q_2 ; q_2^\prime) \: \Phi^\dagger(q_2 ; q_2^\pp) \right]
\notag \\ &\times
\Bigg\{
\mu^2 ( \ul{q_1^\prime} + \ul{q_2^\prime} \: , \: q_1^+ + q_2^+) \:
\mu^2 ( -\ul{q_1^\pp} - \ul{q_2^\pp} \: , \: -q_1^+ - q_2^+) 
\notag \\ & \hspace{0.5cm} \times
\Bigg[ \frac{1}{N_c^2} \, D_{4 \, , \, 4} (\thalf q_1 - \thalf q_1^\prime - \delta k_1^\prime , q_2^\prime - q_2 , q_2^\pp - q_2 , \thalf q_1 - \thalf q_1^\pp - \delta k_1^\pp \: ; \: 
\thalf q_1^\prime - \thalf q_1 - \delta k_1^\prime \, \thalf q_1^\pp - \thalf q_1 - \delta k_1^\pp) \Bigg]
\notag \\ \notag \\ &+
\mu^2 ( \ul{q_1^\prime} - \ul{q_1^\pp} \: , \: 0^+ ) \:
\mu^2 ( \ul{q_2^\prime} - \ul{q_2^\pp} \: , \: 0^+ )
\notag \\ & \hspace{0.5cm} \times
\Bigg[ \Big\langle \hat{D}_2 (\thalf q_1^\pp - \thalf q_1 - \delta k_1^\pp , \thalf q_1^\prime - \thalf q_1 - \delta k_1^\prime ) \, \hat{D}_2 (\thalf q_1 - \thalf q_1^\prime - \delta k_1^\prime , \thalf q_1 - \thalf q_1^\pp - \delta k_1^\pp ) \,
\notag \\ &\hspace{2cm}\times
\left| \hat{D}_2 \right|^2 (q_2^\pp - q_2 , q_2^\prime - q_2) \Big\rangle
\notag \\ & \hspace{0.5cm}
- \frac{1}{N_c^2} \Big\langle 
\hat{D}_4 (\thalf q_1 - \thalf q_1^\prime - \delta k_1^\prime , \thalf q_1^\prime - \thalf q_1 - \delta k_1^\prime , \thalf q_1^\pp - \thalf q_1 - \delta k_1^\pp , \thalf q_1 - \thalf q_1^\pp - \delta k_1^\pp) 
\notag \\ &\hspace{2cm}\times
\left| \hat{D}_2 \right|^2 (q_2^\pp - q_2 , q_2^\prime - q_2) \Big\rangle \Bigg]
\notag \\ \notag \\ &+
\mu^2 ( \ul{q_1^\prime} - \ul{q_2^\pp} \: , \: q_1^+ - q_2^+) \:
\mu^2 ( \ul{q_2^\prime} - \ul{q_1^\pp} \: , \: q_2^+ - q_1^+ )
\notag \\ & \hspace{0.5cm} \times
\Bigg[
\frac{1}{N_c^2} \, 
D_{4 \, , \, 4} (\thalf q_1 - \thalf q_1^\prime - \delta k_1^\prime, q_2 - q_2^\pp , q_2 - q_2^\prime , \thalf q_1 - \thalf q_1^\pp - \delta k_1^\pp \: ; \: 
\thalf q_1^\prime - \thalf q_1 - \delta k_1^\prime  \, \thalf q_1^\pp - \thalf q - \delta k_1^\pp )
\Bigg] \Bigg\} .
\end{align}
Of the four interaction terms remaining in the braces, the third one (quadrupole trace) vanishes after integration over $\delta k_1^\prime \, \delta k_1^\pp$ due to repeated adjacent arguments; the second one (double dipole) is the same as \eqref{e:Wcancel6}; and the first and last ones are further reductions of the double-quadrupole \eqref{e:dblquad1}:
\begin{align} \label{e:dblquad2}
& \int \dtwo \{ \delta k_1^\prime \, \delta k_1^\pp \} \:
D_{4 \, , \, 4} (
\thalf q_1 - \thalf q_1^\prime - \delta k_1^\prime , 
p_1 , 
p_2 ,
\thalf q_1 - \thalf q_1^\pp - \delta k_1^\pp \: ; \: 
\thalf q_1^\prime - \thalf q_1 - \delta k_1^\prime \, 
\thalf q_1^\pp - \thalf q_1 - \delta k_1^\pp)
\notag \\ \notag \\ &=
\int d^2 \{ x_1 \, y_1 \, x_2 \, y_2 \, w_1 \, z_2 \} \,
\int \dtwo \{ \delta k_1^\prime \, \delta k_1^\pp \} \,
e^{-i \left( \thalf \ul{q_1} - \thalf \ul{q_1^\prime} - \ul{\delta k_1^\prime} \right)\cdot \ul{x_1}} \, 
e^{i  \ul{p_1} \cdot \ul{y_1}} \, 
e^{-i \ul{p_2} \cdot \ul{x_2}} \, 
e^{i \left( \thalf \ul{q_1} - \thalf \ul{q_1^\pp} - \ul{\delta k_1^\pp} \right) \cdot \ul{y_2}} \, 
\notag \\ &\hspace{1cm} \times
e^{i \left( \thalf \ul{q_1^\prime} - \thalf \ul{q_1} - \ul{\delta k_1^\prime} \right) \cdot \ul{w_1}} \, 
e^{-i \left( \thalf \ul{q_1^\pp} - \ul{\thalf q_1} - \ul{\delta k_1^\pp} \right) \cdot \ul{z_2}} \, 
\left\langle \Dfourhat{x_1}{y_1}{x_2}{y_2} \, \Dfourhat{y_1}{w_1}{z_2}{x_2} \right\rangle
\notag \\ \notag \\ &=
\int d^2 \{ x_1 \, y_1 \, x_2 \, y_2 \, w_1 \, z_2 \} \,
e^{-i ( \ul{q_1} - \ul{q_1^\prime} )\cdot \ul{x_1}} \, 
e^{i  \ul{p_1} \cdot \ul{y_1}} \, 
e^{-i \ul{p_2} \cdot \ul{x_2}} \, 
e^{i ( \ul{q_1} - \ul{q_1^\pp} ) \cdot \ul{y_2}} \, 
\left\langle \left| \Dfourhat{x_1}{y_1}{x_2}{y_2} \right|^2 \right\rangle
\notag \\ \notag \\ &\equiv
\left| D_4 \right|^2 (\ul{q_1} - \ul{q_1^\prime} , \ul{p_1} , \ul{p_2} , \ul{q_1} - \ul{q_1^\pp} ) ,
\end{align}
which is just the square of the quadrupole.  Thus the double-gluon cross section takes the especially compact form in momentum space
\begin{align} \label{e:GG1}
& \frac{d\sigma^{G G}}{d^2 q_1 \, dy_1 \: d^2 q_2 \, dy_2} =
\left(\frac{1}{2(2\pi)^3}\right)^2 \frac{N_c^4}{4}
\int \dtwo \{ q_1^\prime \, q_1^\pp \, q_2^\prime \, q_2^\pp \} 
\notag \\ &\times
\tr_D \left[ \Phi(q_1 ; q_1^\prime) \: \Phi^\dagger(q_1 ; q_1^\pp) \right] \:
\tr_D \left[ \Phi(q_2 ; q_2^\prime) \: \Phi^\dagger(q_2 ; q_2^\pp) \right] \:
\notag \\ &\times
\Bigg\{
\mu^2 ( \ul{q_1^\prime} + \ul{q_2^\prime} \: , \: q_1^+ + q_2^+) \:
\mu^2 ( -\ul{q_1^\pp} - \ul{q_2^\pp} \: , \: -q_1^+ - q_2^+) \:
\Bigg[ \frac{1}{N_c^2} \, 
\left| D_4 \right|^2 (\ul{q_1} - \ul{q_1^\prime} , \ul{q_2^\prime} - \ul{q_2} , \ul{q_2^\pp} - \ul{q_2} , \ul{q_1} - \ul{q_1^\pp} ) \Bigg]
\notag \\ &+
\mu^2 ( \ul{q_1^\prime} - \ul{q_1^\pp} \: , \: 0^+ ) \:
\mu^2 ( \ul{q_2^\prime} - \ul{q_2^\pp} \: , \: 0^+ ) \:
\Bigg[ \Big\langle 
\left| \hat{D}_2 \right|^2 (q_1^\pp - q_1 , q_1^\prime - q_1)
\left| \hat{D}_2 \right|^2 (q_2^\pp - q_2 , q_2^\prime - q_2) 
\Big\rangle \Bigg]
\notag \\ &+
\mu^2 ( \ul{q_1^\prime} - \ul{q_2^\pp} \: , \: q_1^+ - q_2^+) \:
\mu^2 ( \ul{q_2^\prime} - \ul{q_1^\pp} \: , \: q_2^+ - q_1^+ ) \:
\Bigg[
\frac{1}{N_c^2} \, 
\left| D_4 \right|^2 (\ul{q_1} - \ul{q_1^\prime} , \ul{q_2} - \ul{q_2^\pp} , \ul{q_2} - \ul{q_2^\prime} , \ul{q_1} - \ul{q_1^\pp} )
\Bigg] \Bigg\} .
\end{align}

At this point, we can directly compare the cross section \eqref{e:GG1} against the known expressions in the literature; Ref.~\cite{Altinoluk:2018ogz} gives the $GG$ cross section in momentum space, and Ref.~\cite{Kovchegov:2012nd} gives the cross section in coordinate space.  Here we will perform the transformation to coordinate space to demonstrate exact agreement with Ref.~\cite{Kovchegov:2012nd}.  We have also compared with Ref.~\cite{Altinoluk:2018ogz} in momentum space.

Most of the work in performing the cross-check against Ref.~\cite{Kovchegov:2012nd} comes from unfolding the compact expression \eqref{e:GG1} back into coordinate space.  Inserting the Fourier transforms of the various quantities gives
\begin{align} \label{e:GG3}
& \frac{d\sigma^{G G}}{d^2 q_1 \, dy_1 \: d^2 q_2 \, dy_2} =
\left(\frac{1}{2(2\pi)^3}\right)^2 \frac{N_c^4}{4}
\int d^2\{ x_1 \, y_1 \, x_2 \, y_2 \, x_1^\prime \, y_1^\prime \, x_2^\prime \, y_2^\prime \, b_1 \, b_2 \}
\notag \\ \notag \\ &\times
\Bigg\{
\tr_D \left[ \Phi(x_1 - b_1 ; x_1^\prime - b_1) \: \Phi^\dagger(y_1 - b_1 ; y_1^\prime - b_1) \right] 
\tr_D \left[ \Phi(x_2 - b_2 ; x_2^\prime - b_2) \: \Phi^\dagger(y_2 - b_2 ; y_2^\prime - b_2) \right] \:
\notag \\ &\hspace{1cm}\times
\Big\langle 
\left| \hat{D}_2 \right|^2 (\ul{x_1^\prime} ,\ul{y_1^\prime} )
\left| \hat{D}_2 \right|^2 (\ul{x_2^\prime} ,\ul{y_2^\prime} ) 
\Big\rangle 
\notag \\ &\hspace{1cm}\times
e^{-i q_1 \cdot (\ul{x_1^\prime} - \ul{y_1^\prime} + \ul{x_1} - \ul{y_1})} \,
e^{-i q_2 \cdot (\ul{x_2^\prime} - \ul{y_2^\prime} + \ul{x_2} - \ul{y_2})} \:
\mu^2 ( \ul{b_1} \: , \: 0^+ ) \:
\mu^2 ( \ul{b_2}\: , \: 0^+ ) 
\notag \\ \notag \\ &+
\Bigg[
\tr_D \left[ \Phi(x_1 - b_1 ; x_1^\prime - b_1) \: \Phi^\dagger(y_2 - b_2 ; y_2^\prime - b_2) \right] 
\tr_D \left[ \Phi(x_2 - b_2 ; x_2^\prime - b_2) \: \Phi^\dagger(y_1 - b_1 ; y_1^\prime - b_1) \right] \:
\notag \\ &\hspace{1cm}\times
\frac{1}{N_c^2} \, \left| D_4 \right|^2 ( \ul{x_1^\prime} , \ul{y_1^\prime} , \ul{x_2^\prime} , \ul{y_2^\prime})
\Bigg]
\notag \\ & \hspace{1cm} \times
\Bigg[
e^{-i \ul{q_1} \cdot (\ul{x_1^\prime} - \ul{y_2^\prime} + \ul{x_1} - \ul{y_2} - \ul{b_1} + \ul{b_2})} \,
e^{-i \ul{q_2} \cdot (\ul{x_2^\prime} - \ul{y_1^\prime} + \ul{x_2} - \ul{y_1} - \ul{b_2} + \ul{b_1})} \:
\mu^2 ( \ul{b_1} \: , \: q_1^+ - q_2^+) \:
\mu^2 ( \ul{b_2} \: , \: q_2^+ - q_1^+ ) \:\:
\notag \\ & \hspace{1cm} +
e^{-i \ul{q_1} \cdot (\ul{x_1^\prime} - \ul{y_2^\prime} + \ul{x_1} - \ul{y_2} - \ul{b_1} + \ul{b_2}) } \,
e^{i \ul{q_2} \cdot (\ul{x_2^\prime} - \ul{y_1^\prime} + \ul{x_2} - \ul{y_1} + \ul{b_1} - \ul{b_2})} \:
\mu^2 ( \ul{b_1} \: , \: q_1^+ + q_2^+) \:
\mu^2 ( \ul{b_2} \: , \: -q_1^+ - q_2^+) 
\Bigg]
\Bigg\} .
\end{align}
In arriving at \eqref{e:GG3}, we have used symmetry properties of the squared dipole and the gluon wave functions \eqref{e:GWFsq}.  Inserting \eqref{e:GWF2} in the four wave functions generates 16 terms, each containing 4 delta functions.  As before, after integrating over those delta functions, we drop any remaining primes on the integration variables; this leads to all of the various wave functions in a given set of terms having the same arguments, with the differences residing in the Wilson line interactions.

For the first term, the dependences on coordinates with subscript $1$ and subscript $2$ completely factorize:
\begin{align}
\int d^2 &\{ x_1 \, y_1 \, x_1^\prime \, y_1^\prime \} \:
e^{-i q_1 \cdot (\ul{x_1^\prime} - \ul{y_1^\prime} + \ul{x_1} - \ul{y_1})} \:
\tr_D \left[ \Phi(x_1 - b_1 ; x_1^\prime - b_1) \: \Phi^\dagger(y_1 - b_1 ; y_1^\prime - b_1) \right] 
\left| \hat{D}_2 \right|^2 (\ul{x_1^\prime} ,\ul{y_1^\prime} )
\notag \\ \notag \\ &=
2 \:e^{-i q_1 \cdot (\ul{x_1} - \ul{y_1})} \: \tr[ \phi(x_1 - b_1) \, \phi^\dagger (y_1 - b_1)] \:
\notag \\ &\hspace{1cm} \times
\Bigg\{
\left| \hat{D}_2 \right|^2 (\ul{x_1} ,\ul{y_1} )
- \left| \hat{D}_2 \right|^2 (\ul{x_1} ,\ul{b_1} )
- \left| \hat{D}_2 \right|^2 (\ul{b_1} ,\ul{y_1} )
+ 1 \Bigg\} ,
\end{align}
so that we can just multiply this result by the same expression under the exchange $1 \leftrightarrow 2$.  For the second and third terms which share a set of ``crossed'' wave functions, the expressions do not cleanly factorize, but the delta functions generated do possess the following symmetry:
\begin{align}
&\Bigg\{
\delta^2 (x_1 - b_1) \, \delta^2 (y_2 - b_2) \: \tr[ \phi(x_1^\prime - b_1) \, \phi^\dagger (y_2^\prime - b_2)] 
\notag \\ &\hspace{1cm} 
- \delta^2 (x_1 - b_1) \, \delta^2 (y_2^\prime - b_2) \: \tr[ \phi(x_1^\prime - b_1) \, \phi^\dagger (y_2 - b_2)] 
\notag \\ &\hspace{1cm} 
- \delta^2 (x_1^\prime - b_1) \, \delta^2 (y_2 - b_2) \: \tr[ \phi(x_1 - b_1) \, \phi^\dagger (y_2^\prime - b_2)] 
\notag \\ &\hspace{1cm} 
+ \delta^2 (x_1^\prime - b_1) \, \delta^2 (y_2^\prime - b_2) \: \tr[ \phi(x_1 - b_1) \, \phi^\dagger (y_2 - b_2)] 
\Bigg\} \:\: \times \:\: \Bigg\{ 1 \leftrightarrow 2 \Bigg\} .
\end{align}
The result of these delta functions, after dropping all primes, is to sum over all possible ways to take the squared quadrupole $| D_4 |^2 (x_1 , y_1 , x_2 , y_2)$ and map $x_i \rightarrow b_i$ or $y_i \rightarrow b_i$, generating a minus sign for each such replacement.  Note that some of these permutations will cancel pairs of Wilson lines to generate squared dipoles and, in the last term, the unit operator.  The result is
\begin{align} \label{e:GG4}
& \frac{d\sigma^{G G}}{d^2 q_1 \, dy_1 \: d^2 q_2 \, dy_2} =
\frac{N_c^4}{\left[ 2(2\pi)^3 \right]^2}
\int d^2\{ x_1 \, y_1 \, x_2 \, y_2 \, b_1 \, b_2 \}
\notag \\ \notag \\ &\times
\Bigg\{
e^{-i q_1 \cdot (\ul{x_1} - \ul{y_1})} \, e^{-i q_2 \cdot (\ul{x_2} - \ul{y_2})} \:
\mu^2 ( \ul{b_1} \: , \: 0^+ ) \: \mu^2 ( \ul{b_2}\: , \: 0^+ ) \:
\notag \\ &\hspace{1cm} \times
\tr_D \left[ \phi(x_1 - b_1) \: \phi^\dagger(y_1 - b_1) \right] 
\tr_D \left[ \phi(x_2 - b_2) \: \phi^\dagger(y_2 - b_2) \right] \:
\notag \\ &\hspace{1cm} \times
\Bigg\langle 
\Bigg( 1
- \left| \hat{D}_2 \right|^2 (\ul{x_1} ,\ul{b_1} )
- \left| \hat{D}_2 \right|^2 (\ul{b_1} ,\ul{y_1} )
+ \left| \hat{D}_2 \right|^2 (\ul{x_1} ,\ul{y_1} ) \Bigg)
\notag \\ &\hspace{1cm} \times
\Bigg( 1
- \left| \hat{D}_2 \right|^2 (\ul{x_2} ,\ul{b_2} )
- \left| \hat{D}_2 \right|^2 (\ul{b_2} ,\ul{y_2} )
+ \left| \hat{D}_2 \right|^2 (\ul{x_2} ,\ul{y_2} ) \Bigg)
\Bigg\rangle 
\notag \\ \notag \\ &+
\Bigg[
e^{-i \ul{q_1} \cdot (\ul{x_1} - \ul{y_2})} \, e^{-i \ul{q_2} \cdot (\ul{x_2} - \ul{y_1})} \:
\mu^2 ( \ul{b_1} \: , \: q_1^+ - q_2^+) \: \mu^2 ( \ul{b_2} \: , \: q_2^+ - q_1^+ ) \:\:
\notag \\ & \hspace{1cm} +
e^{-i \ul{q_1} \cdot (\ul{x_1} - \ul{y_2}) } \, e^{i \ul{q_2} \cdot (\ul{x_2} - \ul{y_1})} \:
\mu^2 ( \ul{b_1} \: , \: q_1^+ + q_2^+) \: \mu^2 ( \ul{b_2} \: , \: -q_1^+ - q_2^+) 
\Bigg]
\notag \\ & \hspace{1cm} \times
\tr_D \left[ \phi(x_1 - b_1) \: \phi^\dagger(y_2 - b_2) \right] \:
\tr_D \left[ \phi(x_2 - b_2) \: \phi^\dagger(y_1 - b_1) \right] 
\notag \\ & \times
\frac{1}{N_c^2} \Bigg( 
\left| D_4 \right|^2 ( \ul{x_1} , \ul{y_1} , \ul{x_2} , \ul{y_2})
- \left| D_4 \right|^2 ( \ul{x_1} , \ul{b_1} , \ul{x_2} , \ul{y_2})
- \left| D_4 \right|^2 ( \ul{x_1} , \ul{y_1} , \ul{b_2} , \ul{y_2})
+ \left| D_4 \right|^2 ( \ul{x_1} , \ul{b_1} , \ul{b_2} , \ul{y_2})
\notag \\ & \hspace{1cm}
- \left| D_4 \right|^2 ( \ul{x_1} , \ul{y_1} , \ul{x_2} , \ul{b_2})
+ \left| D_4 \right|^2 ( \ul{x_1} , \ul{b_1} , \ul{x_2} , \ul{b_2})
+ \left| D_2 \right|^2 ( \ul{x_1} , \ul{y_1})
- \left| D_2 \right|^2 ( \ul{x_1} , \ul{b_1})
\notag \\ & \hspace{1cm}
- \left| D_4 \right|^2 ( \ul{b_1} , \ul{y_1} , \ul{x_2} , \ul{y_2})
+ \left| D_2 \right|^2 ( \ul{x_2} , \ul{y_2})
+ \left| D_4 \right|^2 ( \ul{b_1} , \ul{y_1} , \ul{b_2} , \ul{y_2})
- \left| D_2 \right|^2 ( \ul{b_2} , \ul{y_2})
\notag \\ & \hspace{1cm}
+ \left| D_4 \right|^2 ( \ul{b_1} , \ul{y_1} , \ul{x_2} , \ul{b_2})
- \left| D_2 \right|^2 ( \ul{x_2} , \ul{b_2})
- \left| D_2 \right|^2 ( \ul{b_1} , \ul{y_1})
+ 1 \Bigg)
\Bigg\}.
\end{align}
The last step is to insert the explicit form of the gluon emission wave function \eqref{e:GWF} and to convert the squared fundamental dipole and quadrupole into the adjoint dipole and quadrupole using \eqref{e:Wconvert1}.  The final expression for the double-gluon cross section is
\begin{align} \label{e:GG5}
& \frac{d\sigma^{G G}}{d^2 q_1 \, dy_1 \: d^2 q_2 \, dy_2} =
\frac{1}{\left[ 2(2\pi)^3 \right]^2} 
\int d^2\{ x_1 \, y_1 \, x_2 \, y_2 \, b_1 \, b_2 \}
\notag \\ \notag \\ &\times
\Bigg\{ \left( \frac{4 \alpha_s C_f}{\pi} \right)^2 \:
e^{-i q_1 \cdot (\ul{x_1} - \ul{y_1})} \, e^{-i q_2 \cdot (\ul{x_2} - \ul{y_2})} \:
\left[ \frac{2 N_c}{g^2} \: \mu^2 ( \ul{b_1} \: , \: 0^+ ) \right] \: 
\left[ \frac{2 N_c}{g^2} \: \mu^2 ( \ul{b_2}\: , \: 0^+ ) \right] 
\notag \\ &\hspace{1cm} \times
\frac{(\ul{x_1} - \ul{b_1}) \cdot (\ul{y_1} - \ul{b_1})}{(\ul{x}_1 - \ul{b}_1)_T^2 \, (\ul{y}_1 - \ul{b}_1)_T^2} \:\:
\frac{(\ul{x_2} - \ul{b_2}) \cdot (\ul{y_2} - \ul{b_2})}{(\ul{x}_2 - \ul{b}_2)_T^2 \, (\ul{y}_2 - \ul{b}_2)_T^2}
\notag \\ &\hspace{1cm} \times
\Bigg\langle 
\Bigg( 1
- \hat{D}_2^{adj} (\ul{x_1} ,\ul{b_1} )
- \hat{D}_2^{adj} (\ul{b_1} ,\ul{y_1} )
+ \hat{D}_2^{adj} (\ul{x_1} ,\ul{y_1} ) \Bigg)
\notag \\ &\hspace{1cm} \times
\Bigg( 1
- \hat{D}_2^{adj} (\ul{x_2} ,\ul{b_2} )
- \hat{D}_2^{adj} (\ul{b_2} ,\ul{y_2} )
+ \hat{D}_2^{adj} (\ul{x_2} ,\ul{y_2} ) \Bigg)
\Bigg\rangle 
\notag \\ \notag \\ &+
\left( \frac{4 \alpha_s}{\pi} \right)^2
\left( \frac{C_F}{2 N_c} \right)
\Bigg[
e^{-i \ul{q_1} \cdot (\ul{x_1} - \ul{y_2})} \, e^{-i \ul{q_2} \cdot (\ul{x_2} - \ul{y_1})} \:
\left[ \frac{2 N_c}{g^2} \: \mu^2 ( \ul{b_1} \: , \: q_1^+ - q_2^+) \right] \: 
\left[ \frac{2 N_c}{g^2} \: \mu^2 ( \ul{b_2} \: , \: q_2^+ - q_1^+ ) \right]
\notag \\ & \hspace{1cm} +
e^{-i \ul{q_1} \cdot (\ul{x_1} - \ul{y_2}) } \, e^{i \ul{q_2} \cdot (\ul{x_2} - \ul{y_1})} \:
\left[ \frac{2 N_c}{g^2} \: \mu^2 ( \ul{b_1} \: , \: q_1^+ + q_2^+) \right] \: 
\left[ \frac{2 N_c}{g^2} \: \mu^2 ( \ul{b_2} \: , \: -q_1^+ - q_2^+) \right]
\Bigg]
\notag \\ & \hspace{1cm} \times
\frac{(\ul{x_1} - \ul{b_1}) \cdot (\ul{y_2} - \ul{b_2})}{(\ul{x}_1 - \ul{b}_1)_T^2 \, (\ul{y}_2 - \ul{b}_2)_T^2} \:\:
\frac{(\ul{x_2} - \ul{b_2}) \cdot (\ul{y_1} - \ul{b_1})}{(\ul{x}_2 - \ul{b}_2)_T^2 \, (\ul{y}_1 - \ul{b}_1)_T^2}
\notag \\ & \times
\Bigg( 
D_4^{adj} ( \ul{x_1} , \ul{y_1} , \ul{x_2} , \ul{y_2})
- D_4^{adj} ( \ul{x_1} , \ul{b_1} , \ul{x_2} , \ul{y_2})
- D_4^{adj} ( \ul{x_1} , \ul{y_1} , \ul{b_2} , \ul{y_2})
+ D_4^{adj} ( \ul{x_1} , \ul{b_1} , \ul{b_2} , \ul{y_2})
\notag \\ & \hspace{1cm}
- D_4^{adj} ( \ul{x_1} , \ul{y_1} , \ul{x_2} , \ul{b_2})
+ D_4^{adj} ( \ul{x_1} , \ul{b_1} , \ul{x_2} , \ul{b_2})
+ D_2^{adj} ( \ul{x_1} , \ul{y_1})
- D_2^{adj} ( \ul{x_1} , \ul{b_1})
\notag \\ & \hspace{1cm}
- D_4^{adj} ( \ul{b_1} , \ul{y_1} , \ul{x_2} , \ul{y_2})
+ D_2^{adj} ( \ul{x_2} , \ul{y_2})
+ D_4^{adj} ( \ul{b_1} , \ul{y_1} , \ul{b_2} , \ul{y_2})
- D_2^{adj} ( \ul{b_2} , \ul{y_2})
\notag \\ & \hspace{1cm}
+ D_4^{adj} ( \ul{b_1} , \ul{y_1} , \ul{x_2} , \ul{b_2})
- D_2^{adj} ( \ul{x_2} , \ul{b_2})
- D_2^{adj} ( \ul{b_1} , \ul{y_1})
+ 1 \Bigg)
\Bigg\} ,
\end{align}
in exact agreement with Eqs.~(33) and (38) of \cite{Kovchegov:2012nd}.  This matching again employs the dictionary \eqref{e:mumap} to the discrete valence quark model.  We also note that Ref.~\cite{Kovchegov:2012nd} has no dependence on the plus momentum off-forwardness: $\mu^2(\ul{b} , \Delta q^+) = \mu^2(\ul{b})$, such that all three terms of \eqref{e:GG5} add together with the same densities $T(b_1 - B) \, T(b_2 - B)$.  This assumption that $\mu^2 (\ul{q_1} - \ul{q_2} , q_1^+ - q_2^+) = \mu^2 (\ul{q_1} - \ul{q_2})$ corresponds in coordinate space to $\mu^2 (\ul{x} , x^-) = \mu^2 (\ul{x}) \, \delta(x^-)$, so that the projectile is infinitely Lorentz-contracted into a delta function pancake at $x^- = 0$.  Our calculation here relaxes that assumption and is therefore sensitive to the 3D structure of the projectile.  The successful cross-check \eqref{e:GG5} against Ref.~\cite{Kovchegov:2012nd} serves as a validation of the preceding calculations based on the double-pair expressions \eqref{e:saus2} and \eqref{e:Pac2}.

%
\section{Hadronization in the Heavy Flavor Sector}
\label{sec:Heavy} 
%

%
\begin{figure}
\includegraphics[width=\textwidth]{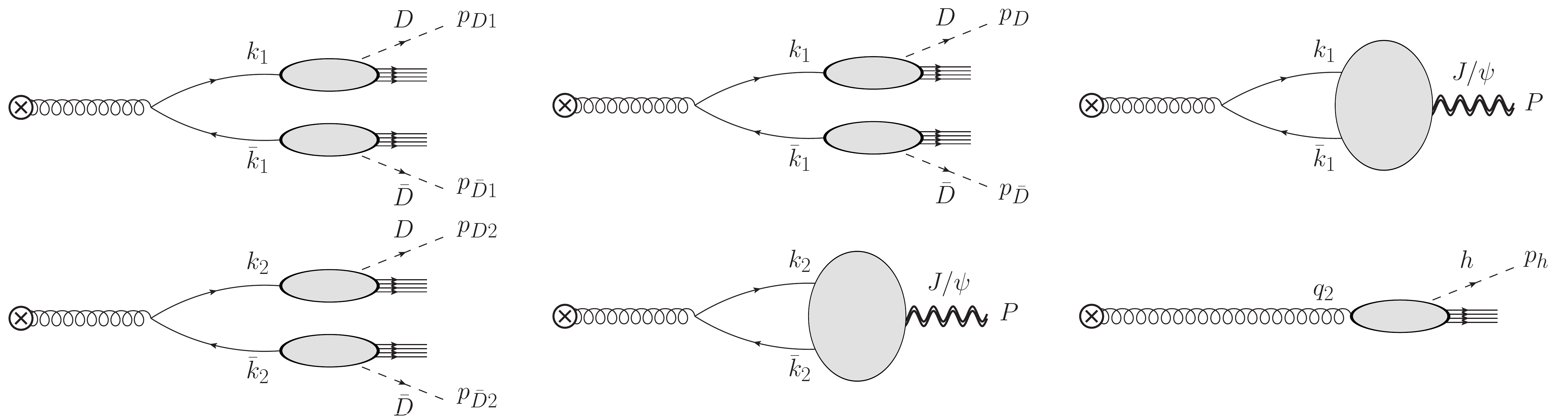}
\caption{A few examples of how the partonic cross sections calculated in this paper can hadronize into the heavy flavor sector.  The partonic $(c \barc) \, (c \barc)$ cross section could hadronize into the open heavy flavor sector such as $(D \barD) \, (D \barD)$ (left panel, Eq.~\eqref{e:DDbarDDbar}) or may include heavy quarkonia such as $(D \barD) \, (J/\psi)$ (center panel, Eq.~\eqref{e:JpsiDDbar}).  The partonic $(c \barc) \, G$ cross section could hadronize similarly, but include a light hadron $h$ produced by the gluon, as in $(J/\psi) \, h$ (right panel, Eq.~\eqref{e:Jpsih}).}
\label{f:Hadroniz}
\end{figure}
%

To connect with the experimental measurements of interest, the partonic cross sections calculated above need to be convoluted with appropriate nonperturbative descriptions of hadronization.  While there is an abundance of hadronic observables which are of interest, one of the cleanest analogs between the partonic and hadronic cross sections is in the heavy flavor sector.  While for light hadron production, many partonic channels can all contribute with comparable weights, heavy flavor production is dominated by the production of heavy quark pairs~\cite{Huang:2015mva,Nahrgang:2013saa}. We can characterize heavy flavor production in either the open or hidden heavy flavor sectors; here for specificity we will consider the production of $D$ and $\bar D$ mesons in the open sector or $J/\psi$ mesons in the hidden sector, and we will generally follow the methods of augmenting CGC calculations with hadronization prescribed in Ref.~\cite{Ma:2018bax}.  The various hadronic cross sections presented in this Section are the final primary result of this paper.

For production of open $D$ mesons, we need to convolute the partonic cross sections with fragmentation functions following the standard techniques \cite{Collins:2011zzd}.  Thus from the single-$c \barc$ cross section expressed in Eqs.~\eqref{e:sglpairxsec} and \eqref{e:sglpair} we can directly compute the $D \bar D$ cross section via the convolution
\begin{align} \label{e:DDbar}
\frac{d\sigma^{D \barD}}{d^2 p_D \, dy \: d^2 p_{\barD} \, d\bary} &=
\int\limits_{\frac{p_D^+}{P_a^+/a}}^1 \frac{dz}{z^2} \:
\int\limits_{\frac{p_\barD^+}{P_a^+/a}}^1 \frac{d\barz}{\barz^2} \:
\left[ \frac{d\sigma^{c \barc}}{d^2 k \, dy \: d^2 \bark \, d\bary} \right] \:
D_{D / c} (z) \: D_{\barD / \barc} (\barz) ,
\end{align}
where the final-state momenta of the $D$ and $\barD$ are given by $p_D^\mu = z k^\mu$ and $p_\barD^\mu = \barz \bark^\mu$, respectively in the limit where the hadrons can be considered massless.  In this limit, the rapidities of the hadrons are equal to the rapidities of the quarks: $y_D = y$ and $y_\barD = \bary$, and the factors of $1/z^2 \, , \, 1/\barz^2$ arise from the change of variables $d^2 p_D = z^2 \, d^2 k$ and $d^2 p_\barD = \barz^2 \, d^2 \bark$.  The fragmentation functions $D_{D / c}$ and $D_{\barD / \barc}$ can be taken from e.g. Ref.~\cite{Braaten:1994bz}.  The lower limits of integration in \eqref{e:DDbar} come from the eikonal approximation for the emission of the soft gluon from the projectile: $k^+ , \bark^+ \ll \tfrac{1}{a} P_a^+$, where $\tfrac{1}{a} P_a^+$ is the average momentum per nucleon of a projectile with $a$ nucleons whose total momentum is $P_a^+$.  

In principle the $c \barc$ production mechanism calculated in Eq.~\eqref{e:sglpair} is only one partonic channel that can contribute to final-state $D \barD$ production, and one should generalize \eqref{e:DDbar} by summing over all such partonic channels.  However, as argued above, in the heavy flavor sector we expect the $c \barc$ channel to make the dominant contribution to $D \barD$ production.  It is also important to note that the assumption of independent fragmentation of the $c$ and $\barc$ quarks relies on the partons being well-separated in phase space (either in transverse momentum or in rapidity) such that nonperturbative interference effects in hadronization can be neglected.  Studies of the boundaries of independent fragmentation have been performed in the context of distinguishing target vs. current fragmentation regions in electron-proton collisions \cite{Boglione:2016bph}.

In the same way, we can convolute the $(c \barc) \, (c \barc)$ cross section expressed in Eqs.~\eqref{e:dblxsec}, \eqref{e:saus2}, and \eqref{e:Pac2} with fragmentations to obtain the four-particle hadronic cross section for $D \barD \, D \barD$ production:
\begin{align} \label{e:DDbarDDbar}
& \frac{d\sigma^{(D \barD) \, (D \barD)}}{d^2 p_{D 1} \, dy_1 \: d^2 p_{\barD 1} \, d\bary_1 \: 
d^2 p_{D 2} \, dy_2 \: d^2 p_{\barD 2} d\bary_2} =
\int\limits_{\frac{p_{D 1}^+}{P_a^+/a}}^1 \frac{dz_1}{z_1^2} \:
\int\limits_{\frac{p_{\barD 1}^+}{P_a^+/a}}^1 \frac{d\barz_1}{\barz_1^2} \:
\int\limits_{\frac{p_{D 2}^+}{P_a^+/a}}^1 \frac{dz_2}{z_2^2} \:
\int\limits_{\frac{p_{\barD 2}^+}{P_a^+/a}}^1 \frac{d\barz_2}{\barz_2^2} \:
\notag \\ & \times
\left[ \frac{d\sigma^{(c \barc) \, (c \barc)}}{d^2 k_1 \, dy_1 \: d^2 \bark_1 \, d\bary_1 \: 
d^2 k_2 \, dy_2 \: d^2 \bark_2 \, d\bary_2} \right] \:
D_{D / c} (z_1) \: D_{\barD / \barc} (\barz_1) \: D_{D / c} (z_2) \: D_{\barD / \barc} (\barz_2) .
\end{align}
Here the final-state hadron momenta are related to the partonic ones by $p_{D 1}^\mu = z_1 \, k_1^\mu$, $p_{\barD 1}^\mu = \barz_1 \, \bark_1^\mu$, $p_{D 2}^\mu = z_2 \, k_2^\mu$, and $p_{\barD 2}^\mu = \barz_2 \, \bark_2^\mu$.  Again, implicit in \eqref{e:DDbarDDbar} is an assumption that the hadrons are well-separated in phase space, such that there is no interference during hadronization which mixes the pairs.  This can be accomplished, for instance, by a rapidity gap
\footnote{Note, however, that if the rapidity gap between any two particles becomes sufficiently large $\Delta y_{i j} > \frac{1}{\alpha_s}$, small-$x$ evolution between the produced particles can become important.  Such considerations are beyond the scope of this paper, and we presently restrict ourselves to smaller rapidity separations $\Delta y_{i j} < \frac{1}{\alpha_s}$.}
between the pairs $y_1 , \bary_1 \gg y_2 , \bary_2$ and a large relative momentum between each corresponding $D$ and $\barD$ meson: $| \ul{p}_{D 1} - \ul{p}_{\barD 1}|_T \, , \, | \ul{p}_{D 2} - \ul{p}_{\barD 2}|_T \gg \Lambda_{QCD}$.  Such kinematics are, in fact, appropriate for exploring the long-range correlations which may exist among the produced hadrons.

Likewise, we can also convolute the $(c \barc) \, G$ cross section given in \eqref{e:mixedxsec} with fragmentation functions to describe the three-particle correlations between a $D$ meson, a $\barD$ meson, and a light hadron $h$.  That expression is given by
\begin{align} \label{e:DDbarh}
\frac{d\sigma^{(D \barD) \, h}}{d^2 p_D \, dy \: d^2 p_{\barD} \, d\bary \: 
d^2 p_h \, dy_h } &=
\int\limits_{\frac{p_{D}^+}{P_a^+/a}}^1 \frac{dz}{z^2} \:
\int\limits_{\frac{p_{\barD}^+}{P_a^+/a}}^1 \frac{d\barz}{\barz^2} \:
\int\limits_{\frac{p_{h}^+}{P_a^+/a}}^1 \frac{dz_h}{z_h^2} \:
\left[ \frac{d\sigma^{(c \barc) \, G}}{d^2 k \, dy \: d^2 \bark \, d\bary \: 
d^2 q \, dy_h } \right] \:
\notag \\ & \times
D_{D / c} (z) \: D_{\barD / \barc} (\barz) \: D_{h / G} (z_h) .
\end{align}
Again, the final-state hadronic momenta are related to the partonic ones by $p_D = z \, k$, $p_\barD = \barz \, \bark$, and $p_h = z_h \, q$, and the usual caveats for independent fragmentation apply.

For the production of heavy quarkonia such as the $J/\psi$, we will follow the procedure of \cite{Ma:2018bax} and employ the Improved Color Evaporation Model (ICEM) \cite{Ma:2016exq} to treat the hadronization of a partonic $c \barc$ pair into a final-state $J/\psi$.  For the hadronization of a single pair, we obtain the hidden heavy flavor analogue of \eqref{e:DDbar}:
\begin{align} \label{e:Jpsi}
\frac{d\sigma^{J/\psi}}{d^2P \, dY} &= F_{J/\psi} \int\limits_{m_{J/\psi}}^{2 m_D} dM \:
\left(\frac{M}{m_{J/\psi}}\right)^2 \int\limits_0^{\sqrt{\tfrac{M^2}{4} - m_c^2}} d(\Delta k) \int\limits_0^{2\pi} d\phi \,
\notag \\ &\times
\frac{(\Delta k) \, \sqrt{M^2 + P_T^2}}
{M \sqrt{\big( m_c^2 + k_T^2 \big) \big(m_c^2 + \bark_T^2\big)} \big|\sinh(y - \bary) \big|} \:
\left[ \frac{d\sigma^{c \barc}}{d^2 k \, dy \: d^2 \bark \, d\bary} \right] ,
\end{align}
with $M = (k + \bark)^2$ the invariant mass of the $c \barc$ pair, $\Delta k$ the relative momentum between the $c \barc$ pair in its rest frame, and $\phi$ the corresponding angle.  The ratio $\frac{M}{m_{J/\psi}}$ is a conversion factor allowing the momentum $P$ of the final $J/\psi$ to differ from the $c \barc$ center-of-mass momentum, and the other factors correspond to the Jacobian from the change of variables.  The single nonperturbative parameter $F_{J/\psi}$ describes the probability for the $c \barc$ pair to hadronize to the $J/\psi$ by randomly emitting soft particles.

Similarly, we can translate the perturbative $(c \barc) (c \barc)$ cross section from Eqs.~\eqref{e:dblxsec}, \eqref{e:saus2}, and \eqref{e:Pac2} into a production cross section for two $J/\psi$ mesons, writing the analogous expression
\begin{align} \label{e:JpsiJpsi}
&\frac{d\sigma^{(J/\psi) \,( J/\psi)}}{d^2P_1 \, dY_1 \: d^2P_2 \, dY_2} = F_{J/\psi}^2 
\int\limits_{m_{J/\psi}}^{2 m_D} dM_1 \, dM_2 \:
\left(\frac{M_1}{m_{J/\psi}} \frac{M_2}{m_{J/\psi}}\right)^2 \,
\int\limits_0^{2\pi} d\phi_1 d\phi_2 \hspace{-0.5cm}
\int\limits_0^{\sqrt{\tfrac{M_1^2}{4} - m_c^2}} \hspace{-0.5cm}  d(\Delta k_1) \hspace{-0.5cm}
\int\limits_0^{\sqrt{\tfrac{M_2^2}{4} - m_c^2}} \hspace{-0.5cm}  d(\Delta k_2)
\notag \\ &\times
\frac{(\Delta k_1) \, \sqrt{M_1^2 + P_{1T}^2}}
{M_1 \sqrt{\big( m_c^2 + k_{1T}^2 \big) \big(m_c^2 + \bark_{1T}^2\big)} \big|\sinh(y_1 - \bary_1) \big|} \,
\frac{(\Delta k_2) \, \sqrt{M_2^2 + P_{2T}^2}}
{M_2 \sqrt{\big( m_c^2 + k_{2T}^2 \big) \big(m_c^2 + \bark_{2T}^2\big)} \big|\sinh(y_2 - \bary_2) \big|} \:
\notag \\ &\times
\left[ \frac{d\sigma^{(c \barc) \, (c \barc)}}{d^2 k_1 \, dy_1 \: d^2 \bark_1 \, d\bary_1 \: d^2 k_2 \, dy_2 \: d^2 \bark_2 \, d\bary_2} \right] .
\end{align}
Now $M_1$ is the invariant mass of the pair $(k_1 , \bark_1)$ and $M_2$ is the invariant mass of the pair $(k_2 , \bark_2)$, with the relative momenta $\Delta k_1 , \Delta k_2$ and angles $\phi_1 , \phi_2$, respectively.  Again, this expression assumes a sufficient kinematic separation in phase space such that we do not have to consider interferences during hadronization.

Last, we can consider hadronic cross sections which involve both the formation of a quarkonium state via the ICEM and fragmentation.  The partonic $(c \barc) \, (c \barc)$ cross section can hadronize into a $J/\psi$ as well as an open $D \barD$ pair, which we write as
\begin{align} \label{e:JpsiDDbar}
&\frac{d\sigma^{(D \barD) \, (J/\psi)}}{d^2 p_D \, dy \: d^2p_\barD \, d\bary \: d^2P \, dY} = F_{J/\psi} \int\limits_{m_{J/\psi}}^{2 m_D} dM_2 \: \left(\frac{M_2}{m_{J/\psi}}\right)^2 
\int\limits_0^{\sqrt{\tfrac{M_2^2}{4} - m_c^2}} d(\Delta k_2) \int\limits_0^{2\pi} d\phi_2 \,
\notag \\ &\times
\int\limits_{\frac{p_D^+}{P_a^+/a}}^1 \frac{dz_1}{z_1^2} \:
\int\limits_{\frac{p_\barD^+}{P_a^+/a}}^1 \frac{d\barz_1}{\barz_1^2} \:
D_{D / c} (z_1) \: D_{\barD / \barc} (\barz_1) 
\frac{(\Delta k_2) \, \sqrt{M_2^2 + P_T^2}}
{M_2 \sqrt{\big( m_c^2 + k_{2T}^2 \big) \big(m_c^2 + \bark_{2T}^2\big)} \big|\sinh(y_2 - \bary_2) \big|} \:
\notag \\ &\times
\left[ \frac{d\sigma^{(c \barc) \, (c \barc)}}{d^2 k_1 \, dy_1 \: d^2 \bark_1 \, d\bary_1 \: 
d^2 k_2 \, dy_2 \: d^2 \bark_2 \, d\bary_2} \right] .
\end{align}
Similarly, we can write the cross section for the hadronization of the partonic $(c \barc) \, G$ state into a $J/\psi$ meson in association with a light hadron $h$:
\begin{align} \label{e:Jpsih}
&\frac{d\sigma^{(J/\psi) \, h}}{d^2P \, dY \: d^2 p_h \, dy_h} = F_{J/\psi} \int\limits_{m_{J/\psi}}^{2 m_D} dM_1 \: \left(\frac{M_1}{m_{J/\psi}}\right)^2 
\int\limits_0^{\sqrt{\tfrac{M_1^2}{4} - m_c^2}} d(\Delta k_1) \int\limits_0^{2\pi} d\phi_1 \,
\int\limits_{\frac{p_h^+}{P_a^+/a}}^1 \frac{dz_h}{z_h^2} \:
D_{h / G} (z_h) 
\notag \\ & \hspace{1cm} \times
\frac{(\Delta k_1) \, \sqrt{M_1^2 + P_T^2}}
{M_1 \sqrt{\big( m_c^2 + k_{1T}^2 \big) \big(m_c^2 + \bark_{1T}^2\big)} \big|\sinh(y_1 - \bary_1) \big|} \:
\left[ \frac{d\sigma^{(c \barc) \, G}}{d^2 k_1 \, dy_1 \: d^2 \bark_1 \, d\bary_1 \: d^2 q_2 \, dy_h} \right] .
\end{align}
The above hadronic cross sections represent all possible ways for the partonic production of a $(c \barc) \, (c \barc)$ state to hadronize into open $D \barD$ pairs or $J/\psi$ quarkonia, as well as all possible ways for a $(c \barc) \, G$ state to hadronize into $D \barD$ or $J/\psi$ along with a light hadron $h$.  These hadronic cross sections open the door to study a wealth of multi-hadron final states in $pA$ and heavy-light ion collisions.

Before concluding, a few comments are in order about the models and assumptions used for hadronization above.  In the inclusive hadronization of a charm quark into a $D$ meson or a gluon into a light hadron, we have employed collinear fragmentation functions.  The use of collinear nonperturbative functions for fragmentation stands in contrast to the description of the initial state, which includes the effects of intrinsic transverse momentum arising from the nonperturbative wave functions of the projectile and target.  This asymmetric treatment of the intrinsic transverse momentum scales is justified by the asymmetric nature of the semi-dilute/dense regime.  The intrinsic transverse momentum of the dense target is characterized by the target saturation scale $Q_{s, t}$ and is resummed to all orders; the intrinsic transverse momentum of the pojectile is characterized by the projectile saturation scale $Q_{s, p}$ and is computed order by order in perturbation theory; and the intrinsic transverse momentum of hadronization is simply of order $\Lambda_{QCD}$ and is not enhanced by any high density scales.  Thus for the semi-dilute / dense regime, we have the hierarchy of scales \eqref{e:heavylight}, for which we can neglect the intrinsic transverse momentum characterized by the fragmentation functions.  This hierarchy of scales is in the same spirit as the framework of hybrid factorization, in which the production of two distinguishable heavy quarkonia has recently been calculated \cite{Boussarie:2018zwg}.

One potential drawback to the asymmetric treatment of the projectile, target, and fragmentation sectors is that the fragmentation functions employed above do not possess an unambiguous factorization scale $\mu_F$.  This feature also applies to the description of fragmentation employed in Ref.~\cite{Ma:2018bax}.  While the particular fragmentation functions given in Ref.~\cite{Braaten:1994bz} do not explicitly refer to a factorization scale $\mu_F$, fragmentation functions in general -- and the available fragmentation functions for light hadrons in particular -- carry such a dependence on an arbitrary scale.  This arbitrary scale dependence in the fragmentation sector is compensated by the scale dependence of the parton distribution functions of the projectile and target, such that the observable cross section is overall invariant under the renormalization-group evolution of its various nonperturbative pieces.  In our case, the projectile, target, and fragmentation sectors have all been treated very differently, such that the scale dependence coming from the projectile and target distributions is ambiguous.  
\footnote{This is not to be confused with the scale dependence on a rapidity regulator; the RG evolution in rapidity is contained within the rapidity dependence of the Wilson line traces.  This dependence is generally characterized by the JIMWLK functional evolution equation.}
In a more complete treatment which puts the nonperturbative projectile, target, and fragmentation sectors on comparable footing, the cancellation of this factorization scale dependence would become explicit.  This is what is seen explicitly in the standard hybrid factorization framework of the dilute / dense regime~\cite{Chirilli:2012jd,Chirilli:2011km}, and we expect that the same would be true for the semi-dilute / dense regime in a treatment such as that of Ref.~\cite{Boussarie:2018zwg}, which formulates hybrid factorization in terms of double parton distributions of the projectile.

Finally, let us note that while the ICEM is a particularly simple and convenient model for describing the hadronization of a quarkonium state from a $q \barq$ pair, it is by no means unique.  A variety of other descriptions of this hadronization process exist, in particular the effective field theory of Non-Relativistic Quantum Chromodynamics (NRQCD)  \cite{Bodwin:1994jh}.  While different hadronization formalisms have their own advantages and disadvantages, NRQCD has the particular advantage of being a self-consistent effective field theory of QCD.  Employing it requires a more detailed treatment than just a simple convolution of the partonic $c \barc$ cross section as done above, using a series of projection operators to select out the quantum numbers of the $c \barc$ state appropriate for a given hadronization channel.  Of these projection operators, the color projections onto singlet and octet $c \barc$ states will require a more detailed implementation because they will modify the Wilson lines which enter the multipole traces.  These various projections are in principle straightforward, but they are beyond the scope of this paper; we leave the incorporation of an NRQCD-based approach to quarkonium production for future work.  It is interesting to note, however, that inclusive $J/\psi$ production at small $x$ in the CGC framework was studied in Ref.~\cite{Ma:2018bax} using both NRQCD and the ICEM for hadronization, concluding that the ICEM is a good approximation to the NRQCD approach due to the dominance of the ${}^3 S_1^{[8]}$ production channel.

%
\section{Conclusions}
\label{sec:concl} 
%

In this paper, we have computed a number of cross sections for the production of multiple particles at mid-rapidity in the semi-dilute / dense regime of the color-glass condensate effective field theory.  At the partonic level, we have computed for the first time the production cross sections for two quark/antiquark pairs $(q \barq) \, (q \barq)$ (Eqs.~\eqref{e:dblxsec}, \eqref{e:saus2}, and \eqref{e:Pac2}) and for one quark/antiquark pair plus a gluon $(q \barq) \, G$ (Eq.~\eqref{e:mixedxsec}).  The double-pair expression significantly extends previous work \cite{Altinoluk:2016vax} in that it is fully differential in all four particles, includes all time orderings and all orders of multiple rescattering in the target fields, and is evaluated with exact $N_c$.  These new partonic cross sections are one of the primary new results of this paper.  

Additionally, we proved a simple mapping \eqref{e:Gmap2} between the production amplitude for a $q \barq$ pair and the production amplitude for a gluon, which we used to obtain the $(q \barq) \, G$ cross section from the $(q \barq) \, (q \barq)$ cross section, and which we validated by cross-checking the single-gluon \eqref{e:Gcheck2} and double-gluon \eqref{e:GG1} production cross sections against the literature \cite{Kovchegov:2012mbw, Kovchegov:2012nd, Altinoluk:2018ogz}.  The mapping \eqref{e:Gmap2} and its application to derive whole classes of cross sections from the multi-pair cross section is the second primary result of this paper.

Finally, in Sec.~\ref{sec:Heavy} we discussed how to translate the partonic cross sections computed here into hadronic ones in the heavy flavor sector through the use of collinear fragmentation functions for open heavy flavor and the Improved Color Evaporation Model \cite{Ma:2016exq} for heavy quarkonia.  This procedure translates each of the partonic cross sections into a range of hadronic observables, allowing us to write down expressions for the production of: $(D \barD) \, (D \barD)$ -- Eq.~\eqref{e:DDbarDDbar}; $(D \barD) \, (J/\psi)$  -- Eq.~\eqref{e:JpsiDDbar}; $(J/\psi) \, (J/\psi)$  -- Eq.~\eqref{e:JpsiJpsi}; $(D \barD) \, h$ -- Eq.~\eqref{e:DDbarh}; and $(J/\psi) \, h$  -- Eq.~\eqref{e:Jpsih}, where $h$ is any light hadron.  These expressions open the door for a wide range of phenomenology to study the production and correlations of many particles in the heavy flavor sector, and they are the third primary result of this paper.

The ability to perform a small number of ab initio calculations in the CGC formalism at the partonic level to simultaneously predict the production cross sections and correlations of a wide variety of hadronic observables has the potential to broadly test the initial-state mechanisms as an explanation for the observed correlations in heavy- and heavy-light ion collisions.  Genuine multiparticle production computed within the CGC framework makes it possible to self-consistently compute higher cumulants such as $v_2\{4\}$ and charge-dependent correlations like $\gamma_{112}$ from purely initial-state mechanisms.  Similarly, the coordinate-space program begun in Ref.~\cite{Martinez:2018ygo} aspires to take partonic correlations such as these as inputs to the initial conditions of subsequent hydrodynamic evolution, including contributions from conserved charges due to quark production.  As we continue to extend this program of genuine multiparticle production, beyond simple approximations like the so-called ``dilute / dilute glasma graphs'' or the large-$N_c$ approximation, we anticipate that it will open up broad opportunities to test the effects of initial-state correlations, with and without the impact of strongly-coupled final-state interactions.

One potential barrier to such a comprehensive program of multi-hadron phenomenology in the CGC approach is that multiparticle cross sections, like the ones calculated here, invoke higher and higher $n$-point correlators of Wilson lines, up to the octupole $D_8$ for double-pair production.  These operators become increasingly difficult to evaluate, even in simple models like the MV model, for which analytic expressions are only available for the $4$-point functions~\cite{Dominguez:2012ad}.  However, a compelling argument summarized in Ref.~\cite{Altinoluk:2018ogz} and attributed to Ref.~\cite{Kovner:2018vec} suggests that, up to corrections suppressed by the large area of the target, $n$-point correlators can in general be factorized into products of 2-point correlators (dipoles), which are well-constrained in theory and phenomenology.  If this argument holds, then the increasingly complex Wilson line structure is no obstacle to the pursuit of multi-hadron phenomenology.

Aside from the applications already discussed above, there are a number of other direct extensions of this method we can pursue in future work.  One is to repeat the double-pair calculation of Sec.~\ref{sec:dblpair} in coordinate space to study the spatial correlations among quarks and antiquarks; as discussed in Ref.~\cite{Martinez:2018ygo}, these double-pair correlations are the dominant effect for same-sign charged particles and for opposite-sign charged particles at distances larger than the inverse quark mass $1/m$.  Another is to extend the calculations performed here for double-pair production to triple-pair production: $(q \barq) \, (q \barq) \, (q \barq)$.  The number of permutations will increase substantially in going to triple-pair production, but the fundamental mechanics of the calculation will not change, and the compact expression \eqref{e:ampl1} in momentum space makes such an extension tractable.  Moreover, the gluonic mapping \eqref{e:Gmap2} will make it possible to immediately translate the triple-pair cross section into a whole family of related cross sections:$(q \barq) \, (q \barq) \, (q \barq)$; $(q \barq) \, (q \barq) \, G$; $(q \barq) \, G \, G$; and $G \, G \, G$.

Last, we note that the expressions for hadronization in the heavy flavor sector we explore in Sec.~\ref{sec:Heavy} can be significantly improved and extended.  The heavy flavor sector is convenient as a justification for the assumption that a given hadron (like a $D$ meson) in the final state is dominated by fragmentation from a given parton (like a $c$ quark).  In principle, a sum over all partonic channels with appropriate fragmentation functions will relax this assumption and make it possible to study fragmentation into several identified hadrons.  As we extend the program to compute multiparticle production in the CGC framework, we will include more and more of these partonic channels, allowing a complete calculation of correlations in inclusive hadron production.  In particular, the CGC contribution to the charge-dependent correlations $\gamma_{112}$ and $\gamma_{123}$ which form the background to the signal for the chiral magnetic effect is of special importance.  While some exploratory work on this subject was done in Ref.~\cite{Kovner:2017gab}, it includes neither scattering in the target fields nor hadronization, both of which are likely to strongly modify the charge-dependent correlations.  However, with the calculation of a range of partonic channels and appropriate charge-dependent fragmentation functions, a robust computation of the hadronic charge correlations becomes possible.  For all of these reasons, we believe that the theoretical advancements presented in this paper represent a significant step toward implementing a comprehensive program of phenomenology to study multi-hadron correlations from the initial state.

%

\acknowledgements{The authors wish to thank N. Armesto, D. Pitonyak, J. Noronha-Hostler, V. Skokov, P. Tribedy, and R. Venugopalan for useful discussions. This work is supported in part by the U.S. Department of Energy grant DE-FG02-03ER41260 and the BEST (Beam Energy Scan Theory) DOE Topical Collaboration (MM), DOE Contract No. DE-AC52-06NA25396 and the DOE Early Career Program (MS), 
the European Research Council 39 grant HotLHC ERC-2011-StG-279579, Ministerio de Ciencia e Innovaci\'on of Spain under project FPA2014-58293-C2-1-P and Unidad de Excelencia Mar\'ia de Maeztu under project MDM-2016-0692, Xunta de Galicia (Conseller\'ia de Educaci\'on) and FEDER (DW).}
%

\appendix

%
\section{Color Averaging in the Projectile and Target}
\label{app:Averaging} 
%

In calculating cross sections, we will need to compute squares and interferences of the elementary building block \eqref{e:ampl1} and average them over various fluctuating quantities.  Aside from the averaging over the quantum numbers of the produced particles, which is performed in the usual way, the event averaging covers three distinct types of fluctuations.  These are fluctuations in the color fields of the projectile, fluctuations in the color fields of the target, and fluctuations over the global collision geometry.  These three types of averaging factorize from one another, such that we can average the sources $\rho$ in the color fields of the projectile, which we denote $\langle \cdots \rangle_{\mathrm{proj}}$, separately from the averaging of the Wilson lines over the color fields in the target, which we denote $\langle \cdots \rangle_{\mathrm{tgt}}$ or simply $\langle \cdots \rangle$ when there is no ambiguity.  The collision geometry is characterized by an impact parameter $\ul{B}$ between the centers of the projectile and target, which can be either held fixed at the cross section level or integrated out at the end of the calculation.  Similarly, there may be other parameters describing the overall collision geometry, such as the angular orientation of a non-spherical nucleus like uranium; these global parameters will also be integrated out at the cross section level.

In this paper, we make no particular assumption about the nature of the averaging in the color fields of the target; our final expressions for the cross sections will involve various traces of Wilson lines \eqref{e:Wdef} corresponding to color dipole, quadrupole, sextupole, and octupole operators.  We denote those corresponding operators by
\begin{subequations} \label{e:multipoles}
\begin{align}
\Dtwohat{x}{y} &\equiv \frac{1}{N_c} \tr[ V_{\ul x} V_{\ul y}^\dagger ] \\
\Dfourhat{x}{y}{z}{w} &\equiv \frac{1}{N_c} \tr[ V_{\ul x} V_{\ul y}^\dagger V_{\ul z} V_{\ul w}^\dagger ] \\
\Dsixhat{x}{y}{z}{w}{u}{v} &\equiv \frac{1}{N_c} \tr[ V_{\ul x} V_{\ul y}^\dagger V_{\ul z} V_{\ul w}^\dagger 
V_{\ul u} V_{\ul v}^\dagger] \\
\Deighthat{x}{y}{z}{w}{u}{v}{r}{s} &\equiv \frac{1}{N_c} \tr[ V_{\ul x} V_{\ul y}^\dagger V_{\ul z} V_{\ul w}^\dagger 
V_{\ul u} V_{\ul v}^\dagger V_{\ul r} V_{\ul s}^\dagger].
\end{align}
\end{subequations}
For quantities that have already been averaged we drop the hat over the operator, writing e.g. $\Dtwo{x}{y} = \langle \Dtwohat{x}{y} \rangle$.  When computing similar traces with adjoint Wilson lines, we will denote the operator with the superscript ``$adj$'', writing e.g. 
\begin{align*}
\hat{D}_2^{adj} (\ul{x} , \ul{y}) \equiv \frac{1}{N_c^2 - 1} U_{\ul x}^{a b} \left( U^\dagger_{\ul y} \right)^{b a} .
\end{align*}
And we will maintain the same notation whether invoking Wilson lines in coordinate space or momentum space, writing e.g. $\Dtwohat{p}{q} = \frac{1}{N_c} \tr[ V(p) V^\dagger (q) ]$ in momentum space.

It is also important to note that the (averaged) Wilson line traces from Eqs.~\eqref{e:multipoles} implicitly depend on a rapidity scale $Y$.  This rapidity scale regulates the light-cone divergences associated with higher-order corrections to these operators, and there are a range of different schemes available to regulate these divergences.  Physically, we can think of this scale as being set by the total rapidity interval $Y \propto \ln\frac{s}{\bot^2}$ of the collision, and the quantum evolution with the running of this scale is given by the JIMWLK evolution equations \cite{JalilianMarian:1996xn,JalilianMarian:1997gr,Iancu:2001ad} (or the large-$N_c$ analogue, the BK equation~\cite{Kovchegov:1999yj,Balitsky:1995ub}).  This evolution is triggered when the rapidity interval is parametrically large, $Y \sim \tfrac{1}{\alpha_s}$.  On the other hand, we restrict ourselves here to the case when the produced particles are close enough in rapidity that we do not need to consider quantum evolution in the rapidity interval between the particles.  Formally, this means $\Delta y_{i j} < \tfrac{1}{\alpha_s}$ for the rapidity interval $\Delta y_{i j}$ between any two particles $i, j$ tagged at mid-rapidity.

For the average $\langle \cdots \rangle_{\mathrm{proj}}$ over projectile color fields, we'll use a Gaussian averaging procedure inspired by the McLerran-Venugopalan (MV) model \cite{McLerran:1993ni}.  Gaussian averaging corresponds to limiting the interaction of a source particle to two gluons, which is the lowest order in perturbation theory that preserves color neutrality.  For Gaussian color averaging, the expectation value of products of several $\rho$'s factorizes into a sum over all possible pairwise ``contractions,'' such that it is only necessary to specify the two-point function $\langle \rho \, \rho \rangle$ to fully specify the result of the averaging procedure:
\begin{align}
\langle \rho^a (x) &\, \rho^b (y) \, \rho^{c *} (z) \, \rho^{d *} (w) \rangle_{\mathrm{proj}} =
\left\langle \rho^a (x) \, \rho^b (y) \right\rangle_{\mathrm{proj}} \, 
\left\langle \rho^{c *} (z) \, \rho^{d *} (w) \right\rangle_{\mathrm{proj}} 
\notag \\ &+
\left\langle \rho^a (x) \, \rho^{c *} (z) \right\rangle_{\mathrm{proj}} \, 
\left\langle \rho^b (y) \, \rho^{d *} (w) \right\rangle_{\mathrm{proj}} +
\left\langle \rho^a (x) \, \rho^{d *} (w) \right\rangle_{\mathrm{proj}} \, 
\left\langle \rho^b (y) \, \rho^{c *} (z) \right\rangle_{\mathrm{proj}} .
\end{align}

The original MV model \cite{McLerran:1993ni} has been generalized in a number of different ways in the literature; for our purposes, it is useful to enumerate three distinct physical assumptions about the two-point function:
\begin{subequations} \label{e:avgs}
\begin{align}
\left\langle \rho^a (x) \, \rho^{b \, *} (y) \right\rangle_{\mathrm{proj}} =
\begin{cases}
\delta^{a b} \, \delta^2 ( \ul{x} - \ul{y} ) \, \delta(x^- - y^-) \: \mu^2 (\ul{x} , x^-) & 
\mathrm{Locality}
\\
\delta^{a b} \, \delta(x^- - y^-) \: \mu^2 ( |x - y|_T^2 , x^-) & 
\mathrm{2D \: Transl. \: Inv.}
\\
\delta^{a b} \, \delta^2 ( \ul{x} - \ul{y} ) \, \delta(x^- - y^-) \: \mu^2 (x^-) & 
\mathrm{Both}
\end{cases}
\\ \notag \\
\left\langle \rho^a (q_1) \, \rho^{b \, *} (q_2) \right\rangle_{\mathrm{proj}} =
\begin{cases}
\delta^{a b} \: \mu^2 (\ul{q_1} - \ul{q_2} , q_1^+ - q_2^+) & 
\mathrm{Locality}
\\
\delta^{a b} \, (2\pi)^2 \delta^2 (\ul{q_1} - \ul{q_2}) \: \mu^2 (q_{1T}^2 , q_1^+ - q_2^+) & 
\mathrm{2D \: Transl. \: Inv.}
\\
\delta^{a b} \, (2\pi)^2 \delta^2 (\ul{q_1} - \ul{q_2}) \: \mu^2 (q_1^+ - q_2^+)  & 
\mathrm{Both} .
\end{cases}
\end{align}
\end{subequations}
For averages with both of the color fields in the amplitude $\langle \rho \, \rho \rangle$, the momentum of the second source is reversed: $q_2 \rightarrow - q_2$, and for averages with both of the color fields in the complex conjugate amplitude $\langle \rho^* \, \rho^* \rangle$, the momentum of the first source is reversed: $q_1 \rightarrow - q_1$.

The first case in Eqs.~\eqref{e:avgs}, ``Locality,'' is the one we will employ throughout this paper.  This assumption describes color fluctuations characterized by Gaussian random noise: they are only correlated locally at the same point, with different points in space having totally uncorrelated random fluctuations.  The second case, ``2D Translational Invariance,'' does not require locality, but does assume that the average distribution of source charges is uniform in the transverse plane; this assumption is employed in references such as \cite{Altinoluk:2015uaa} and \cite{Altinoluk:2016vax}.  The simplest case uses ``Both'' locality and 2D translational invariance; this assumption is the one employed by the original MV model \cite{McLerran:1993ni} and generalized somewhat in Ref.~\cite{Lappi:2007ku}.  An interesting argument about the general structure of the two-point function requiring only very weak assumptions about color neutrality was recently given in Ref.~\cite{Kovner:2017ssr}; for additional discussion about the properties of the two-point function see e.g. Refs.~\cite{Lappi:2007ku} and \cite{Altinoluk:2018ogz}.

Whatever physical assumptions are made about the two-point function in the Gaussian averaging, the color charge density fluctuations are characterized by the scale $\mu^2$, which is related to the saturation scale of the projectile $Q_{s, \mathrm{proj}}^2 \propto \mu^2$~\cite{Kovchegov:2012mbw, Kovchegov:1996ty}.  As defined in \eqref{e:avgs}, the scale $\mu^2$ effectively contains a factor of the coupling $g^2$ associated with radiating a soft gluon from the color sources; some references prefer to write this factor explicitly, but we will use the conventions of \eqref{e:avgs} in which that coupling is contained within $\mu^2$.  Note also that the dimensions of $\mu^2$ as written in \eqref{e:avgs} are different among the different physical assumptions.

It should also be emphasized that the averaging performed here for the projectile, denoted $\langle \cdots \rangle_{\mathrm{proj}}$, refers only to averaging over color configurations at a fixed collision geometry.  The scale $\mu^2$ in \eqref{e:avgs} is written with a dependence on the transverse position $\ul{x}$, which implicitly keeps fixed the global impact parameter $\ul{B}$ between the projectile and target.  In translating back from the continuous color charge densities used in \eqref{e:avgs} to the discrete valence quark distributions used for example in Ref.~\cite{Martinez:2018ygo}, the corresponding dictionary is
\begin{align} \label{e:mumap}
\mu^2 (\ul{b_1})\cdots \mu^2 (\ul{b_n}) \rightarrow
\int d^2 B \: \left[ \frac{g^2}{2 N_c} T_{\mathrm{proj}} (\ul{b_1} - \ul{B}) \right] \cdots
\left[ \frac{g^2}{2 N_c} T_{\mathrm{proj}} (\ul{b_n} - \ul{B}) \right] ,
\end{align}
where the factor of $g^2$ accounts for the coupling constant in the emission of the soft gluon from the valence quark source and the $\frac{1}{2 N_c}$ arises from averaging over the color states of the valence quark: $\frac{1}{N_c} \tr_c[ t^a t^b ] = \frac{1}{2 N_c} \delta^{a b}$.

%
\section{Wilson Line Color Algebra in Momentum Space}
\label{app:WLines} 
%

The cancellation of Wilson lines which is trivial in coordinate space takes on a more subtle form in momentum space:
\begin{align} \label{e:Wcancel1}
1 = V_{\ul x} V_{\ul x}^\dagger = \int \frac{d^2 p}{(2\pi)^2} \frac{d^2 q}{(2\pi)^2} \, e^{i (\ul{p} - \ul{q}) \cdot \ul{x}} \: V(p) \, V^\dagger (q)
\end{align}
Clearly we can't just cancel momentum-space Wilson lines of equal argument on the right-hand side: $V(p) V^\dagger (p) \neq 1$.  In fact, there are two actions taking place on the momentum-space Wilson lines resulting in the cancellation: a Fourier transform over the relative momentum $(p-q)$ and integral over the center-of-mass momentum $\frac{p + q}{2}$.  The only way for these two integrals to lead to unity on the left-hand is if one of these actions results in a delta function, which the other one picks up.  But at this level it is not clear which operation should generate the delta function.

On the other hand, we can engineer a cancellation of Wilson lines in momentum space by explicitly generating Wilson lines in coordinate space with the same argument:
\begin{align}
V(p) V^\dagger (p + q) = \int d^2 x \, d^2 y \, e^{-i \ul{p} \cdot \ul{x}} \, e^{i (\ul{p} + \ul{q}) \cdot \ul{y}} \: V_{\ul x} V_{\ul y}^\dagger .
\end{align}
Clearly if we integrate over the shared momentum $p$, we generate a delta function that sets $\ul{x} = \ul{y}$ and cancels the Wilson lines.  The remaining integral over $d^2 y$ then generates a new delta function:
\begin{align} \label{e:Wcancel2}
\int \frac{d^2 p}{(2\pi)^2} V(p) V^\dagger (p + q) = 
\int d^2 y \, e^{i \ul{q} \cdot \ul{y}} \: V_{\ul y} V_{\ul y}^\dagger = (2\pi)^2 \delta^2 (\ul{q}) .
\end{align}
This condition is necessary, and as it turns out, it is also sufficient to guarantee \eqref{e:Wcancel1}.  Using \eqref{e:Wcancel2} in \eqref{e:Wcancel1} we obtain
\begin{align}
V_{\ul x} V_{\ul x}^\dagger &= \int \frac{d^2 p}{(2\pi)^2} \frac{d^2 q}{(2\pi)^2} \, e^{- i \ul{q} \cdot \ul{x}} \: V(p) \, V^\dagger (p + q)
\notag \\ &=
\int \frac{d^2 q}{(2\pi)^2} \, e^{- i \ul{q} \cdot \ul{x}} \: (2\pi)^2 \delta^2 (\ul{q})
\notag \\ &=
1 ,
\end{align}
so we can conclude that the two conditions are in fact equivalent:
\begin{align} \label{e:Wcancel3}
\Big[ \int \frac{d^2 p}{(2\pi)^2} V(p) V^\dagger (p + q) = (2\pi)^2 \delta^2 (\ul{q}) \Big]
\leftrightarrow
\Big[ V_{\ul x} V_{\ul x}^\dagger = 1 \Big] .
\end{align}

Another important feature is the conversion between Wilson line traces in the fundamental and adjoint representations.  In coordinate space, the squares of fundamental dipoles and quadrupoles can be directly converted into adjoint dipoles and quadrupoles, plus a constant term which is $N_c$ suppressed:
\begin{subequations} \label{e:Wconvert1}
\begin{align}
\hat{D}_2^{adj} (\ul{x} , \ul{y}) &\equiv \frac{1}{N_c^2 - 1} U_{\ul x}^{a b} \left( U_{\ul y}^\dagger\right)^{b a}
=
\frac{N_c}{2 C_F} \left| \hat{D}_2 (\ul{x} , \ul{y}) \right|^2 - \frac{1}{N_c^2 - 1}
\\ \notag \\
\hat{D}_4^{adj} (\ul{x} , \ul{y} , \ul{z} , \ul{w}) &\equiv \frac{1}{N_c^2 - 1} 
U_{\ul x}^{a b} \left( U_{\ul y}^\dagger \right)^{b c} U_{\ul z}^{c d} \left( U_{\ul w}^\dagger \right)^{d a}
=
\frac{N_c}{2 C_F} \left| \hat{D}_4 (\ul{x} , \ul{y} , \ul{z} , \ul{w}) \right|^2 - \frac{1}{N_c^2 - 1} .
\end{align}
\end{subequations}
In going to momentum space, that additive constant becomes instead proportional to delta functions:
\begin{subequations} \label{e:Wconvert2}
\begin{align}
\hat{D}_2^{adj} (p , q) &= \frac{N_c}{2 C_F} \left| \hat{D}_2 \right|^2 (p , q) - \frac{1}{N_c^2 - 1} \,
(2\pi)^4 \, \delta^2 (p) \, \delta^2 (q)
\\
\hat{D}_4^{adj} (p , q , p' , q') &= \frac{N_c}{2 C_F} \left| \hat{D}_4 \right|^2 (p , q , p' , q') - \frac{1}{N_c^2 - 1} \,
(2\pi)^8 \, \delta^2 (p) \, \delta^2 (q) \, \delta^2(p') \, \delta^2(q') .
\end{align}
\end{subequations}

%
%

\end{document}